\documentclass[a4paper,11pt]{article}
\usepackage{jheppub} 
\DeclareMathAlphabet\bfcal{OMS}{cmsy}{b}{n}

\usepackage{dcolumn}
\usepackage{bm}
\usepackage{epsfig}
\usepackage{graphicx}
\usepackage{amsmath}
\usepackage{amssymb}
\usepackage{mathrsfs}
\usepackage{color}
\usepackage[usenames,dvipsnames]{xcolor}
\usepackage{hyperref}
\usepackage{feynmp-auto}
\usepackage{tikz}
\usepackage{braket}
\usepackage[caption=false]{subfig}
\usepackage{multirow}
\usepackage{array}
\usepackage{slashed}
\usepackage{rotating}
\usepackage{caption}
\usepackage{subcaption}
\usepackage{bbold}
\usepackage{xcolor}

\usepackage[normalem]{ulem}
\usepackage{marginnote}

\newcommand\identity{1\kern-0.25em\text{l}}
\newcommand\vv{\text{v}}

\newcommand\pv{{\bf p}}
\newcommand\qv{{\bf q}}

\newcommand\Dv{{\bf D}}
\newcommand\bv{{\bf b}}

\newcommand\rv{{\bf r}}
\newcommand\Pc{{\mathcal P}}

\newcommand{\Otsosing}{\braket{\mathcal{O}^{V}(^3S_1^{[1]})}}
\newcommand{\Otsooct}{\braket{\mathcal{O}^{V}(^3S_1^{[8]})}}
\newcommand{\Oosz}{\braket{\mathcal{O}^{V}(^1S_0^{[8]})}}
\newcommand{\Otpz}{\braket{\mathcal{O}^{V}(^3P_J^{[8]})}}

\DeclareMathOperator*{\sumint}{%
\mathchoice%
  {\ooalign{$\displaystyle\sum$\cr\hidewidth$\displaystyle\int$\hidewidth\cr}}
  {\ooalign{\raisebox{.14\height}{\scalebox{.7}{$\textstyle\sum$}}\cr\hidewidth$\textstyle\int$\hidewidth\cr}}
  {\ooalign{\raisebox{.2\height}{\scalebox{.6}{$\scriptstyle\sum$}}\cr$\scriptstyle\int$\cr}}
  {\ooalign{\raisebox{.2\height}{\scalebox{.6}{$\scriptstyle\sum$}}\cr$\scriptstyle\int$\cr}}
}

\title{\boldmath Factorizing quarkonium production matrix elements using effective field theory}

\author[a]{Marston Copeland,}
\author[a]{Ivan Vitev}

\affiliation[a]{Theoretical Division, MS. B283, Los Alamos National Laboratory,\\  Los Alamos, NM 87545, USA}

\emailAdd{pmcopeland@lanl.gov}

\abstract{We use effective field theory to factorize production matrix elements that appear in quarkonium cross sections in NRQCD. By applying a Hubbard–Stratonovich transformation we show that the soft and ultrasoft sectors of NRQCD can be decoupled from the heavy quark and antiquark fields in a hybrid vNRQCD/pNRQCD Lagrangian at leading order in the velocity power counting. This enables us to separate quarkonium production matrix elements in terms of matrix elements of color-singlet composite fields, which we can write as the wavefunction at the origin, and state independent vacuum correlators of chromo-electric and chromo-magnetic gluon fields. This approach verifies powerful connections between the LDMEs of different S-wave vector quarkonium states, originally derived using pNRQCD.  Additionally, we find new operator contributions for the color-octet P-wave mechanism, which satisfy a similar set of relationships. Finally, this approach allows us to factorize the production matrix elements that appear in the transverse momentum dependent factorization framework, known as TMD soft transition functions, in terms of state independent gluon correlators. This work restores some universality for TMD production operators and dramatically improves the predictive power of NRQCD in the TMD framework.}

\begin{document}

\noindent\makebox[\dimexpr\linewidth-1cm\relax][r]{LA-UR-26-24327}

\maketitle
\flushbottom

\section{Introduction}
Since the discovery of the $J/\psi$ meson in 1974~\cite{SLAC-SP-017:1974ind, E598:1974sol}, quarkonia have been instrumental in advancing our understanding of quantum chromodynamics (QCD). Their relatively simple internal structure makes them  ideal candidates for studying bound state formation under strong interactions. In particular, quarkonia are non-relativistic bound states composed of heavy quarks with  large masses and slow relative motion within the system. This means that quarkonium bound state formation is a process that contains many different energy scales - the heavy quark masses ($2m$), the relative momentum of the $Q\bar{Q}$ ($m\vv$), and the binding energy, ($m \vv^2$). When the relative velocity of the heavy quarks is small, $\vv \ll1$, as it is expected to be for a near-Coulombic system like quarkonium, these scales are widely separated. This enables the application of effective field theory (EFT). 

One of the first attempts to exploit this separation of scales came from the EFT known as non-relativistic QCD (NRQCD) \cite{Bodwin:1994jh}. This effective theory, while wildy successful in describing some aspects of quarkonium physics, treated all IR physics below the ``hard" scale, $2m$, as equivalent. It did not properly delineate between the ``soft" ($m\vv$) and ``ultrasoft" ($m \vv^2$) scales, which are in principle also widely separated. This led to a number of difficulties and, in particular, it made the systematic calculation of corrections to the Coulomb potential challenging. 
 The question of how to properly separate the soft and ultrasoft scales in NRQCD led to two new EFTs, ``potential NRQCD" (pNRQCD) \cite{Brambilla:1999xf} and ``velocity NRQCD" (vNRQCD) \cite{Luke:1999kz, Manohar:1999xd}. The pNRQCD approach is to successively integrate out the scales of the theory, starting by integrating out the hard scale and matching onto NRQCD, and then integrating out the soft scale to match onto pNRQCD, which is written in terms of composite quark and antiquark fields. The vNRQCD approach differs in that it simultaneously keeps both the soft and ultrasoft degrees of freedom in the low-energy theory, but separates gluon fields whose components scale homogeneously as soft or ultrasoft into distinct fields via a mode expansion. These two formalisms, while different in their approach, have thus far been shown to be equivalent for all calculations \cite{Manohar:1999xd, Fleming:2005pd, Peset:2015vvi, Hoang:2002yy}. 

Both of these improved versions of NRQCD were developed to systematically compute corrections to the potential between heavy quarks. In fact, the vNRQCD formalism has the advantage of the ``velocity renormalization group equation" which allows one to simultaneously sum large logarithms of the soft and ultrasoft scales to all orders in perturbation theory, therefore improving the perturbative calculations of the potential \cite{Manohar:1999xd}. Historically however, other aspects of quarkonium physics - such as production and decays - have been neglected from pNRQCD and vNRQCD. Instead, the original NRQCD has been the main formalism for studying these processes. In part, this is because the production/decay of a $Q\bar{Q}$ really only concerns physics at, or above, the hard scale $2m$. Physics below the hard scale, i.e. physics occurring at the soft or ultrasoft scales, is related to the bound state formation and hadronization dynamics of the $Q\bar{Q}$ into the final state quarkonium, and can not dramatically change say, the $P_T$ or energy spectrum of a quarkonium state (if the $P_T$ or energy of the quarkonium is large). More specifically, in production, if the $P_T$ of the measured quarkonium is greater than or equal to the mass of the heavy $Q\bar{Q}$ pair (as it would be in a collinear factorization framework), then this $P_T$ must have been generated during the hard process - i.e., the production of the $Q\bar{Q}$. Hence, soft and ultrasoft radiation, with momentum on the order of $m\vv, m\vv^2 \ll 2m$, cannot considerably change final $P_T$ of the quarkonium. Therefore, the production cross section can be determined by matching onto a theory that describes IR physics below the scale $2m$, but does nothing else; i.e., NRQCD. In the original line of thinking, it did not seem necessary to go further and study the dynamics of the soft and ultrasoft degrees of freedom, so the pNRQCD and vNRQCD approaches were not necessary. 

The application of NRQCD to quarkonium production has had a long and storied history by this point - with mixed success \cite{Fleming:1997fq, Yuan:2000cn, Beneke:1998re, Chu:2024fpo, Maxia:2024cjh,Bodwin:2010fi, Braaten:2002fi, Hagiwara:2003cw, Bodwin:2008nf,Braaten:1994xb, Braaten:1994vv,Baumgart:2014upa,Bain:2016clc,Bain:2017wvk,Kang:2017yde,Dai:2017cjq,Wang:2025drz, Copeland:2025osx,Flett:2024htj,Ivanov:2004vd,Chen:2019uit,Blask:2025jua,Sharma:2012dy,Yao:2018nmy, Yao:2020kqy, Yang:2024ejk,Hoang:2002ae, Brambilla:2010cs}. The main innovation of NRQCD for production cross sections was that it allowed for ``color-octet" production mechanisms, which solved a number of problems that stemmed from the much simpler color-singlet model. These color-octet production mechanisms allowed for $Q\bar{Q}$ pairs to be generated during the hard process with quantum numbers that differed from the final state quarkonium (notably, it allowed for the pairs to be produced with color, while obviously the final state quarkonium is color-less). These different mechanisms are weighted by non-perturbative production matrix elements in the effective theory known as the long distance matrix elements (LDMEs) of NRQCD. The LDMEs describe the probability for $Q\bar{Q}$ pairs in some initial configuration to transition into the final state quarkonium particle. For spin triplet S-wave quarkonia, like the $J/\psi$, $\psi(2S)$, and $\Upsilon(nS)$ states, the leading order matrix element is the $\Otsosing$ (where $V$ indicates the vector quarkonium state). This describes the probability for a $^3S_1^{[1]}$ $Q\bar{Q}$ pair to transition to a $^3S_1^{[1]}$ quarkonium. Since the quantum numbers of the $Q\bar{Q}$ already match those of the quarkonium, power-suppressed spin/color/angluar momentum-flipping operator insertions in the EFT are not necessary, so $\Otsosing$ is leading in the $\vv$ power-counting of NRQCD. S-wave vector quarkonia then have three subleading LDMEs which appear via the aforementioned color-octet mechanisms. These are denoted by $\Otsooct$, $\Oosz$, and $\Otpz$. Unlike for the $^3S_1^{[1]}$ LDME, these matrix elements experience power suppression because NRQCD operators must be inserted via a time-ordered product with the effective interaction Lagrangian in order to place the $Q\bar{Q}$ in a state that matches the final quantum numbers of the quarkonium. This is necessary for hadronization to occur. This quantum number flipping transition process is non-perturbative, however, so it suffices to treat these LDMEs as (universal) parameters in the theory. This has been the underlying philosophy of quarkonium production in NRQCD for the last 30 or so years. A major goal of this approach has been to constrain the LDMEs through a comparison with data, since not only are these matrix elements constant, but they are also process-independent. If they can be constrained, then only process-dependent (perturbative) partonic cross sections describing the production of a $Q\bar{Q}$ are needed in order to make first principles predictions. 

Recently, however, there has been a paradigm shift. It has been realized in several distinct instances that soft and ultrasoft dynamics can play a critical role during quarkonium production. Most obviously, in the TMD factorization framework \cite{Boussarie:2023izj}, when a final state quarkonium is produced with $P_T \sim m\vv$, the transverse momentum created during the production process stems almost entirely from soft gluon radiation. This has led several authors to write down new $P_T$ dependent color-octet production operators that replace the LDMEs, known as TMD shape functions (TMDShFs) \cite{Boer:2023zit, Maxia:2025zee, Maxia:2024cjh, Echevarria:2019ynx,Echevarria:2024idp, Echevarria:2025oab}. In order to derive these operators from first principles and prove the factorization theorems that they appear in, one needs to use a combination of vNRQCD and soft collinear effective theory \cite{Echevarria:2019ynx,Echevarria:2024idp}. Traditional NRQCD is not sufficient. Moreover, a more careful investigation of the soft radiation showed that the TMDShFs are not even the dominant production operators. Completely new operators are necessary to encode the transverse momentum generated from the quantum number flipping transitions that place the $Q\bar{Q}$ in a color singlet. This leads to different production matrix elements known as TMD soft transition functions (TMDSTFs) \cite{Copeland:2025vop}. One problem with both the TMDShFs and TMDSTFs, however, is that they both encode soft radiation from other parts of the collision which cannot be separated from the soft radiation coming from the $Q\bar{Q}$. Therefore, these matrix elements have some inherent process dependence which spoils universality. This problem only arises if the hard scale of the collision $Q$ is similar to the mass of the quarkonium and the $P_T$ of the quarkonium is on the order of the soft scale of NRQCD. If $P_T \sim 2m \ll Q$, then the TMD production operators are matched onto the usual LDMEs \cite{Echevarria:2020qjk, Copeland:2023wbu, Copeland:2023qed, Echevarria:2023dme} and universality holds. 

In the collinear factorization framework, an analysis using the pNRQCD formalism led to results that are even more 
unorthodox~\cite{Brambilla:2022ayc}. In this work, it was shown that, by integrating out the soft dynamics in the LDMEs and matching onto the composite fields of pNRQCD, one can rewrite the matrix elements in terms of universal vacuum correlators of chromo-magnetic and chromo-electric field operators times the S-wave quarkonia's wave function at the origin. This relationship implies a factorization of scales in the LDMEs - showing that, within the pNRQCD framework, the soft and ultrasoft physics can be separated in the production matrix elements. If true, this result is very powerful and leads to strong constraints and relations between the different LDMEs of S-wave quarkonia, reducing the number of free parameters in NRQCD production cross sections from 12 to 3 \cite{Brambilla:2022ayc, Brambilla:2024iqg, Brambilla:2022rjd}. 


Upon a naive inspection, it is not clear how these results can be replicated using the vNRQCD approach, which is concerning given that these two approaches are supposedly equivalent. This is because the soft fields exist simultaneously with the potential and ultrasoft scaling fields in the vNRQCD formalism, so the heavy quarks and antiquarks cannot be easily decoupled from soft gluons. This prevents the soft and ultrasoft Hilbert spaces in production matrix elements from being separated when using vNRQCD. This problem is also directly related to the apparent process dependence that appears in TMD production matrix elements for quarkonium, like the TMDShFs and the TMDSTFs \cite{Echevarria:2024idp, Copeland:2025vop}. 


In this work, we address the apparent tension between the pNRQCD and vNRQCD formalisms and ``factorize" the LDMEs into soft and ultrasoft components using vNRQCD. To do this, we use a technique first introduced to study heavy di-quark systems in vNRQCD \cite{Fleming:2005pd}, performing a Hubbard-Stratonovich transformation that introduces composite color-singlet and color-octet fields into the vNRQCD Lagrangian. We find that the soft-sector can only be decoupled in production matrix elements due to a unique feature of the calculation - the quark/antiquark fields are produced at a point. 

With this motivation, the rest of the paper is organized as follows. In section~\ref{sec: background} we introduce the necessary background material, discuss quarkonium production NRQCD, and review the vNRQCD framework. In section~\ref{sec: decouple} we review the Hubbard-Stratonovich transformation technique in vNRQCD, first introduced in Ref.~\cite{Fleming:2005pd}, and show how it can be applied to production. In section~\ref{sec: factorization} we use these techniques to re-derive the important results that factorize the LDMEs, first shown using the pNRQCD approach, and obtain new operator contributions for the P-wave states. In section~\ref{sec: discussion} we discuss  applications of our calculation and show that ultrasoft power-corrections are smaller that previously estimated and should not seriously change our results. We then apply these techniques to decouple the soft and ultrasoft scales in the TMDSTFs, creating new relations between the leading TMD production matrix elements for different S-wave quark quarkonium states. Finally, in section~\ref{sec: conclusion}, we conclude and discuss next steps. Selected definitions and technical aspects are collected in the Appendix. 

\section{Background}
\label{sec: background}
%
In this section we review the fundamental assumptions of quarkonium production in NRQCD. We also review the vNRQCD formalism and the soft and ultrasoft scales within the effective theory. 

\subsection{NRQCD}
NRQCD is an EFT developed to study non-relativistic bound states of heavy quarks in QCD. The power-counting parameter of NRQCD is the small relative velocity of the $Q\bar{Q}$ pair, $\vv \ll 1$. This means that higher order operators in the EFT should be systematically suppressed by powers of $\vv$ with respect to the leading order kinetic operators, allowing the relevant non-perturbative effects to be systematically organized. For charmonium systems $\vv \sim 0.5$, and for bottomonium $\vv\sim 0.3$ \cite{Fleming:2000ib}. As such, the leading order operators in NRQCD are more effective at describing bottomonium systems, like $\Upsilon$ states, than charmonium like the $J/\psi$. Power corrections in $\vv$ for charmonium systems are typically found to be large \cite{Blask:2025jua}. The first few terms in the NRQCD Lagrangian are given below \cite{Bodwin:1994jh}
\begin{equation}
    \begin{aligned}
        {\mathcal L}_{\text{NRQCD}} = & \; \psi^\dagger\bigg(iD_t + \frac{{\bf D}^2}{2m}\bigg)\psi + \chi^\dagger \bigg(iD_t - \frac{{\bf D}^2}{2m}\bigg)\chi\\
        &\, + \frac{c_F}{2m}[\psi^\dagger(g{\bf B}\cdot \boldsymbol{\sigma})\psi - (\psi \rightarrow \chi)] \\
        & \; + \frac{c_2}{8m^2}[\psi^\dagger(\Dv \cdot g{\bf E}-g{\bf E}\cdot \Dv)\psi + (\psi \rightarrow \chi)] \\
        & \; + \frac{c_3}{8m^2}[\psi^\dagger(i\Dv\times g{\bf E} - g{\bf E}\times i\Dv) \cdot \boldsymbol{\sigma}\psi + (\psi \rightarrow \chi)] \\
        & \; + \frac{c_1}{8m^3}[\psi^\dagger(\Dv^2)^2\psi - (\psi \rightarrow \chi)] +\cdots \, .
    \end{aligned}
\end{equation}

As explained in the introduction, one area that NRQCD has been frequently applied to is quarkonium production. This is because in quarkonium production cross sections the creation of a heavy $Q\bar{Q}$ pair has to occur at least at the scale $2m$, which is much greater than hadronization dynamics which occur around $\Lambda_{QCD}$. Therefore, matching production cross sections onto NRQCD inherently ``factorizes" the cross section into a perturbatively calculable partonic cross section describing the hard production of the $Q\bar{Q}$ (this is the matching coefficient in the EFT) and low-energy, non-perturbative matrix elements which describe the hadronization of the $Q\bar{Q}$ into the final quarkonium state. For the production of S-wave vector quarkonia (denoted by $V$), one can generically write this factorization as,

\begin{equation}
\label{eq: NRQCD fact}
    {\rm d}\sigma_{A+B\to V+X}\ = \sum_n {\rm d}\hat{\sigma}_{A+B\to Q\bar{Q}(n)+X} \braket{\mathcal{O}^{V}(n)} \, .
\end{equation} 
This is usually referred to as the ``NRQCD factorization conjecture". In eq.~(\ref{eq: NRQCD fact}), ${\rm d}\hat{\sigma}$ represents the perturbatively calculable partonic cross section describing the production of a $Q\bar{Q}$ state in the quantum number configuration denoted using spectroscopic notation, $n = ~^{2s+1}L_J^{[c]}$. The partonic cross sections are weighted by a corresponding LDME, denoted by $\braket{\mathcal{O}^{V}(n)}$. The LDMEs are non-perturbative vacuum production matrix elements of NRQCD quark/antiquark operators in the quantum number configuration, $n$. Generically, for the quarkonium state $V$, they have the form 
\begin{equation}
    \braket{\mathcal{O}^{V}(n)} = \sum_X \bra{0}\chi^\dagger(0) [{\cal K} \cdot \Gamma]_n \psi (0)\ket{V, X}\bra{V,X} \psi^\dagger(0) [{\cal K} \cdot \Gamma]_n \chi(0)\ket{0}.
\label{eq: gen LDME}
\end{equation}
where the spin and color projectors $[{\cal K} \cdot \Gamma]_n$ place the operator in the configuration $n$. The LDMEs are completely non-perturbative parameters, but in the framework they are effectively universal constants to be fit to data. 

Several collaborations have attempted to constrain the LDMEs through global fits to quarkonium production data. In particular, $J/\psi$ production is experimentally  appealing to measure because of it's clean decay to two leptons, so there is a wealth of data to compare against the theoretical results. Different LDME fits to $J/\psi$ production data are presented in table \ref{tab: LDMEs}.

\begin{table}[htbp]
\centering
\begin{tabular}{c|c|c|c|c}
\hline
& $\begin{aligned}\Otsosing \\ \times \, \text{GeV}^3\end{aligned}$ & $\begin{aligned}\Otsooct \\ \times 10^{-2} \, \text{GeV}^3\end{aligned}$ & $\begin{aligned}\Oosz \\ \times 10^{-2} \, \text{GeV}^3\end{aligned}$ & $\begin{aligned}\Otpz/m_c^2 \\ \times 10^{-2} \, \text{GeV}^3\end{aligned}$\\
\hline
B \& K \cite{Butenschoen:2011yh,Butenschoen:2012qr} & $1.32 \pm 0.20$ & $0.224 \pm 0.59$ & $4.97 \pm 0.44$ & $-0.72 \pm 0.88$\\
Chao et al.~\cite{Chao:2012iv} & $1.16\pm 0.20$ & $0.30 \pm 0.12$ & $8.9 \pm 0.98$ & $0.56 \pm 0.21$ \\
Bodwin et al.~\cite{Bodwin:2014gia} & $1.32 \pm 0.20$ & $1.1 \pm 1.0$ & $9.9 \pm 2.2$ & $0.49 \pm 0.44$ \\
Brambilla et al.~\cite{Brambilla:2024iqg} & $1.16 \pm 0.20$ & $1.05 \pm 0.12$ & $0.07 \pm 0.25$ & $1.88 \pm 0.26$ \\
\hline
\end{tabular}
\caption{Different fits for the NRQCD LDMEs for $J/\psi$ production.\label{tab: LDMEs}}
\end{table}

Interestingly, each of the fits in table \ref{tab: LDMEs} find remarkably different results depending on the range of $p_T$ used, whether logarithms of $p_T$ are re-summed, and whether heavy quark symmetry constraints are applied. Perhaps more perplexing is the order of magnitude difference between the $\braket{{\cal O}^{J/\psi}(^1S_0^{[8]})}$ matrix element and the other color-octet LDMEs observed in each of the fits when, according to traditional NRQCD power counting, these matrix elements should all be around the same size.

In NRQCD, the power counting arguments for the LDMEs go as follows. Heavy quarks in color-octet configurations transition to color-singlet states via emitting additional ultrasoft gluons in the effective theory. Each emission corresponds to a time-ordered insertion of a subleading NRQCD operator in the LDME, creating additional v-suppression. For S-wave vector quarkonium states, the LDMEs have the following scalings:
\begin{itemize}
    \item $\Otsosing$: scales as $\vv^3$, requires no operator insertions;
    \item $\Otsooct$: scales as $\vv^7$, requires two chromoelectric transitions, i.e.~insertions of ${\bf A}\cdot \boldsymbol{\nabla}$, or one insertion of ${\bf A}^2$;
    \item $\Oosz$: scales as $\vv^7$, requires a single chromomagnetic transition, i.e.~an insertion of $\boldsymbol{\sigma} \cdot {\bf B}$;
    \item $\Otpz$: scales as $\vv^7$, requires a single chromoelectric transition and has further suppression due to its initial angular momentum quantum number.
\end{itemize}
In this analysis, we will revisit the traditional assumptions made for quarkonium production operators in an attempt to shed light on the discrepancies between the size of the matrix elements. 

\subsection{vNRQCD}
%
In the ordinary theory of quarkonium production, scales below $2m$ are not carefully considered. This means that the LDMEs contain physics in both the soft ($mv$) and ultrasoft ($mv^2$) sectors of the theory. In large part, this is because both of these scales are non-perturbative, and hence so are the low-energy operators. This makes it is difficult to make any analytical statements about the LDMEs. However, as we will see, a careful delineation of the scales in an effective field theory enables powerful factorization theorems to be derived within the production matrix elements. The delineation of these scales can be handled either using the pNRQCD framework \cite{Brambilla:1999xf} or the vNRQCD framework \cite{Luke:1999kz}. Since the vNRQCD framework explicitly maintains the quark and antiquark degrees of freedom in the Lagrangian, it is much easier to connect this EFT to prior quarkonium production work completed in NRQCD. Additionally, vNRQCD has the advantage of simultaneously keeping the soft and ultrasoft scales in the theory, which allows it to be easily paired with SCET \cite{Bauer:2000ew, Bauer:2000yr} to derive the factorization theorems for quarkonium cross sections. For these reasons, we find it important to understand how the soft and ultrasoft degrees of freedom can be separated in the context of the vNRQCD framework. 

In the theory of vNRQCD, gluons are separated into different ``modes", which scale as the momentum they carry. The two gluons that appear in the theory are those that can satisfy the onshell condition, $p^2 = 0$. These are the soft gluons, which scale as 
\begin{equation}
    A_\pv^\mu \sim (m\vv, m{\bf v}) ,
\end{equation}
and the ultrasoft gluons,
\begin{equation}
    A_{us}^\mu \sim (m\vv^2, m{\bf v}^2).
\end{equation}
Quarks have potential momentum scaling
\begin{equation}
    p_c \sim (m\vv^2, m{\bf v}),
\end{equation}
so as to satisfy a non-relativistic disperion relation. Gluons with potential momentum scaling are off shell since gluons are necessarily relativistic. Therefore, they are integrated out of the theory. vNRQCD uses a label formalism, where soft momentum is phased out via a field redefinition,
\begin{equation}
\begin{aligned}
    \psi(x)  = \sum_\pv e^{-i\pv\cdot {\bf x}} \psi_\pv(x), \\
    A^\mu(x) = \sum_\qv e^{-iq\cdot x}A_q^\mu(x) ,
\end{aligned}
\label{eq: labeled fields}
\end{equation}
and denoted by the label subscript. Ultrasoft gluons do not have a label. The label momentum operator $\mathcal{P}$ then projects out the soft momentum of a field:
\begin{equation}
    \boldsymbol{\mathcal P}^i \psi_{\pv} = {\bf p}^i \psi_\pv \, , \qquad \mathcal{P}^\mu A_q ^\nu = q^\mu A_q^\nu \; .
\label{eq: projector}
\end{equation}
The covariant derivative $D$ only acts on the ultrasoft momentum of fields and carries ultrasoft gluon fields, $iD=i\partial -gA$. Therefore,
\begin{equation}
    D^\mu \psi_\pv \sim (mv^2) \times \psi_\pv .
\end{equation}
Using the covariant derivative, one forms $G_{us}^{\mu\nu}=\frac{1}{ig}[D^\mu,D^\nu]$ and ${\bf B}_{us}^k = -\frac12\epsilon^{ijk}G_{us}^{ij}$.
With these pieces, the vNRQCD Lagrangian is defined as \cite{Luke:1999kz}
\begin{equation}
    \begin{aligned}
        {\mathcal L}_{\text{vNRQCD}} = & \; -\frac14 G_{us}^{\mu \nu}G_{us, \mu \nu} + \sum_p | p^\mu A_p^\nu - p^\nu A_p^\mu|^2 \\
        & \; + \sum_\pv \psi^\dagger_\pv \bigg( iD^0 - \frac{(\boldsymbol{\mathcal P}-i\Dv)^2}{2m} + \frac{\boldsymbol{\mathcal{P}}^4}{8m^3} + \frac{c_Fg}{2m}{\bf B} _{us}\cdot \boldsymbol{\sigma}\bigg) \psi_\pv + (\psi \rightarrow \chi) \\
        & \; -g^2 \sum_{q,q^\prime,\pv,\pv^\prime} \frac12 U_{\mu \nu}(\pv, \pv^\prime, q, q^\prime) \psi^\dagger_{\pv^\prime}[A_{q^\prime}^\mu, A_q^\nu] \psi_\pv  + (\psi \leftrightarrow \chi) \\
        & \; -g^2 \sum_{q,q^\prime,\pv,\pv^\prime} \frac12 W_{\mu \nu}(\pv, \pv^\prime, q, q^\prime) \psi^\dagger_{\pv^\prime}\{A_{q^\prime}^\mu, A_q^\nu\} \psi_\pv  + (\psi \leftrightarrow \chi) \\
        & \; - \sum_{\pv, \qv} V(\pv, \qv) \psi^\dagger_\qv T^a \psi_\pv \chi^\dagger_{-\pv}\bar{T}^a\chi_{-\qv}  \, .
    \end{aligned}
\label{eq: vNRQCD L}
\end{equation}
The coefficients $U_{\mu\nu}$ and $W_{\mu\nu}$ are derived from matching the Coulomb gluon exchange graphs in the full theory onto vNRQCD and can be found to ${\cal O}(\vv^2)$ in ref.~\cite{Manohar:1999xd}. The potential $V$ comes from integrating out off-shell gluons with Coulomb momentum scaling in gluon exchange graphs in the full theory \cite{Manohar:1999xd,Luke:1999kz}. One slight difference in our notation from the origianl vNRQCD Lagrangian of ref.~\cite{Luke:1999kz} is that in this work, $\chi_\qv$ is defined to create an antiquark, as opposed to $\chi^\dagger_\qv$. This is to align better with the notation used in NRQCD~\cite{Bodwin:1994jh}, which will be convenient when matching the production operators onto vNRQCD.
%

\subsection{A comment on quarkonium production operators}
\label{sec: zero radial}
When matching quarkonium production cross sections onto NRQCD, the production matrix elements of the effective theory are written as local quark-antiquark operators, as in eq.~(\ref{eq: gen LDME}).
The locality of the production matrix element can be understood by the matching procedure, where full QCD amplitudes describing the production of a $Q\bar{Q}$ pair are expanded order-by-order in $\vv$ so hard propagators carrying momentum at the scale $2m$ (or larger) are shrunk to a point and quark/antiquark spinors are also expanded, which can be matched onto operators that have been multipole expanded. Importantly, this means that the fields in the production operators have zero radial separation. While this result is universal, the steps to show that this is true are subtly different in the collinear factorization vs. TMD factorization approaches. For clarity, we will review the matching procedures in both approaches. 

\subsubsection{Collinear factorization}
In the collinear factorization framework the produced quarkonium necessarily carries hard momentum at or above the scale $M_{V} \sim 2m$. In light-cone coordinates, where any vector $v^\mu$ can be decomposed as  
$v^+ = (v^0+v^3)/\sqrt{2}, v^- = (v^0-v^3)/\sqrt{2},$ and ${\bf v}^i_T = (v^1, v^2)$, we can always boost to a frame where the components of the quarkonia's momentum scale as 
\begin{equation}
    (P^+_V,P^-_V,{\bf P}_{V,T} ) \gtrsim (m,m,m ).
\end{equation}
In general, any quarkonium production process will require at least one heavy quark and antiquark of the same flavor to be produced. Therefore, the partonic matrix element describing the production can be written in the form,
\begin{equation}
    \bra{0}T\{\overline{\Psi}_i (x)  {\cal O}_{ij}^{\mu_1...\mu_n} \Psi_j (y) \}\ket{Q\bar{Q}(n)+X},
\end{equation}
where we've used the capital psi, $\Psi$, to denote the relativistic heavy quark field. Here, $ {\cal O}_{ij}^{\mu_1...\mu_n} $ is a placeholder for the intermediate Dirac structures, fields, and propagators that connect $\overline{\Psi}(x)$ and $\Psi(y)$. The Lorentz indices come from intermediate gluon and photon vertices and propagators, and the spin indices $i,j$ come from Dirac structures that appear in the amplitudes at each order in the NRQCD power-counting. We have suppressed color indices for clarity. 

The NRQCD framework requires us to match the quark and antiquark fields onto their NRQCD counterparts by integrating out the large components of the momentum. In general, the full QCD field $\Psi$ will have both quark and antiquark components, but because this particular matrix element in the example is describing the production of a $Q\bar{Q}$ pair, it is sufficient to replace
\begin{equation}
\begin{aligned}
    \Psi(x) \to \sum_v e^{-imv\cdot x}\psi(x) ,\\
    \bar{\Psi}(x) \to \sum_v e^{-imv \cdot x}\chi^\dagger(x) ,
\end{aligned}
\label{eq: NRQCD field matching}
\end{equation}
at leading order, where $v$ is the direction of motion for the $Q\bar{Q}$ pair. Now, since the quarkonium carries hard momentum in the collinear factorization framework, $P\sim m$, each component of the coordinates $x$ and $y$ must scale as $1/m$. Therefore, the NRQCD fields in eq.~(\ref{eq: NRQCD field matching}) should be multipole expanded
\begin{equation}
\begin{aligned}
    &\psi(x) = e^{x\cdot \partial}  \psi (0) e^{-x \cdot \partial} \approx \psi(0) +{\cal O}(\vv),\\
    &\chi^\dagger(x) = e^{x\cdot \partial} \chi^\dagger (0)e^{-x \cdot \partial} \approx \chi^\dagger(0) +{\cal O}(\vv),
\end{aligned}
\label{eq: multipole expan}
\end{equation}
where the partial derivative acts on any residual momentum of the field that scales as $m\vv$, so that the combination $x\cdot \partial \sim m\vv/m =\vv$.

Therefore, after expanding the connecting operator in powers of $\vv$, the quarkonium production operators in the collinear factorization framework reduce to an object with heavy $Q\bar{Q}$ fields located at the origin,

\begin{equation}
    C(Q,M) \chi^\dagger(0) {\cal K} \cdot \Gamma \psi(0).
\end{equation}
These fields have zero radial separation. 

\subsubsection{TMD factorization}

In TMD factorization, we can work in a frame where the quarkonium's momentum scales as 
\begin{equation}
    P_{V}^\mu = (P^+_V,P^-_V,{\bf P}_{V,T} ) \sim (m ,m , m\vv ).
\label{eq: TMD momentum scaling}
\end{equation}
As in collinear factorization, the production of a $Q\bar{Q}$ can be described in the full theory via a partonic matrix element,
\begin{equation}
    \bra{0}T\{\overline{\Psi}_i (x)  {\cal O}_{ij}^{\mu_1...\mu_n}(x-y) \Psi_j (y) \}\ket{Q\bar{Q}(n)+X}.
\label{eq: TMD partonic ME}
\end{equation}
where the operator, ${\cal O}(x-y)$, will generically be some combination of propagators and vertices that connect the two quark fields in the amplitude. 

Again, we can match the quark and antiquark fields onto their NRQCD counterparts,
\begin{equation}
\begin{aligned}
    \Psi(x) \to \sum_v e^{-imv\cdot x}\psi(x) , \\
    \bar{\Psi}(x) \to \sum_v e^{-imv \cdot x}\chi^\dagger(x) .
\end{aligned}
\label{eq: NRQCD field matching}
\end{equation}
After matching onto the NRQCD fields, due to the distinct momentum scaling in the TMD framework (given by eq.~(\ref{eq: TMD momentum scaling})), only the plus and minus components of the quark field's positions can be multipole expanded away,
\begin{equation}
\begin{aligned}
    &\psi(x) = e^{(x^+  \partial^- + x^-\partial^+)} \psi (x_T) e^{-(x^+  \partial^- + x^-\partial^+)} \approx \psi(x_T) +{\cal O}(\vv), \\
    &\chi^\dagger(y) = e^{(x^+  \partial^- + x^-\partial^+)} \chi^\dagger (y_T) e^{-(x^+  \partial^- + x^-\partial^+)} \approx \chi^\dagger(y_T) +{\cal O}(\vv) .
\end{aligned}
\end{equation}

The hard partonic matching coefficient for matching onto the effective theory operators will come from expanding the amplitude generated from the matrix element in eq.~(\ref{eq: TMD partonic ME}) in the limit $m \gg m\vv, m\vv^2$. According to eq.~(\ref{eq: TMD momentum scaling}), $p_T\sim m\vv$, so the connecting operator should be expanded so that it only contains hard momentum,
\begin{equation}
\begin{aligned}
    {\cal O}_{ij}^{\mu_1...\mu_n}&(x-y) = \int\frac{d^4 p_H}{(2\pi)^4}  \tilde{{\cal O}}_{ij}^{\mu_1...\mu_n}(p_H) e^{i p \cdot(x-y)} \\
    \approx&  \int\frac{d p_H^+ dp^-_H}{(2\pi)^2} \tilde{{\cal O}}_{ij}^{\mu_1...\mu_n}(p^+_H, p_H^-) e^{i p^+ (x-y)^- +i p^-(x-y)^+} \delta^{(2)}(x_T-y_T) + {\cal O}\bigg(\frac{p_T^2}{m^2}\bigg).
\end{aligned}
\end{equation}
Therefore, we see that this momentum expansion forces $y_T \to x_T$ in the effective TMD production operator. Therefore we can write the final NRQCD operator in the TMD factorization framework as
\begin{equation}
    C'(Q,M) \chi^\dagger(x_T) {\cal K} \cdot \Gamma \psi(x_T).
\end{equation}
While this operator is slightly different than the production operator in the collinear factorization framework, (the matching coefficient is also distinct), it too has zero radial separation.


\section{Hubbard-Stratonovich transformations in vNRQCD}
%
\label{sec: decouple}

In this section, we review a powerful technique introduced by Fleming and Mehen in ref.~\cite{Fleming:2005pd} that allows one to match the vNRQCD framework onto the pNRQCD Lagrangian at leading order. We discuss how the soft and ultrasoft scales can be separated within this formalism.

\subsection{pNRQCD from vNRQCD}

%
vNRQCD and pNRQCD are two distinct EFTs for bound states of heavy quarks that describe the low-energy dynamics far below the scale $2m$. While supposedly describing the same physics, they are written in terms of completely different degrees of freedom - \mbox{vNRQCD} contains soft and ultrasoft gluons and maintains the explicit quark and antiquark degrees of freedom, while pNRQCD integrates out the soft scale so that only composite color-singlet ($S_\rv$) or color-octet ($O_\rv$) fields and the ultrasoft gluons remain. These composite fields, which are bosons, describe the dynamics of the quark/antiquark bound state, not the individual particles themselves. Since these EFTs supposedly describe the same physics, it is natural to ask: how can they be related?

Ref.~\cite{Fleming:2005pd} showed that the answer to this question is via a Hubbard-Stratonovich transformation. This useful technique, which is often used in quantum many-body calculations, rewrites any four-fermion interaction in a Lagrangian in terms of auxiliary bosonic degrees of freedom. In general, this transformation takes the form
\begin{equation}
    \psi^\dagger \psi\psi^\dagger\psi \to -\frac14\sigma^2+ \sigma \psi^\dagger \psi,
\end{equation}

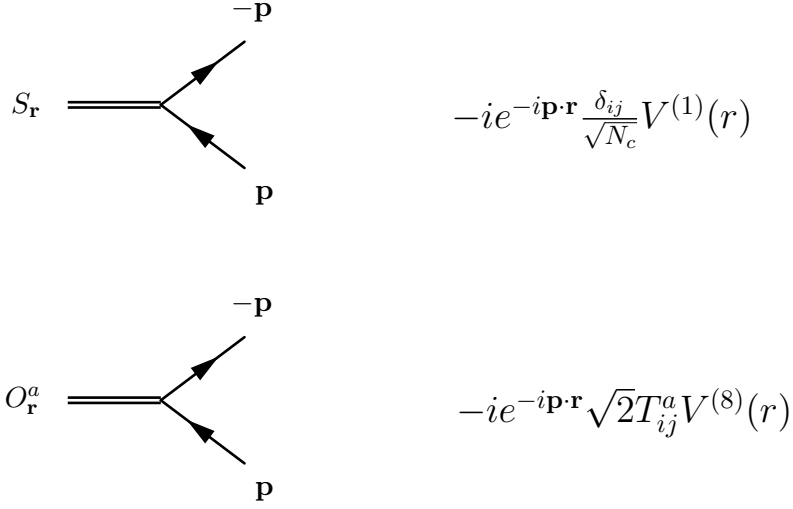
\begin{figure}[t]
    \centering
    \begin{tabular}{ c c }
\hspace{-30mm}\begin{fmffile}{Svertex}
     \begin{fmfgraph*}(70,95)
     \fmfipair{l',l,t,x,a,b,c,d,m,n,e,f,i,j,y,z,v}
    \fmfiequ{l'}{(0.1w,.525h)}
    \fmfiequ{l}{(0w,.5h)}
    \fmfiequ{t}{(w,.5h)}
    \fmfiequ{x}{(w,.525h)}
    \fmfiequ{a}{(.5w,.7h)}
    \fmfiequ{b}{(.5w,.5h)}
    \fmfiequ{c}{(.75w,.25h)}
    \fmfiequ{d}{(.5w, .3h)}
    \fmfiequ{e}{(.8w,.85h)}
    \fmfiequ{f}{(.2w,.85h)}
    \fmfiequ{y}{(.95w,.75h)}
    \fmfiequ{z}{(.95w,.25h)}
    \fmfiequ{i}{(.95w,h)}
    \fmfiequ{j}{(.35w,.6h)}
    \fmfiequ{v}{(.65w,.6h)}
    \fmfiequ{m}{(.5w,.75h)}
    \fmfiequ{n}{(.5w,.25h)}
    \fmfi{double}{l--b}
    %
    \fmfi{fermion}{b .. y}
    \fmfi{fermion}{z .. b}
    \fmfiv{l.d=8,l.a=70,l=$-\pv$}{y}
    \fmfiv{l.d=8,l.a=-60,l=$\pv$}{z}
    \fmfiv{l.d=10,l.a=180,l.s=15,l=${\Huge S_\rv}$}{l}
  \end{fmfgraph*}
\end{fmffile}
 &\hspace{20mm}\begin{minipage}{.75cm}\centering
     {\Large $-ie^{-i\pv\cdot\rv}\frac{\delta_{ij}}{\sqrt{N_c}}V^{(1)}(r)$} \vspace{25mm}
 \end{minipage} \vspace{-1.1cm}\\
 
\hspace{-30mm}\begin{fmffile}{Overtex}
     \begin{fmfgraph*}(70,95)
    \fmfipair{l',l,t,x,a,b,c,d,m,n,e,f,i,j,y,z,v}
    \fmfiequ{l'}{(0.1w,.525h)}
    \fmfiequ{l}{(0w,.5h)}
    \fmfiequ{t}{(w,.5h)}
    \fmfiequ{x}{(w,.525h)}
    \fmfiequ{a}{(.5w,.7h)}
    \fmfiequ{b}{(.5w,.5h)}
    \fmfiequ{c}{(.75w,.25h)}
    \fmfiequ{d}{(.5w, .3h)}
    \fmfiequ{e}{(.8w,.85h)}
    \fmfiequ{f}{(.2w,.85h)}
    \fmfiequ{y}{(.95w,.75h)}
    \fmfiequ{z}{(.95w,.25h)}
    \fmfiequ{i}{(.95w,h)}
    \fmfiequ{j}{(.35w,.6h)}
    \fmfiequ{v}{(.65w,.6h)}
    \fmfiequ{m}{(.5w,.75h)}
    \fmfiequ{n}{(.5w,.25h)}
    \fmfi{double}{l--b}
    %
    \fmfi{fermion}{b .. y}
    \fmfi{fermion}{z .. b}
    \fmfiv{l.d=8,l.a=70,l=$-\pv$}{y}
    \fmfiv{l.d=8,l.a=-60,l=$\pv$}{z}
    \fmfiv{l.d=10,l.a=180,l.s=15,l=${\Huge O^a_\rv}$}{l}
  \end{fmfgraph*}
\end{fmffile}
&
\hspace{20mm}
\begin{minipage}{.75cm}\centering
     {\Large $-ie^{-i\pv\cdot\rv}\sqrt{2}T_{ij}^aV^{(8)}(r)$} \vspace{25mm}
 \end{minipage}
  \vspace{-1.3cm}
\end{tabular}
    \caption{Feynman rules for composite field interactions \cite{Fleming:2005pd}.}
    \label{fig: composite feynman}
\end{figure}
\noindent where the original operator can be recovered by integrating out the bosonic field, $\sigma$. This technique is convenient for vNRQCD because a four-fermion operator naturally arises as the potential interaction term. To cast the vNRQCD Lagrangian of eq.~(\ref{eq: vNRQCD L}) in a form suitable for this transformation, the authors of ref.~\cite{Fleming:2005pd} Fourier transformed the Coulomb potential and used a Fierz decomposition to write the Coulomb potential in terms of color singlet and color-octet four-fermion operators. After doing this, the Hubbard-Stratonovich transformation for vNRQCD takes the form of 
\begin{equation}
\begin{aligned}
    \Delta{\cal L} = &\int d^3\rv V^{(1)}(r) \bigg(S_\rv^\dagger(x)-\sum_\qv e^{-i\qv\cdot \rv } \frac{1}{\sqrt{N_c}} \psi^\dagger_\qv (x)\chi_{-\qv} (x)\bigg) \\  & \hspace{4cm}\times \bigg( S_\rv (x)- \sum_\pv e^{i\pv \cdot \rv } \frac{1}{\sqrt{N_c}}\chi^\dagger_\pv(x) \psi_\pv(x)\bigg)\\
    &
    +  \int d^3\rv V^{(8)}(r) \bigg(O_\rv^\dagger(x)-\sum_\qv e^{-i\qv\cdot \rv } \sqrt{2}\psi^\dagger_\qv (x)T^a\chi_{-\qv}(x) \bigg) \\  & \hspace{4cm}\times \bigg( O_\rv (x)- \sum_\pv e^{i\pv \cdot \rv } \sqrt{2}\chi^\dagger_\pv(x) T^a\psi_\pv(x)\bigg) , 
\end{aligned}
\end{equation}
to be added to the Lagrangian. Here the color-octet and color-singlet potentials are
\begin{equation}
    V^{(1)}(r) = - C_F \frac{\alpha_s}{r},  \hspace{1cm}  V^{(8)}(r) = \bigg(\frac{C_A}{2} - C_F\bigg)\frac{\alpha_s}{r} ,
\end{equation}
and $S_\rv$ and $O_\rv$ are the bosonic color-octet and color-singlet auxiliary fields. After enforcing reparameterization invariance (RPI), the new hybrid vNRQCD/pNRQCD Lagrangian, including both the auxiliary fields and the quark/antiquark fields, is given by  
\begin{equation}
\begin{aligned}
    {\cal L}_{{\rm vNRQCD'}} =& -\frac14 F^{\mu\nu}F_{\mu\nu} + \sum_\pv \psi^\dagger_\pv \bigg(iD_0 - \frac{({\bfcal P} - i{\bf D})^2}{2m}\bigg)\psi_\pv +\sum_\pv \chi^\dagger_\pv \bigg(iD_0 - \frac{({\bfcal P} - i{\bf D})^2}{2m}\bigg)\chi_\pv\\
    & + \int d^3\rv V^{(1)}(r) \bigg(S_\rv^\dagger S_\rv \\
    & - S^\dagger_\rv \sum_\pv e^{i\pv\cdot\rv} \frac{1}{\sqrt{N_c}}\chi^\dagger_{-\pv}({\bf x} -\rv/2)W_\rv({\bf x}-\rv/2, {\bf x} + \rv/2)\psi_\pv({\bf x} + \rv/2) \\
    &- S_\rv \sum_\qv e^{-i\qv \cdot \rv} \frac{1}{\sqrt{N_c}} \psi^\dagger_\qv ({\bf x} -\rv/2)W_\rv({\bf x}-\rv/2, {\bf x} + \rv/2)\chi_{-\qv}({\bf x} + \rv/2)\bigg)\\
    &+ \int d^3\rv V^{(8)}(r) \bigg(O_\rv^\dagger O_\rv \\
    & - O^\dagger_\rv \sum_\pv e^{i\pv\cdot\rv} \sqrt{2}\chi^\dagger_{-\pv}({\bf x} -\rv/2)W_\rv({\bf x}-\rv/2, {\bf x} + \rv/2)\psi_\pv({\bf x} + \rv/2) \\
    &- O_\rv \sum_\qv e^{-i\qv \cdot \rv} \sqrt{2} \psi^\dagger_\qv ({\bf x} -\rv/2)W^\dagger_\rv({\bf x}-\rv/2 T^a W({\bf x} + \rv/2)\chi_{-\qv}({\bf x} + \rv/2)\bigg)\\
     & \; -g^2 \sum_{q,q^\prime,\pv,\pv^\prime} \frac12 U_{\mu \nu}(\pv, \pv^\prime, q, q^\prime) \psi^\dagger_{\pv^\prime}[A_{q^\prime}^\mu, A_q^\nu] \psi_\pv  + (\psi \leftrightarrow \chi) \\
        & \; -g^2 \sum_{q,q^\prime,\pv,\pv^\prime} \frac12 W_{\mu \nu}(\pv, \pv^\prime, q, q^\prime) \psi^\dagger_{\pv^\prime}\{A_{q^\prime}^\mu, A_q^\nu\} \psi_\pv  + (\psi \leftrightarrow \chi).
\end{aligned}
\label{eq: composite Lagrangian}
\end{equation}
Solving for the equations of motion for the auxiliary fields make it obvious that these objects are composite quark/antiquark fields~\cite{Fleming:2005pd}. Thus, we will refer to them as composite fields. 
The Feynman rules for interactions of quarks and antiquarks with the composite fields are given in fig.~\ref{fig: composite feynman}.

\begin{figure}[htbp]
    \centering
        \begin{fmffile}{SrLoop}
  \begin{fmfgraph*}(130,100)
    \fmfipair{l',l,t,x,a,b,c,d,m,n,e,f,i,j,y,z,v}
    \fmfiequ{l'}{(0.1w,.525h)}
    \fmfiequ{l}{(0w,.5h)}
    \fmfiequ{t}{(w,.5h)}
    \fmfiequ{x}{(w,.525h)}
    \fmfiequ{a}{(.5w,.7h)}
    \fmfiequ{b}{(.35w,.5h)}
    \fmfiequ{c}{(.75w,.25h)}
    \fmfiequ{d}{(.5w, .3h)}
    \fmfiequ{e}{(.8w,.95h)}
    \fmfiequ{f}{(.2w,.95h)}
    \fmfiequ{y}{(.65w,.5h)}
    \fmfiequ{z}{(.75w,.25h)}
    \fmfiequ{i}{(.95w,h)}
    \fmfiequ{j}{(.35w,.6h)}
    \fmfiequ{v}{(.65w,.6h)}
    \fmfiequ{m}{(.5w,.75h)}
    \fmfiequ{n}{(.5w,.25h)}
    \fmfi{double}{l--b}
    \fmfi{double}{t--y}
    %
    \fmfi{fermion}{b .. a .. y}
    \fmfi{fermion}{y .. d .. b}
    \fmfiv{l.d=7,l.a=-90,l=$-\pv$}{d}
    \fmfiv{l.d=5,l.a=90,l=$\pv$}{a}
    \fmfiv{l.d=5,l.a=180,l.s=15,l=${\Huge S_\rv}$}{l}
    \fmfiv{l.d=5,l.a=0,l=${\Huge S_{\rv'}}$}{t}
  \end{fmfgraph*}
\end{fmffile}
\hspace{1.5cm}
    \centering
        \begin{fmffile}{OrLoop}
  \begin{fmfgraph*}(130,100)
    \fmfipair{l',l,t,x,a,b,c,d,m,n,e,f,i,j,y,z,v}
    \fmfiequ{l'}{(0.1w,.525h)}
    \fmfiequ{l}{(0w,.5h)}
    \fmfiequ{t}{(w,.5h)}
    \fmfiequ{x}{(w,.525h)}
    \fmfiequ{a}{(.5w,.7h)}
    \fmfiequ{b}{(.35w,.5h)}
    \fmfiequ{c}{(.75w,.25h)}
    \fmfiequ{d}{(.5w, .3h)}
    \fmfiequ{e}{(.8w,.95h)}
    \fmfiequ{f}{(.2w,.95h)}
    \fmfiequ{y}{(.65w,.5h)}
    \fmfiequ{z}{(.75w,.25h)}
    \fmfiequ{i}{(.95w,h)}
    \fmfiequ{j}{(.35w,.6h)}
    \fmfiequ{v}{(.65w,.6h)}
    \fmfiequ{m}{(.5w,.75h)}
    \fmfiequ{n}{(.5w,.25h)}
    \fmfi{double}{l--b}
    \fmfi{double}{t--y}
    %
    \fmfi{fermion}{b .. a .. y}
    \fmfi{fermion}{y .. d .. b}
    \fmfiv{l.d=7,l.a=-90,l=$-\pv$}{d}
    \fmfiv{l.d=7,l.a=90,l=$\pv$}{a}
    \fmfiv{l.d=5,l.a=180,l.s=15,l=${\Huge O_\rv}$}{l}
    \fmfiv{l.d=5,l.a=0,l=${\Huge O_{\rv'}}$}{t}
  \end{fmfgraph*}
\end{fmffile}
    \caption{Leading order quark-antiquark loops. Evaluating the loops and integrating out the quark/antiquark fields reproduces the kinetic terms in the pNRQCD Lagrangian \cite{Fleming:2005pd}.}
    \label{fig: composite loops}
\end{figure}
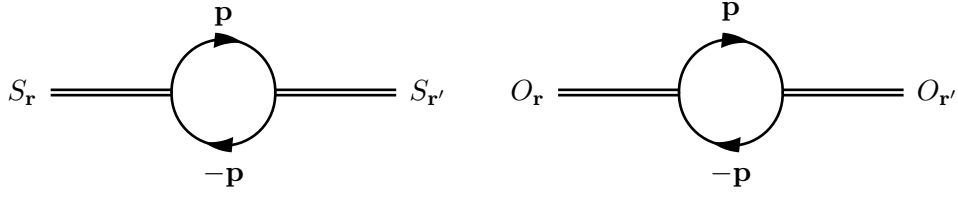

Integrating out the composite fields returns us to the original vNRQCD Lagrangian. Integrating out the quarks and antiquarks reproduces the pNRQCD Lagrangian. For example, at leading order, integrating out the quark/antiquark loops in fig.~\ref{fig: composite loops} yields the kinetic terms for the singlet and octet fields
\begin{equation}
    {\cal L}^{(0)} = \int d^3 \rv S^\dagger_\rv \bigg(i\partial_0 + \frac{\nabla^2_\rv}{m}-V^{(1)}(r)\bigg)S_\rv + \int d^3 \rv O_\rv^{a \dagger} \bigg(i\partial_0 + \frac{\nabla_\rv^2}{m}-V^{(8)}(r)\bigg)O_\rv^a ,
\label{eq: pNRQCD}
\end{equation}
which agrees exactly with the pNRQCD Lagrangian at leading order. Coupling gluons to the quark and antiquark loops leads to effective gluon interactions with the composite fields at ${\cal O}(\vv)$ and reproduces the next-to-leading order pNRQCD Lagrangian \cite{Fleming:2005pd}.

The intermediate Lagrangian in eq.~(\ref{eq: composite Lagrangian}) serves as a useful tool for understanding the dynamics of boundstate formation. Since both the composite fields (which are closely related to the physical quarkonium states) and the quark and antiquark degrees of freedom are explicit, this formulation is a good springboard for studying the full range of physics and scales involved in quarkonium production - which begins with the production of a heavy $Q\bar{Q}$ pair and ends with a fully hadronized quarkonium particle in the final state. In particular, this technique will be useful for studying how scales can be separated from each other at each step in the production process. 


\subsection{Decoupling the ultrasoft gluons}

In the composite Lagrangian of eq.~(\ref{eq: composite Lagrangian}), there are many different field ``modes" at play. There are interactions between the ultrasoft gluons and the potential scaling quark/antiquarks, and there are also interactions with soft gluons and the quark/antiquarks. Moreover, there are interactions with the composite fields, which carry only ultrasoft momentum, with the quark and antiquark sector. These many interactions make it difficult to disentangle the effects of different scales on quarkonium production. 

For example, in the composite field formulation presented in the previous section, there is an interaction between the ultrasoft $A^0$ gluon and the quark and antiquark fields which is the same order as the kinetic term in the power counting. Therefore, this interaction does not experience a power suppression and can appear at all orders in $\alpha_s(m\vv^2)$. Fortunately, this interaction can be easily decoupled from the Lagrangian via a BPS field redefinition on the fields~\cite{Fleming:2005pd}. This redefinition is given by
\begin{equation}
    \psi_\pv(x) \to U_v(x) \psi'_\pv(x),~~~~~~~~~U_v(x) = {\cal P}\exp\bigg(-ig \int_{-\infty}^0 ds A_0 (x+s)\bigg).
\label{eq: BPS field redef}
\end{equation}
Under the field redefinition, the leading pieces of the ultrasoft Lagrangian reduce to
\begin{equation}
\begin{aligned}
    {\cal L}^{(0), us}_c =&  \sum_\pv  \psi^{'\dagger}_\pv\bigg(i\partial_0  - \frac{{\bfcal P}^2}{2m}\bigg)\psi'_\pv +\sum_\pv \chi^{'\dagger}_\pv \bigg(i\partial_0  + \frac{{\bfcal P}^2}{2m}\bigg)\chi'_\pv \\
    &+ \int d^3\rv V^{(1)}(r) \bigg(S_\rv^\dagger S_\rv - S^\dagger_\rv \sum_\pv e^{i\pv\cdot\rv} \frac{1}{\sqrt{N_c}}\chi^{'\dagger}_{-\pv}  \psi_\pv  - S_\rv \sum_\qv e^{-i\qv \cdot \rv} \frac{1}{\sqrt{N_c}} \psi^{'\dagger}_\qv \chi'_{-\qv}\bigg)\\
    &+ \int d^3\rv V^{(1)}(r) \bigg(\tilde{O}_\rv^{\dagger,a} \tilde{O}^a_\rv - \tilde{O}^{\dagger,b}_\rv \sum_\pv e^{i\pv\cdot\rv}\sqrt{2}\chi^{'\dagger}_{-\pv}  T^b\psi_\pv  - \tilde{O}^b_\rv \sum_\qv e^{-i\qv \cdot \rv} \sqrt{2} \psi^{'\dagger}_\qv T^b \chi'_{-\qv}\bigg) \, .
\end{aligned}
\label{eq: WAA Lagrangian}
\end{equation}
%
where $\tilde{O}^{b} = {\cal U}_v^{ba} O^a$ absorbs the adjoint Wilson line defined by ${\cal U}_v^{ba}T^b = U^\dagger_v T^a U_v$. Importantly, this field redefinition also modifies the power corrections to the Lagrangian as well. At NLP, the ultrasoft interactions are modified to \cite{Fleming:2005pd}
\begin{equation}
    {\cal L}^{(1), us} =  \frac{1}{2m}\psi_\pv^{'\dagger} \bigg(2\pv \cdot {\bfcal E}_{us} + {\bfcal E}_{us}^2\bigg)\psi_\pv'  +  \frac{c_F}{2m}\psi_\pv^{'\dagger}\boldsymbol{\sigma} \cdot {\bfcal B}_{us}\psi'_\pv + (\psi \to \chi) , 
\end{equation}
where the gauge invariant chromo-magnetic and chromo-electric fields are defined as 
\begin{equation}
\begin{aligned}
    {\bfcal B}^k_{us} =& \frac12 U_v^\dagger \epsilon^{ijk} G^{ij}_{us} U_v , \\
    {\bfcal E}_{us}^i = &-\frac1g U_v^\dagger({\boldsymbol{\nabla}}^i - g {\bf A}_{us}^i)U_v ,
\end{aligned}
\label{eq: gauge inv electric and magnetic}
\end{equation}
with  ${\bf E}_{us} = iv\cdot \partial {\bfcal E}_{us}$. After making these field redefinitions, integrating out the quark and antiquark fields yields the interaction Lagrangian for the composite fields
\begin{equation}
\begin{aligned}
    {\cal L}_c^{(1)} =& i\frac{g}{\sqrt{2N_c}}\int d^3 \rv\bigg[-\frac{2}{m}[\nabla_\rv \tilde{O}^a_\rv]S_\rv \cdot {\bfcal E}^a_{us} + (V^{(1)}(r) -V^{(8)}(r)) \tilde{O}^a_\rv S_\rv \rv \cdot {\bfcal E}_{us}^a\bigg] . 
\end{aligned}
\end{equation}
After some simplification, this can be written in the form of the classic subleading pNRQCD Lagrangian \cite{Brambilla:1999xf, Fleming:2005pd}
\begin{equation}
    {\cal L}_p = g\sqrt{\frac{2}{N_c}}\int d^3\rv {\rm Tr}[O^\dagger \rv \cdot {\bf E}_{us}]S_\rv + h.c. + ...
\end{equation}

Notice that this field redefinition will cause Wilson lines in the soft sector as well. If one wishes to disentangle these effects, the redefinition
\begin{equation}
    A_\qv^\mu \to U_v^\dagger A_\qv^{'\mu}U_v
\label{eq: soft BPS}
\end{equation}
will leave the soft Lagrangian  unchanged: 
\begin{equation}
\begin{aligned}
    {\cal L}_s = 
        & \; -g^2 \sum_{q,q^\prime,\pv,\pv^\prime} \frac12 U_{\mu \nu}(\pv, \pv^\prime, q, q^\prime) \psi^{'\dagger}_{\pv^\prime}[A_{q^\prime}^{'\mu}, A_q^{'\nu}] \psi'_\pv  + (\psi \leftrightarrow \chi, \, T \leftrightarrow \bar{T}) \\
        & \; -g^2 \sum_{q,q^\prime,\pv,\pv^\prime} \frac12 W_{\mu \nu}(\pv, \pv^\prime, q, q^\prime) \psi^{'\dagger}_{\pv^\prime}\{A_{q^\prime}^{'\mu}, A_q^{'\nu}\} \psi'_\pv  + (\psi \leftrightarrow \chi, \, T \leftrightarrow \bar{T}).
\end{aligned}
\end{equation}
It is important to note that, the field redefinitions in eq.~(\ref{eq: BPS field redef}) and~(\ref{eq: soft BPS}) remove the leading order $A_0$ interaction from the lowest order terms in the Lagrangian, but this interaction is not removed from the theory. Instead, it is moved into sub-leading terms in the Lagrangian and the currents involved in the matching. The utility of this, however, is that it allows the ultrasoft Hilbert space to be separated at leading order from the potential and soft sectors,
\begin{equation}
    \ket{X_{Q/\bar{Q}}, X_{s}, X_{us}} = \ket{X_{Q/\bar{Q}}, X_{s}}\otimes\ket{ X_{us}} .
\end{equation}
This is a necessary step to factorize the ultrasoft sector from the other pieces of the Lagrangian. In the section that follows, we will not be concerned with factorizing the ultrasofts, but instead, we will want to factorize the soft sector from the quarks and antiquarks, which is a considerably more challenging task.

\subsection{A soft sector in pNRQCD?}

\label{sec: soft decouple}

The soft sector of vNRQCD contains an interaction between two soft-gluons and the quarks (or antiquarks) that also scales as $\vv^5$ in the power-counting. This interaction forms a ``seagull vertex" and does not experience any power suppression with respect to the kinetic term. This means it can appear at all orders in perturbation theory without any penalty. Why doesn't this interaction appear in the pNRQCD framework? 

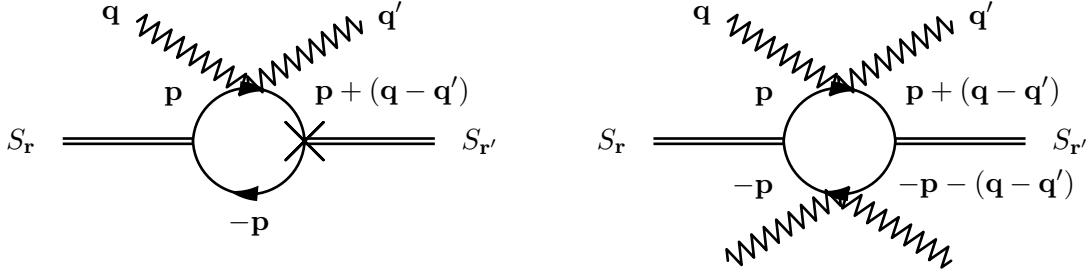
\begin{figure}[htbp]
    \centering
        \begin{fmffile}{DSV}
  \begin{fmfgraph*}(140,100)
    \fmfipair{l',l,t,x,a,b,c,d,m,n,e,f,i,j,y,z,v}
    \fmfiequ{l'}{(0.1w,.525h)}
    \fmfiequ{l}{(0w,.5h)}
    \fmfiequ{t}{(w,.5h)}
    \fmfiequ{x}{(w,.525h)}
    \fmfiequ{a}{(.5w,.7h)}
    \fmfiequ{b}{(.35w,.5h)}
    \fmfiequ{c}{(.75w,.25h)}
    \fmfiequ{d}{(.5w, .3h)}
    \fmfiequ{e}{(.8w,.95h)}
    \fmfiequ{f}{(.2w,.95h)}
    \fmfiequ{y}{(.65w,.5h)}
    \fmfiequ{z}{(.75w,.25h)}
    \fmfiequ{i}{(.95w,h)}
    \fmfiequ{j}{(.35w,.6h)}
    \fmfiequ{v}{(.65w,.6h)}
    \fmfiequ{m}{(.5w,.75h)}
    \fmfiequ{n}{(.5w,.25h)}
    \fmfi{double}{l--b}
    \fmfi{double}{t--y}
    %
    \fmfi{zigzag}{a--e}
    \fmfi{zigzag}{f--a}
    \fmfi{fermion}{b .. a .. y}
    \fmfi{fermion}{y .. d .. b}
    \fmfiv{l.d=7,l.a=160,l=$\qv$}{f}
    \fmfiv{l.d=7,l.a=20,l=$\qv'$}{e}
    \fmfiv{l.d=7,l.a=-90,l=$-\pv$}{d}
    \fmfiv{l.d=5,l.a=140,l=$\pv$}{j}
    \fmfiv{l.d=5,l.a=40,l=$\pv+(\qv-\qv')$}{v}
    \fmfiv{d.sh=cross,d.f=full,d.si=20}{y}
    \fmfiv{l.d=10,l.a=180,l.s=15,l=${\Large S_\rv}$}{l}
    \fmfiv{l.d=10,l.a=0,l=${\Large S_{\rv'}}$}{t}
  \end{fmfgraph*}
\end{fmffile}
\hspace{2.25cm}
    \centering
        \begin{fmffile}{4SV}
  \begin{fmfgraph*}(140,100)
    \fmfipair{l',l,t,x,a,b,c,d,m,n,e,f,i,j,y,z,v}
    \fmfiequ{l'}{(0.1w,.525h)}
    \fmfiequ{l}{(0,.5h)}
    \fmfiequ{t}{(w,.5h)}
    \fmfiequ{x}{(w,.525h)}
    \fmfiequ{a}{(.5w,.7h)}
    \fmfiequ{b}{(.35w,.5h)}
    \fmfiequ{c}{(.75w,.25h)}
    \fmfiequ{d}{(.5w, .3h)}
    \fmfiequ{e}{(.8w,.95h)}
    \fmfiequ{f}{(.2w,.95h)}
    \fmfiequ{y}{(.65w,.5h)}
    \fmfiequ{j}{(.35w,.6h)}
    \fmfiequ{i}{(.35w,.4h)}
    \fmfiequ{z}{(.65w,.4h)}
    \fmfiequ{v}{(.65w,.6h)}
    \fmfiequ{m}{(.2w,.05h)}
    \fmfiequ{n}{(.8w,.05h)}
    \fmfi{double}{l--b}
    \fmfi{double}{t--y}
    %
    \fmfi{zigzag}{a--e}
    \fmfi{zigzag}{f--a}
    \fmfi{zigzag}{d--m}
    \fmfi{zigzag}{n--d}
    \fmfi{fermion}{b .. a .. y}
    \fmfi{fermion}{y .. d .. b}
    \fmfiv{l.d=7,l.a=160,l=$\qv$}{f}
    \fmfiv{l.d=7,l.a=20,l=$\qv'$}{e}
    \fmfiv{l.d=5,l.a=-140,l=$-\pv$}{i}
    \fmfiv{l.d=5,l.a=140,l=$\pv$}{j}
    \fmfiv{l.d=5,l.a=40,l=$\pv+(\qv-\qv')$}{v}
    \fmfiv{l.d=1,l.a=-40,l=$-\pv-(\qv-\qv')$}{z}
    \fmfiv{l.d=10,l.a=180,l.s=15,l=${\Large S_\rv}$}{l}
    \fmfiv{l.d=10,l.a=0,l=${\Large S_{\rv'}}$}{t}
  \end{fmfgraph*}
\end{fmffile}
    \caption{Left: Example of a double soft gluon interaction on a quark loop that cannot couple to the composite field. Right: Soft gluon interactions on a quark loop that can be reduced to an effective soft-composite interaction vertex. }
    \label{fig: soft gluon vertex}
\end{figure}

When matching the vNRQCD framework onto pNRQCD using the Hubbard-Stratonovich transformation, ref.~\cite{Fleming:2005pd} showed that the seagull interaction can indeed keep intermediate quark (or antiquark) near it's mass shell, but it will change the three momentum of the heavy quark so that the quark/antiquark pair no longer have equal and opposite momentum. Therefore, the authors argue that the seagull vertex should not couple to the composite field and form a bound state. This is illustrated in the left panel of fig.~\ref{fig: soft gluon vertex}.

One exception exists when the seagull interaction transfers a net zero soft momentum to the quark or antiquark, i.e. $\qv = \qv'$. However, in this case the matching coefficient is divergent since the coefficient for the seagull vertex is proportional to $1/(\qv - \qv')^2$. This momentum region is identically removed from the theory as a zero-bin subtraction. Therefore, the authors of ref.~\cite{Fleming:2005pd} reason that a composite field coupling soft gluons cannot exist in the pNRQCD Lagrangian.

The true story is actually more subtle than this. Consider the case where two soft gluons also scatter off the antiquark and induce the same momentum change as on the quark. This is illustrated in the right panel of fig.~\ref{fig: soft gluon vertex}. Here, the final state quark and antiquark pair can have equal and opposite momentum. Hence, they can recombine to a composite field - the relative momentum has just shifted by $\qv-\qv'$. Therefore, while one insertion of the seagull soft gluon vertex (left side of fig.~\ref{fig: soft gluon vertex}) cannot appear in a theory with composite fields, multiple insertions of this vertex could, in-principle, be allowed. Furthermore, although it appears at $\alpha_s(mv)^2$, the loop in right diagram in fig.~\ref{fig: soft gluon vertex} scales as $\alpha_s(mv)^2\vv^3 \sim \vv^5$ and is not power suppressed with respect to the leading order diagrams in fig.~\ref{fig: composite loops}. This implies that soft-gluon interactions can appear at leading power in the pNRQCD framework. This is a stark contrast to the typical ideology of pNRQCD where the soft scale is completely integrated out from the theory. These  interactions prevent the soft sector and composite fields from being completely decoupled. 

\begin{figure}[htbp]
    \centering
        \begin{fmffile}{4SVprod}
  \begin{fmfgraph*}(160,130)
    \fmfipair{l',l,t,x,a,b,c,d,m,n,e,f,i,j,y,z,v}
    \fmfiequ{l'}{(0.1w,.525h)}
    \fmfiequ{l}{(0,.5h)}
    \fmfiequ{t}{(w,.5h)}
    \fmfiequ{x}{(w,.525h)}
    \fmfiequ{a}{(.5w,.7h)}
    \fmfiequ{b}{(.35w,.5h)}
    \fmfiequ{c}{(.75w,.25h)}
    \fmfiequ{d}{(.5w, .3h)}
    \fmfiequ{e}{(.8w,.95h)}
    \fmfiequ{f}{(.2w,.95h)}
    \fmfiequ{y}{(.65w,.5h)}
    \fmfiequ{j}{(.35w,.6h)}
    \fmfiequ{i}{(.35w,.4h)}
    \fmfiequ{z}{(.65w,.4h)}
    \fmfiequ{v}{(.65w,.6h)}
    \fmfiequ{m}{(.2w,.05h)}
    \fmfiequ{n}{(.8w,.05h)}
    \fmfi{double}{t--y}
    %
    \fmfi{zigzag}{a--e}
    \fmfi{zigzag}{f--a}
    \fmfi{zigzag}{d--m}
    \fmfi{zigzag}{n--d}
    \fmfi{fermion}{b .. a .. y}
    \fmfi{fermion}{y .. d .. b}
    \fmfiv{l.d=7,l.a=160,l=$\qv$}{f}
    \fmfiv{l.d=7,l.a=20,l=$\qv'$}{e}
    \fmfiv{l.d=5,l.a=-140,l=$-\pv$}{i}
    \fmfiv{l.d=5,l.a=140,l=$\pv$}{j}
    \fmfiv{l.d=5,l.a=40,l=$\pv+(\qv-\qv')$}{v}
    \fmfiv{l.d=1,l.a=-40,l=$-\pv-(\qv-\qv')$}{z}
    \fmfiv{d.sh=circle,d.f=empty,d.si=12}{b}
    \fmfiv{d.sh=cross,d.f=full,d.si=12}{b}
  \end{fmfgraph*}
\end{fmffile}
    \caption{Example of soft gluon interactions on a quark loop in production. Quark and antiquark fields from the production operator vertex interact with soft gluons and are coupled to the composite color singlet field.}
    \label{fig: four soft gluon}
\end{figure}
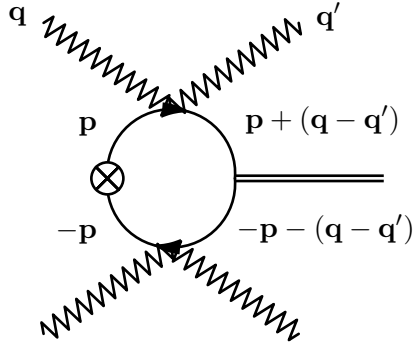

For quarkonium production, however, there is a way out of this predicament. As discussed in sec.~\ref{sec: zero radial}, heavy $Q\bar{Q}$ pairs produced at the scale $2m$ are matched onto local quark-antiquark operators due to the multipole expansion. These operators are not radially separated, which is a property that can be exploited. Consider, the production matrix element which describes the transition of a local $^3S_1^{[1]}$ $Q\bar{Q}$ pair to a vector quarkonium state. In vNRQCD, such a matrix element is given by,
\begin{equation}
    \sum_\qv \bra{0} \chi^\dagger_{-\qv}  \sigma^i \psi_\qv  \ket{V +X} ,
\end{equation}
where other final states, $X$, will be summed over in the matrix element squared. Evaluating the matrix element in time-ordered perturbation theory with two insertions of the seagull vertex and a coupling to the color-singlet composite field yields diagrams such as fig.~\ref{fig: four soft gluon}. Using the gauge-invariant formulation of the soft interaction to simplify the representation of the gluons \cite{Rothstein:2018dzq}, this diagram is given by the expression
\begin{equation}
\begin{aligned}
   &\frac{\alpha_s(mv)^2}{4\sqrt{N_c}}  \bra{0}\sigma^i \bigg[i\int d^3 \rv\int\frac{d^4 p}{(2\pi)^4} \frac{1}{E+\bar{q}_{us}^0+p^0 - \frac{(\pv+\bar{{\bf q}})^2}{2m}+i\epsilon}\frac{1}{E-\bar{q}_{us}^0-p^0 - \frac{(\pv+\bar{{\bf q}})^2}{2m}+i\epsilon}\\
   &\times\frac{1}{E+p^0 - \frac{{\bf p}^2}{2m}+i\epsilon}\frac{1}{E -p^0 - \frac{{\bf p}^2}{2m}+i\epsilon}  \bigg[-if^{abc}T^{c}\frac{2q^0}{\bar{\qv}^2} {\bfcal E}^{a}_\qv \cdot {\bfcal E}^{b}_{\qv'}\bigg]\\
   &\times \bigg[-if^{a'b'c'}\bar{T}^{c_1}\frac{2q^0}{\bar{\qv}^2} {\bfcal E}^{a'}_{-\qv} \cdot {\bfcal E}^{b'}_{-\qv'}\bigg] V^{(1)}(r) e^{i(\pv+\bar{\qv}) \cdot \rv} S_\rv^\dagger\bigg]\ket{V +X}, \\
\end{aligned}
\end{equation}
where $\bar{q} = q-q'$ and 
\begin{equation}
    {\bfcal E}_q^i = -\frac1g S_v^\dagger({\bfcal P}^i - g {\bf A}_q^i)S_v
\label{eq: gauge inv electric}
\end{equation}
is the gauge invariant soft gluon building block in vNRQCD~\cite{Rothstein:2018dzq}. Note that, since $v = (1,0,0,0)$ in the quarkonium rest frame, $v_\mu G^{\mu i,a} \equiv G^{0 i}$ , which is the standard definition of the electric field. Therefore, for simplicity, we also refer to this as the gauge invariant chromo-electric field. Readers should beware that ${\bfcal E}^i$ does not have a one-to-one correspondence with the real chromo-electric field because of the factor of $1/v\cdot{\cal P}$. 

We can integrate out the quark loops from the diagram to determine the effective coupling to the composite field. Here, there are two $p^0$ poles in the lower half of the complex plane, one at $p^0 = E- \pv^2/2m$ and one at $p^0 = E-(\pv+\bar{\qv})^2/2m$. Using residue theorem
\begin{equation}
\begin{aligned}
   &\frac{\alpha_s(mv)^2}{4\sqrt{N_c}}   \bra{0}\sigma^ii\int d^3 \rv\int\frac{d^3 \pv}{(2\pi)^3}  \bigg[\frac{1}{E_{Q\bar{Q}} - \frac{(\pv+\bar{{\bf q}})^2}{m}}\frac{1}{E -\bar{q}^0_{us}-\frac{(\pv+\bar{\qv})^2}{2m} - \frac{{\bf p}^2}{2m}}\\
   &\times\frac{1}{\bar{q}^0_{us}+\frac{(\pv+\bar{\qv})^2}{2m} - \frac{{\bf p}^2}{2m}} +\frac{1}{E_{Q\bar{Q}} -\frac{\pv^2}{m}} \frac{1}{E_{Q\bar{Q}}+\bar{q}_{us}^0-\frac{\pv^2}{2m} - \frac{(\pv+\bar{{\bf q}})^2}{2m}}\frac{1}{-\bar{q}_{us}^0+\frac{\pv^2}{2m} - \frac{(\pv+\bar{{\bf q}})^2}{2m}} \bigg]\\
   &\times \bigg[-if^{abc}T^{c}\frac{2q^0}{\bar{\qv}^2} {\bfcal E}^{a}_\qv \cdot {\bfcal E}^{b}_{\qv'}\bigg]\bigg[-if^{a'b'c'}\bar{T}^{c_1}\frac{2q^0}{\bar{\qv}^2} {\bfcal E}^{a'}_{-\qv} \cdot {\bfcal E}^{b'}_{-\qv'}\bigg]  V^{(1)}(r) e^{i(\pv+\bar{\qv}) \cdot \rv} S_\rv^\dagger\bigg]\ket{V +X} . \\
\end{aligned}
\end{equation}
Here, we define $E_Q = E_{\bar{Q}} = E_{Q\bar{Q}}/2$. Using, $(\pv+\bar{\qv})^2 e^{i(\pv+\bar{\qv})\cdot \rv} = -\nabla^2 e^{i(\pv+\bar{\qv})\cdot \rv} $ and $\pv^2 e^{i(\pv+\bar{\qv})\cdot \rv} = (-i\nabla-\qv)^2 e^{i(\pv+\bar{\qv})\cdot \rv} $ we can write this as
\begin{equation}
\begin{aligned}
   &\frac{\alpha_s(mv)^2}{4\sqrt{N_c}}  \bra{0}\sigma^ii\int d^3 \rv\int\frac{d^3 \pv}{(2\pi)^3}  \bigg[\frac{1}{E_{Q\bar{Q}} + \frac{\nabla^2}{m}}\frac{1}{E_{Q\bar{Q}} -\bar{q}^0_{us}+\frac{\nabla^2}{2m} - \frac{(-i\nabla-\qv)^2}{2m}} \frac{1}{\bar{q}^0_{us}-\frac{\nabla^2}{2m} + \frac{(-i\nabla -\bar{\qv})^2}{2m}}\\
   & +\frac{1}{E_{Q\bar{Q}} -\frac{(-i\nabla -\bar{\qv})^2}{m}} \frac{1}{E_{Q\bar{Q}}+\bar{q}_{us}^0-\frac{(-i\nabla -\bar{\qv})^2}{2m} + \frac{\nabla^2}{2m}}\frac{1}{-\bar{q}_{us}^0+\frac{(-i\nabla -\bar{\qv})^2}{2m} + \frac{\nabla^2}{2m}} \bigg]\\
   &\times \bigg[-if^{abc}T^{c}\frac{2q^0}{\bar{\qv}^2} {\bfcal E}^{a}_\qv \cdot {\bfcal E}^{b}_{\qv'}\bigg]\bigg[-if^{a'b'c'}\bar{T}^{c_1}\frac{2q^0}{\bar{\qv}^2} {\bfcal E}^{a'}_{-\qv} \cdot {\bfcal E}^{b'}_{-\qv'}\bigg] V^{(1)}(r) e^{i(\pv+\bar{\qv}) \cdot \rv} S_\rv^\dagger\bigg]\ket{V +X}.\\
\end{aligned}
\end{equation}
Since this expression is now independent of $\pv$, we can evaluate the $\pv$ integral, which sets $\rv \to 0$.
\begin{equation}
\begin{aligned}
   &\lim_{{\bf a}\to 0}\frac{\alpha_s(mv)^2}{4\sqrt{N_c}}   \bra{0}\sigma^ii\int d^3 \rv\int\frac{d^3 \pv}{(2\pi)^3}  \bigg[\frac{1}{E_{Q\bar{Q}} + \frac{\nabla^2}{m}}\frac{1}{E_{Q\bar{Q}} -\bar{q}^0_{us}+\frac{\nabla^2}{2m} - \frac{(-i\nabla-\qv)^2}{2m}}\frac{1}{\bar{q}^0_{us}-\frac{\nabla^2}{2m} + \frac{(-i\nabla -\bar{\qv})^2}{2m}} \\
   & +\frac{1}{E_{Q\bar{Q}} -\frac{(-i\nabla -\bar{\qv})^2}{m}} \frac{1}{E_{Q\bar{Q}}+\bar{q}_{us}^0-\frac{(-i\nabla -\bar{\qv})^2}{2m} + \frac{\nabla^2}{2m}}\frac{1}{-\bar{q}_{us}^0+\frac{(-i\nabla -\bar{\qv})^2}{2m} + \frac{\nabla^2}{2m}} \bigg]\\
   &\times \bigg[-if^{abc}T^{c}\frac{2q^0}{\bar{\qv}^2} {\bfcal E}^{a}_\qv \cdot {\bfcal E}^{b}_{\qv'}\bigg]\bigg[-if^{a'b'c'}\bar{T}^{c_1}\frac{2q^0}{\bar{\qv}^2} {\bfcal E}^{a'}_{-\qv} \cdot {\bfcal E}^{b'}_{-\qv'}\bigg]  V^{(1)}(r) \delta^{(3)}(\rv -{\bf a})e^{i\bar{\qv} \cdot \rv} S_\rv^\dagger\bigg]\ket{V +X}.\\
\end{aligned}
\end{equation}
Finally, using the equation of motion for the color singlet field to set \cite{Fleming:2005pd}
\begin{equation}
    \bigg(E_{Q\bar{Q}} + \frac{\nabla^2}{m}\bigg)  = V^{(1)}(r)
\end{equation}
gives
\begin{equation}
\begin{aligned}
   &\lim_{a\to 0}\frac{\alpha_s(mv)^2}{4\sqrt{N_c}}  \sum_{X_s, X_{us}} \bra{0}\sigma^ii\int d^3 \rv\int\frac{d^3 \pv}{(2\pi)^3}  \bigg[\frac{1}{V^{(1)}(r)}\frac{1}{V^{(1)}(r) -\bar{q}^0_{us} - \frac{-2i\nabla\cdot\qv+\qv^2}{2m}} \frac{1}{\bar{q}^0_{us}-\frac{\nabla^2}{2m} + \frac{(-i\nabla -\bar{\qv})^2}{2m}}\\
   & +\frac{1}{V^{(1)}(r) -\frac{-2i\nabla\cdot\bar{\qv} + \bar{\qv}^2}{m}} \frac{1}{V^{(1)}(r)+\bar{q}_{us}^0-\frac{-2i\nabla\cdot \qv + \bar{\qv}^2}{2m} }\frac{1}{-\bar{q}_{us}^0+\frac{(-i\nabla -\bar{\qv})^2}{2m} + \frac{\nabla^2}{2m}} \bigg]\\
   &\times \bigg[-if^{abc}T^{c}\frac{2q^0}{\bar{\qv}^2} {\bfcal E}^{a}_\qv \cdot {\bfcal E}^{b}_{\qv'}\bigg]\bigg[-if^{a'b'c'}\bar{T}^{c_1}\frac{2q^0}{\bar{\qv}^2} {\bfcal E}^{a'}_{-\qv} \cdot {\bfcal E}^{b'}_{-\qv'}\bigg]V^{(1)}(r) \delta^{(3)}(\rv -{\bf a})e^{i\bar{\qv} \cdot \rv} S_\rv^\dagger\bigg]\ket{V +X}.\\
\end{aligned}
\label{eq: dsg production final}
\end{equation}

Evaluating the limit ${\bf a} \to 0$ forces eq.~(\ref{eq: dsg production final}) to 0. This is because the potential is singular at $\rv = 0$ and we have more factors of $V(r)$ in the denominator than the numerator. Therefore, we reason that all composite field interactions with soft external fields during production vanish due to the $\rv = 0$ condition. This fact should be true at all orders in perturbation theory. So we conclude that the seagull vertex contributions do not play a role in production matrix elements. This allows the soft sector to be decoupled from the composite fields when $\rv = 0.$

In general, for $r\neq 0$, we find that such an insertion can exist and cannot be decoupled from the composite fields. Therefore, if it can be shown that the radius deviates from zero for production, then it is not valid to neglect these seagull diagrams. For the purposes of this work, however, we rely on the $\rv = 0$ condition and will separate the soft sector from the composite fields to ``factorize" production matrix elements in the following sections. 

Lastly, we comment that BPS field re-definition techniques, which sum interactions to all orders and transform away the resulting Wilson-line structures to decouple different sectors in the Lagrangian, are not applicable here because non-relativistic quark and antiquark propagators don't eikonalize. Therefore the resulting transformation is not unitary. A thorough discussion of this interaction can be found in Appendix \ref{app: WAA prop}.

\section{Factorizing soft contributions to production matrix elements}
\label{sec: factorization}

\begin{figure}[htbp]
    \centering
\begin{tabular}{>{\centering\arraybackslash}m{1.4in}>{\centering\arraybackslash}m{2.2in}}
        
         {\Large $ \{^1S_0^{[8]}, ^3S_1^{[8]}, ^3P_J^{[8]}\}$} & 
       \includegraphics[width = 1.1\linewidth]{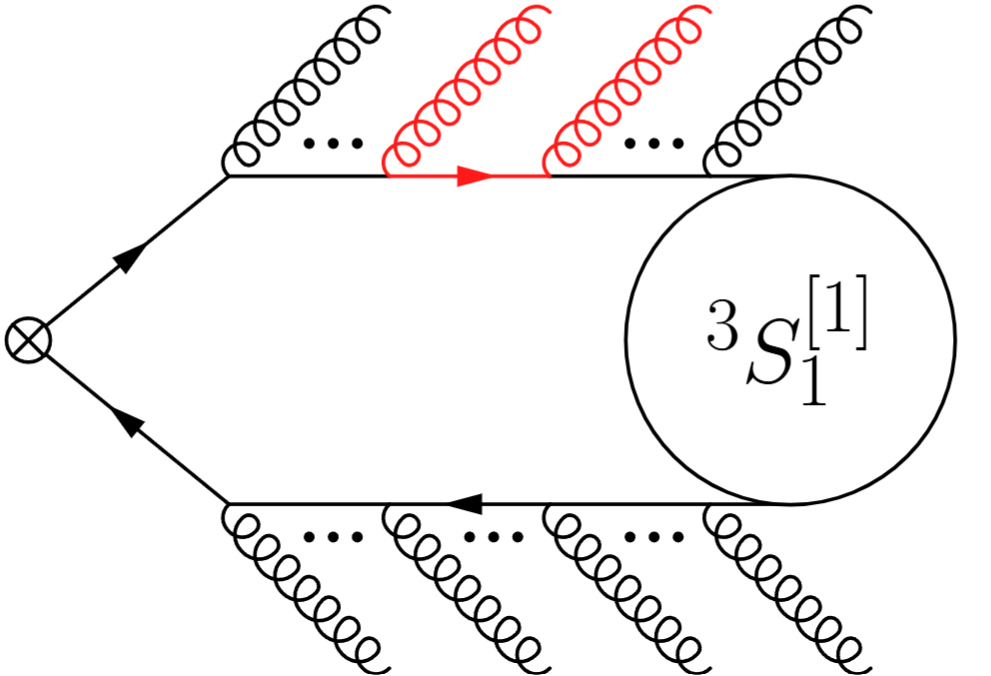} 
       \end{tabular}    \caption{Examples of the transition of a $c\bar{c}$ pair in a color-octet configuration to a $^3S_1^{[1]}$ state via soft gluon radiation at next-to-leading power in the $\vv$ expansion. Operator insertion mediating the $n \to ~^3S_1^{[1]}$ transition is indicated by the red propagator and gluons. Figure take from ref.~\cite{Copeland:2025vop}. }
    \label{fig: subleading transition}
\end{figure}

Within NRQCD, quarkonium production matrix elements describe the probability for a $Q\bar{Q}$ pair (with quantum numbers denoted by $n$) to transition to a final state quarkonium. In vNRQCD, however, the leading production operators are more explicit. When matching onto vNRQCD, soft gluon radiation can occur at the scale $m\vv$ to knock the quark and antiquark far off shell \cite{Copeland:2025vop, Fleming:2019pzj, Echevarria:2024idp, Echevarria:2019ynx}. Not only can this radiation generate Wilson lines, but at sub-leading power this radiation changes the quantum numbers of the heavy $Q\bar{Q}$ pair. The radiation that ultimately places the pair in a color singlet generates the leading contributions for the color-octet mechanisms generated at the hard scale. Diagrammatically, this is represented by fig.~\ref{fig: subleading transition}. The process of transitioning from a color-octet configuration to a color-singlet is encoded by vNRQCD operators where the soft fields are explicitly written with the quark and antiquark fields. 
The dominant vNRQCD operators for color-octet S-wave initial states were written down for the first time in ref.~\cite{Copeland:2025vop} and are given by
\begin{equation}
\begin{aligned}
   \mathcal{O}_3({\cal K}\cdot \Gamma) = &\frac{-1}{2m}\psi^\dagger_{\pv_Q} \bigg[\frac{1}{v\cdot{\cal P}} \big(g{\bfcal E}^\dagger_q \cdot g{\bfcal E}^\dagger_{q'} +g{\bfcal B}_q\cdot\boldsymbol{\sigma} \big)   - g \boldsymbol{\sigma}\cdot {\bfcal E}^\dagger_q  S_v^\dagger  \frac{\boldsymbol{\sigma}\cdot {\bfcal P}^{\dagger} }{v\cdot{\cal P}^\dagger} S_v \bigg] S_v^\dagger ({\cal K} \cdot \Gamma) S_v\chi_{\pv_{\bar{Q}}} \, \\
    & - \frac{1}{2m}g\psi^\dagger_{\pv_Q}  {\boldsymbol \sigma} \cdot  {\bfcal E}_q^\dagger S_v^\dagger ({\cal K} \cdot \Gamma) S_v \chi_{\pv_{\bar{Q}}}\,,
\label{eq: vNRQCD SL charm}
\end{aligned}
\end{equation}
and 
\begin{equation}
\begin{aligned}
   \mathcal{O}_4({\cal K}\cdot \Gamma) = &\frac{-1}{2m}\psi^\dagger_{\pv_Q}   S_v^\dagger ({\cal K} \cdot \Gamma) S_v  \bigg[\frac{1}{v\cdot{\cal P}} \big(g{\bfcal E}_q \cdot g{\bfcal E}_{q'} +g{\bfcal B}_q\cdot\boldsymbol{\sigma} \big)   - g S_v^\dagger \frac{\boldsymbol{\sigma}\cdot {\bfcal P} }{v\cdot{\cal P}}S_v \boldsymbol{\sigma}\cdot {\bfcal E}_q    \bigg]\chi_{\pv_{\bar{Q}}} \, \\
    & - \frac{1}{2m} \psi^\dagger_{\pv_Q}S_v^\dagger ({\cal K} \cdot \Gamma) S_v g{\boldsymbol \sigma} \cdot {\bfcal E}_q\chi_{\pv_{\bar{Q}}} .
\label{eq: vNRQCD SL anticharm}
\end{aligned}
\end{equation}
where ${\cal O}_3$ and ${\cal O}_4$ capture the subleading soft radiation off the quark and antiquark, respectively. The Wilson line composed of soft gluons in the $v$-direction is defined by 
\begin{equation}
    S_v(x) =  {\cal P}\exp\bigg(-ig \int_{-\infty}^0 ds v\cdot A_\qv (x+s)\bigg).
\end{equation}
The gauge-invariant chromo-electric field is defined in eq.(\ref{eq: gauge inv electric}) and the soft gauge-invariant chromo-magnetic field is defined as
\begin{equation}
\begin{aligned}
    {\bfcal B}^k_q =& \frac12 S_v^\dagger \epsilon^{ijk} G^{ij} S_v .
\end{aligned}
\label{eq: gauge inv magnetic}
\end{equation}

The operators in eqs. (\ref{eq: vNRQCD SL charm}) and (\ref{eq: vNRQCD SL anticharm}) depend on the color and spin structures, ${\cal K} \cdot \Gamma$, from the initial hard matching vertex. In practice, these are the same vertices that appear in the LDMEs of NRQCD,
\begin{equation}
    ({\cal K} \cdot \Gamma) \in \{ \Gamma^a_{^1S_0^{[8]}}, \Gamma^{\ell, a}_{^3S_1^{[8]}} ,\Gamma^{\ell j, a}_{^3P_{J}^{[8]}},...\} ,
\end{equation}
where the leading color-octet vertices are given by
\begin{equation}
\begin{aligned}
    &\Gamma^a_{^1S_0^{[8]}} = T^a  , \\ 
    &\Gamma^{\ell, a}_{^3S_1^{[8]}} = \sigma^\ell T^a ,\\
    &\Gamma^{\ell j, a}_{^3P_{J}^{[8]}} =\sigma^\ell \qv^j T^a.
\end{aligned}
\label{eq: production vertex}
\end{equation}

To find the leading contribution, we must project out the $~^3S_1^{[1]}$ configuration of each operator, since this matches the quantum numbers for S-wave vector quarkonium states. Procedurally, one should substitute the vertices of eq.~(\ref{eq: production vertex}) and then project out the $^3S_1^{[1]}$ configuration. We will refer to the resulting objects as ``transition operators". Each color-octet LDME should be matched onto a unique transition operator at the scale $m\vv$. The S-wave color-octet transition operators were also derived in ref.~\cite{Copeland:2025vop} using the axial $v\cdot A_q = 0$ gauge.
The P-wave transition operators are more subtle and we will derive them in a future section. For now, we present a gauge invariant formulation of the S-wave transition operators here.

For an initial $^1S_0^{[8]}$ vertex (given in eq.~(\ref{eq: production vertex})) the dominant operator contains a soft chromo-magnetic dipole operator,
%
%
%
%
%
        \begin{equation}
        \begin{aligned}
            & \psi^\dagger_{\bf p_Q} P_{^3S_1^{[1]}}\big[{\cal O}_{3+4}(\Gamma^a_{^1S_0^{[8]}}) \big] \chi_{\bf p_{\bar{Q}}} =   \frac{-g}{2N_cm} \psi^\dagger_{\bf p_Q} {\cal S}_v^{ab}\bigg[ \frac{1}{v\cdot\Pc}c_F{\bfcal B}^b\cdot\boldsymbol{\sigma}\bigg] \chi_{\bf p_{\bar{Q}}}\,,\\
        \label{eq: 1S0 transition 1}
        \end{aligned}
        \end{equation}
where $P_{^3S_1^{[1]}}$ indicates that we are projecting out the color-singlet ${}^3S_1$ state and the adjoint soft Wilson line is defined as
\begin{equation}
    S^\dagger_v T^aS_v = {\cal S}_v^{ab}T^b.
\end{equation}
Note we've included a factor of $c_F$, which is the Wilson coefficient to this operator that comes from integrating out the scale $m$. This is the same as the coefficient for the chromo-magnetic dipole operator in NRQCD and HQET and is given by \cite{Manohar:1999xd,Falk:1990pz}
\begin{equation}
    c_F(m,\mu) = \bigg(\frac{\alpha_s(\mu)}{\alpha_s(m)}\bigg)^{-C_A/\beta_0}
\end{equation}
where $\beta_0 = 11C_A/3 - 2n_f/3$. Likewise, substituting in the $^3S_1^{[8]}$ vertex and projecting out the $^3S_1^{[1]}$ component gives a double chromo-electric field operator,
%
\begin{equation}
        \begin{aligned}
            &\psi^\dagger_{\bf p_Q} P_{^3S_1^{[1]}}\big[{\cal O}_{3+4}(\Gamma^{\ell,a}_{^3S_1^{[8]}}) \big]\chi_{\bf p_{\bar{Q}}} = \frac{-g^2}{4N_cm} d^{gbc} \psi^\dagger_{\bf p_Q} {\cal S}_v^{ag}\bigg[\frac{1}{v\cdot\Pc}  {\bfcal E}_{q}^{ b} \cdot {\bfcal E}_{q'}^{c} \sigma^\ell \bigg] \chi_{\bf p_{\bar{Q}}}\,.
        \end{aligned}
    \label{eq: 3S18 trans operator}
\end{equation}

In ref.~\cite{Copeland:2025vop}, it was conjectured that these vNRQCD operators could be related to the chromo-magnetic and chromo-electric correlators that appear from matching the LDMEs onto the pNRQCD framework \cite{Brambilla:2022ayc}, but the exact connection was not obvious. This is because the soft scale cannot be factorized from the quark and antiquarks in the pure vNRQCD framework. In the following sections we will show that, because the Hubbard-Stratonovich transformation of sec. \ref{sec: decouple} allows the soft and ultrasoft scales in the Lagrangian to be separated during production, the transition operators of eqs.~(\ref{eq: 1S0 transition 1}) and~(\ref{eq: 3S18 trans operator}) can be further factorized into universal chromo-electric and chromo-magnetic vacuum correlators and the wave function at the origin. A result which mirrors the pNRQCD analysis.

\subsection{S-wave matrix elements}
We will begin with the three S-wave LDMEs, $\Otsosing, \Oosz,$ and $\Otsooct$, since their leading transition operators were derived in ref.~\cite{Copeland:2025vop}.

\subsubsection{A test run: Separating scales in $\Otsosing$}
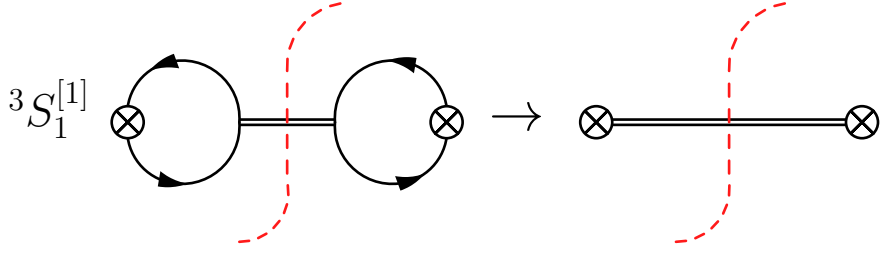
\begin{figure}
    \centering
    \begin{tabular}{ c c c}
    \begin{fmffile}{QL}
  \begin{fmfgraph*}(120,90)
    \fmfipair{l',l,t,x,a,b,c,d,m,n,e,f,i,j,y,z,v}
    \fmfiequ{l'}{(0,.525h)}
    \fmfiequ{l}{(0,.5h)}
    \fmfiequ{t}{(w,.5h)}
    \fmfiequ{x}{(w,.525h)}
    \fmfiequ{a}{(.125w,.75h)}
    \fmfiequ{b}{(.35w,.5h)}
    \fmfiequ{c}{(.75w,.25h)}
    \fmfiequ{d}{(.125w, .25h)}
    \fmfiequ{e}{(.875w,.75h)}
    \fmfiequ{f}{(.875w,.25h)}
    \fmfiequ{y}{(.65w,.5h)}
    \fmfiequ{z}{(.75w,.25h)}
    \fmfiequ{i}{(.95w,h)}
    \fmfiequ{j}{(.35w,0)}
    \fmfiequ{v}{(.7w,h)}
    \fmfiequ{m}{(.5w,.75h)}
    \fmfiequ{n}{(.5w,.25h)}
    \fmfi{double}{b--y}
    %
    \fmfi{fermion}{l .. d .. b}
    \fmfi{fermion}{b .. a .. l}
    \fmfi{fermion}{y .. f .. t}
    \fmfi{fermion}{t .. e .. y}
    \fmfi{dashes, foreground=(1,,0.1,,0.1)}{j{right} .. n -- m .. {right}v}
    \fmfiv{d.sh=circle,d.f=empty,d.si=12}{l}
    \fmfiv{d.sh=cross, l.d=15,l.a=160, l={\LARGE $^3S_1^{[1]}$},d.f=full,d.si=12}{l}
    \fmfiv{d.sh=circle,d.f=empty,d.si=12}{t}
    \fmfiv{d.sh=cross, l.d=15,l.a=20,d.f=full,d.si=12}{t}
  \end{fmfgraph*}
\end{fmffile}
 & \begin{minipage}{.75cm}\centering
     {\huge $\rightarrow$} \vspace{25mm}
 \end{minipage}
 &\begin{fmffile}{3S1compVert}
 \begin{fmfgraph*}(100,90)
    \fmfipair{l',l,t,x,a,b,c,d,m,n,e,f,i,j,y,z,v}
    \fmfiequ{l'}{(0,.525h)}
    \fmfiequ{l}{(0,.5h)}
    \fmfiequ{t}{(w,.5h)}
    \fmfiequ{x}{(w,.525h)}
    \fmfiequ{a}{(.125w,.65h)}
    \fmfiequ{b}{(.35w,.5h)}
    \fmfiequ{c}{(.75w,.25h)}
    \fmfiequ{d}{(.125w, .35h)}
    \fmfiequ{e}{(.875w,.65h)}
    \fmfiequ{f}{(.875w,.35h)}
    \fmfiequ{y}{(.65w,.5h)}
    \fmfiequ{z}{(.75w,.25h)}
    \fmfiequ{i}{(.95w,h)}
     \fmfiequ{j}{(.3w,0)}
    \fmfiequ{v}{(.75w,h)}
    \fmfiequ{m}{(.5w,.75h)}
    \fmfiequ{n}{(.5w,.25h)}
    \fmfi{double}{l--t}
    %
    \fmfiv{d.sh=circle,d.f=empty,d.si=12}{l}
    \fmfiv{d.sh=cross, l.d=15,l.a=160,d.f=full,d.si=12}{l}
    \fmfiv{d.sh=circle,d.f=empty,d.si=12}{t}
    \fmfiv{d.sh=cross, l.d=15,l.a=20,d.f=full,d.si=12}{t}
     \fmfi{dashes, foreground=(1,,0.1,,0.1)}{j{right} .. n -- m .. {right}v}
  \end{fmfgraph*}
\end{fmffile}
\end{tabular}
    \caption{Integrating out the quark/antiquark loops in $^3S_1^{[1]}$ LDME. Double lines represent the color-singlet composite field.}
    \label{fig:LDME quark loops}
\end{figure}

First, consider the $^3S_1^{[1]}$ color singlet LDME, $\Otsosing$, which does not need to be matched onto an intermediate transition operator. For an S-wave vector quarkonia state, $\ket{V}$, this matrix element in vNRQCD is defined as 
\begin{equation}
\begin{aligned}
    \Otsosing& = \sum_\qv \sum_{X_s} \bra{0} \chi^\dagger_{-\qv}  \sigma^i \psi_\qv  \ket{V ,X_s} \bra{V ,X_s} \psi^\dagger_\qv  \sigma^i  \chi_{-\qv} \ket{0}.
\end{aligned}
\end{equation}
We can match this operator onto the color-singlet composite fields by taking a time-ordered product of the color-singlet + quark/antiquark interaction from eq.~(\ref{eq: composite Lagrangian}). This is shown in fig. \ref{fig:LDME quark loops}. 
Using the Feynman rules in fig. \ref{fig: composite feynman}, we find that the LDME reduces to
\begin{equation}
\begin{aligned}
   & \frac1{N_c} \sum_{X_s} \bra{0}\sigma^i \bigg[\int d^3 \rv \int\frac{d^3\qv}{(2\pi)^3}S_\rv^\dagger \frac{V^{(1)}(r)}{E_{Q\bar{Q}} - \frac{\nabla_\rv^2}{m}}  e^{i\qv \cdot \rv} \bigg]\ket{V ,X_s}\\
   &\qquad \qquad  \times\bra{V ,X_s}  \bigg[\int d^3 \rv'\int\frac{d^3\qv'}{(2\pi)^3} S_{\rv}\frac{V^{(1)}(r')}{E_{Q\bar{Q}} - \frac{\nabla_{\rv'}^2}{m}} e^{i\qv' \cdot \rv'} \bigg]  \sigma^i  \ket{0} . \\
\end{aligned}
\end{equation}
where again $E_Q = E_{\bar{Q}} = E_{Q\bar{Q}}/2$. Now using the equations of motion for the composite field to set $V^{(1)}(r) = E_{Q\bar{Q}} - \nabla^2_\rv/m$, $\Otsosing$ reduces to 
%
%
\begin{equation}
\begin{aligned}
   \Otsosing = &N_c\sum_{X_s} \bra{0} [\sigma^i S_0^\dagger(x)]\ket{V ,X_s}\bra{V ,X_s} [S_0(x)  \sigma^i]  \ket{0} ,\\
\end{aligned}
\end{equation}
where $S_0(x)$ is the composite color singlet field evaluated at $r = 0$ and the Pauli matrices act on on the composite fields.  Once quarks and antiquarks are completely integrated out of the theory (which happens as we go from eq.~(\ref{eq: composite Lagrangian}) to eq.~(\ref{eq: pNRQCD})), $S_0$ cannot interact with soft fields and is decoupled from the soft sector in the effective Lagrangian, for reasons described in sec. \ref{sec: soft decouple}. Therefore, we can ``factorize" the Hilbert space accordingly,
\begin{equation}
    \ket{V ,X_s} = \ket{V }\otimes \ket{X_s} ,
\label{eq: fact Hilbert}
\end{equation}
so that the LDME becomes
\begin{equation}
\begin{aligned}
   \Otsosing = &N_c \sum_{X_s} \braket{0|X_s}\braket{X_s|0}  \otimes\sum_{ X_{us}} \bra{0} [\sigma^i S_0^\dagger(x)]\ket{V , X_{us}}\bra{V , X_{us}} [S_0(x)  \sigma^i]  \ket{0} \,\\
   = & N_c \bra{0} [\sigma^i S_0^\dagger(x)]\ket{V }\bra{V } [S_0(x)  \sigma^i]  \ket{0} \, .\\
\end{aligned}
\label{eq: factorized 3S11}
\end{equation}
This result is rather simple since the soft matrix element evaluates to one. However the techniques used here will be very useful for studying the color-octet LDMEs in the following sections. 

The remaining matrix elements are related to the vector quarkonium's radial wave function at the origin since the composite field is evaluated at $r = 0$ \cite{Bodwin:1994jh},
\begin{equation}
    \bra{0} [\sigma^i S_0^\dagger(x)]\ket{V } = \sqrt{\frac{1}{2\pi}
    }R_V^{(0)}(0) .
\label{eq: WF origin}
\end{equation}
Thus, eq.~(\ref{eq: factorized 3S11}) reproduces the classic NRQCD prediction \cite{Bodwin:1994jh,Brambilla:2022rjd},
\begin{equation}
    \Otsosing  = \frac{3N_c}{2\pi} |R_V^{(0)}(0)|^2 ,
\end{equation}
where the factor of 3 comes from summing over polarizations. 

\subsection{$\Oosz$}
\begin{figure}
    \centering
    \begin{tabular}{ c c c c c}
    \begin{fmffile}{QL1S0}
  \begin{fmfgraph*}(120,90)
    \fmfipair{l',l,t,x,a,b,c,d,m,n,e,f,i,j,y,z,v}
    \fmfiequ{l'}{(0,.525h)}
    \fmfiequ{l}{(0,.5h)}
    \fmfiequ{t}{(w,.5h)}
    \fmfiequ{x}{(.5w,.8h)}
    \fmfiequ{a}{(.125w,.75h)}
    \fmfiequ{b}{(.35w,.5h)}
    \fmfiequ{c}{(.75w,.25h)}
    \fmfiequ{d}{(.125w, .25h)}
    \fmfiequ{e}{(.875w,.75h)}
    \fmfiequ{f}{(.875w,.25h)}
    \fmfiequ{y}{(.65w,.5h)}
    \fmfiequ{z}{(.75w,.25h)}
    \fmfiequ{i}{(.95w,h)}
    \fmfiequ{j}{(.35w,0)}
    \fmfiequ{v}{(.7w,h)}
    \fmfiequ{m}{(.5w,.75h)}
    \fmfiequ{n}{(.5w,.25h)}
    \fmfi{double}{b--y}
    %
    \fmfi{fermion}{l .. d .. b}
    \fmfi{fermion}{b .. a .. l}
    \fmfi{fermion}{y .. f .. t}
    \fmfi{fermion}{t .. e .. y}
    \fmfi{zigzag}{l .. x .. t}
    \fmfi{plain}{l .. x .. t}
    \fmfi{dashes, foreground=(1,,0.1,,0.1)}{j{right} .. n -- m .. {right}v}
    \fmfiv{d.sh=circle,d.f=empty,d.si=12}{l}
    \fmfiv{d.sh=cross, l.d=15,l.a=160,d.f=full,d.si=12}{l}
    \fmfiv{d.sh=circle,d.f=empty,d.si=12}{t}
    \fmfiv{d.sh=cross, l.d=15,l.a=20,d.f=full,d.si=12}{t}
  \end{fmfgraph*}
\end{fmffile}
 & \begin{minipage}{.75cm}\centering
     {\huge $\rightarrow$} \vspace{25mm}
 \end{minipage}
 \begin{minipage}{.2cm}
     \scalebox{4}{(} \vspace{30mm}
 \end{minipage}
 &\begin{fmffile}{1S0compVert}
 \begin{fmfgraph*}(85,90)
    \fmfipair{l',l,t,x,a,b,c,d,m,n,e,f,i,j,y,z,v}
    \fmfiequ{l'}{(0,.525h)}
    \fmfiequ{l}{(0,.5h)}
    \fmfiequ{t}{(w,.5h)}
    \fmfiequ{x}{(.5w,.8h)}
    \fmfiequ{a}{(.125w,.65h)}
    \fmfiequ{b}{(.35w,.5h)}
    \fmfiequ{c}{(.75w,.25h)}
    \fmfiequ{d}{(.125w, .35h)}
    \fmfiequ{e}{(.875w,.65h)}
    \fmfiequ{f}{(.875w,.35h)}
    \fmfiequ{y}{(.65w,.5h)}
    \fmfiequ{z}{(.75w,.25h)}
    \fmfiequ{i}{(.95w,h)}
     \fmfiequ{j}{(.3w,0)}
    \fmfiequ{v}{(.75w,h)}
    \fmfiequ{m}{(.5w,.75h)}
    \fmfiequ{n}{(.5w,.25h)}
    \fmfi{double}{l--t}
    %
    \fmfiv{d.sh=circle,d.f=empty,d.si=12}{l}
    \fmfiv{d.sh=cross, l.d=15,l.a=160,d.f=full,d.si=12}{l}
    \fmfiv{d.sh=circle,d.f=empty,d.si=12}{t}
    \fmfiv{d.sh=cross, l.d=15,l.a=20,d.f=full,d.si=12}{t}
     \fmfi{dashes, foreground=(1,,0.1,,0.1)}{j{right} .. n -- m .. {right}v}
  \end{fmfgraph*}
\end{fmffile} & \begin{minipage}{.2cm}
     \scalebox{4}{)} \vspace{30mm}
 \end{minipage} &
\begin{fmffile}{magcorr}
 \begin{fmfgraph*}(85,90)
    \fmfipair{l',l,t,x,a,b,c,d,m,n,e,f,i,j,y,z,v}
    \fmfiequ{l'}{(0,.525h)}
    \fmfiequ{l}{(0,.5h)}
    \fmfiequ{t}{(w,.5h)}
    \fmfiequ{x}{(.5w,.8h)}
    \fmfiequ{a}{(.125w,.65h)}
    \fmfiequ{b}{(.35w,.5h)}
    \fmfiequ{c}{(.75w,.25h)}
    \fmfiequ{d}{(.125w, .35h)}
    \fmfiequ{e}{(.875w,.65h)}
    \fmfiequ{f}{(.875w,.35h)}
    \fmfiequ{y}{(.65w,.5h)}
    \fmfiequ{z}{(.75w,.25h)}
    \fmfiequ{i}{(.95w,h)}
     \fmfiequ{j}{(.3w,0)}
    \fmfiequ{v}{(.75w,h)}
    \fmfiequ{m}{(.5w,.75h)}
    \fmfiequ{n}{(.5w,.25h)}
    \fmfi{zigzag}{l--t}
    \fmfi{plain}{l--t}
    %
    \fmfiv{d.sh=circle,d.f=empty,d.si=12}{l}
    \fmfiv{d.sh=cross, l.d=11,l.a=160,d.f=full,d.si=12}{l}
    \fmfiv{d.sh=circle,d.f=empty,d.si=12}{t}
    \fmfiv{d.sh=cross, l.d=11,l.a=20,d.f=full,d.si=12}{t}
     \fmfi{dashes, foreground=(1,,0.1,,0.1)}{j{right} .. n -- m .. {right}v}
  \end{fmfgraph*}
\end{fmffile}
\end{tabular}
\vspace{-2cm}
    \caption{Factorizing out the soft chromo-magnetic dipole contributions in the $^1S_0^{[8]}$ channel. Zigzag with a line through the center represents the chromo-magnetic field. Double lines represent the color-singlet composite field.}
    \label{fig: 1S08 quark loops}
\end{figure}
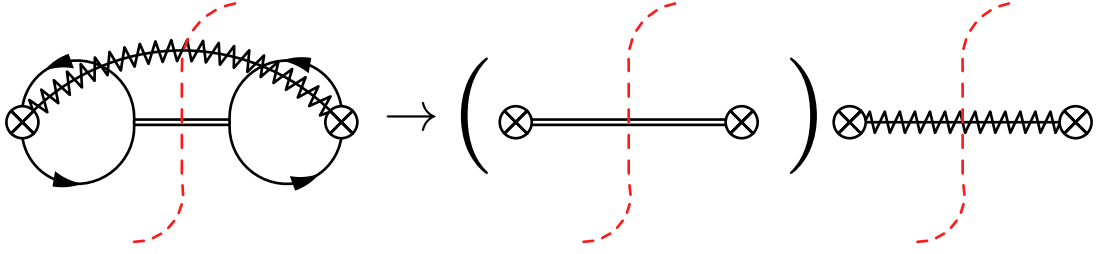

As discussed in sec. \ref{sec: factorization}, the quark-antiquark bilinear in $\Oosz$ matches onto the transition operator given by eq.~(\ref{eq: 1S0 transition 1}). This transition operator in vNRQCD encodes a single soft chromo-magnetic dipole emission  that flips the $^1S_0^{[8]}$ to a $^3S_1^{[1]}$ state. We find that $\Oosz$ can be written as
\begin{equation}
\begin{aligned}
    \Oosz= &\frac{1}{N_c^2M^2}\sum_{\qv}\sumint_{X_s} \bra{0} \big[\chi^\dagger_{-\qv} {\cal S}_v^{\dagger,ab}  \bigg[\frac{1}{v\cdot\Pc} g c_F\sigma \cdot {\bfcal B}^b_s  \bigg] \psi_\qv \ket{V, X_s}\\
    &\times\bra{V,X_s} \psi^\dagger_\qv {\cal S}_v^{ac}\bigg[\frac{1}{v\cdot\Pc} g c_F \sigma \cdot {\bfcal B}^{c}_s\bigg]\chi_{-\qv}\ket{0}.
\end{aligned}
\label{eq: collinear transition function}
\end{equation}
Employing the same techniques as before, we can couple the quark and antiquark fields to the composite color singlet field, shown in fig. \ref{fig: 1S08 quark loops}. 

The quark loop integrals are identical to the $\Otsosing$ calculation from the previous section, so we can write the $\Oosz$ as 
\begin{equation}
\begin{aligned}
   \Otsosing = &\frac{1}{N_cM^2} \sum_{X_s} \bra{0}  {\cal S}_v^{\dagger,ab}\bigg[\frac{1}{v\cdot\Pc} g c_F {\bfcal B}^{i,b}_q(x)  \bigg][\sigma^i S_0^\dagger(x)]\ket{V ,X_s}\\
   &\times\bra{V ,X_s} [S_0(x)  \sigma^j]  {\cal S}_v^{ac}\bigg[\frac{1}{v\cdot\Pc} g c_F {\bfcal B}^{j,c}_q (x) \bigg]  \ket{0} \, .\\
\end{aligned}
\end{equation}
Here we take a non-trivial step. For reasons explained above, the soft Hilbert space separates from the composite field Hilbert space in production processes. Therefore, we can ``factorize" $\Oosz$ so that it takes the following form, 
\begin{equation}
\begin{aligned}
   \Oosz =
    &\frac{1}{3N_cM^2} |\bra{0} [\sigma^i S_0^\dagger(x)]\ket{V }|^2 \otimes \bra{0} \bigg|{\cal S}_v^{ab}\bigg[\frac{1}{v\cdot\Pc} g c_F {\bfcal B}^b_q  \bigg]\bigg|^2 (x) \ket{0} 
   , \\
\end{aligned}
\end{equation}
where we use spin and rotational symmetry to simplify the composite field matrix element. This factorization is represented by the right side of fig. \ref{fig: 1S08 quark loops}. Note, the matrix element with the chromo-magnetic field depends only on the soft scale and the matrix element with the composite color singlet field only depends on the ultrasoft scale. Defining
\begin{equation}
    \braket{{\bfcal B}} = \bra{0} \bigg|{\cal S}_v^{ab}\bigg[g\frac{c_F}{v\cdot\Pc}   {\bfcal B}^b_q  \bigg]\bigg|^2 (x) \ket{0} 
\label{eq: Bs}
\end{equation}
and using eq.~(\ref{eq: WF origin}) from the previous section, we arrive at the compact final expression
\begin{equation}
\begin{aligned}
   \Oosz = &\frac{1}{3 N_cM^2} \braket{{\bfcal B}}\frac{3|R_V(0)|^2}{2\pi}  \big(1 + {\cal O}(\vv^2) \big) , \\
\end{aligned}
\end{equation}
which agrees with the pNRQCD result exactly \cite{Brambilla:2022rjd,Brambilla:2022ayc}.

\subsection{$\Otsooct$}
\label{sec: 3S18}
In a similar fashion, we can match the $\Otsooct$ onto the transition operator given in eq.~(\ref{eq: 3S18 trans operator}) and then integrate out the quark and antiquark fields to separate the scales. We find the matching to be given by
\begin{equation}
\begin{aligned}
     \Otsooct =&\sum_\qv\frac{1}{4N_c^2M^2} \bra{0}\chi_{\bf -q}^\dagger \bigg[\frac{1}{v\cdot\Pc}  \bigg(d^{abc} g^2{\bfcal E}^{ b} \cdot {\bfcal E}^{c} \sigma^{\ell}\bigg)\bigg]\psi_{\bf q}\ket{V,X_s ,X_{us}}\\
     &\times\bra{V,X_s}\psi_\qv^\dagger \bigg[\frac{1}{v\cdot\Pc}  \bigg( d^{ab'c'} g^2{\bfcal E}^{ b'} \cdot {\bfcal E}^{c'} \sigma^{\ell})\bigg]\chi_{\bf -q}\ket{0} \, .
\end{aligned}
\end{equation}
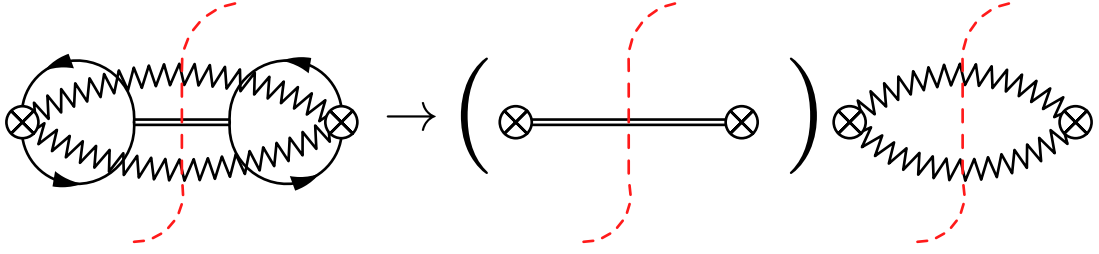
\begin{figure}
    \centering
    \begin{tabular}{ c c c c c}
    \begin{fmffile}{QL3S18}
  \begin{fmfgraph*}(120,90)
    \fmfipair{l',l,t,x,a,b,c,d,m,n,e,f,i,j,y,z,v}
    \fmfiequ{l'}{(0,.525h)}
    \fmfiequ{l}{(0,.5h)}
    \fmfiequ{t}{(w,.5h)}
    \fmfiequ{x}{(.5w,.7h)}
    \fmfiequ{a}{(.125w,.75h)}
    \fmfiequ{b}{(.35w,.5h)}
    \fmfiequ{c}{(.75w,.25h)}
    \fmfiequ{d}{(.125w, .25h)}
    \fmfiequ{e}{(.875w,.75h)}
    \fmfiequ{f}{(.875w,.25h)}
    \fmfiequ{y}{(.65w,.5h)}
    \fmfiequ{z}{(.5w,.3h)}
    \fmfiequ{i}{(.95w,h)}
    \fmfiequ{j}{(.35w,0)}
    \fmfiequ{v}{(.7w,h)}
    \fmfiequ{m}{(.5w,.75h)}
    \fmfiequ{n}{(.5w,.25h)}
    \fmfi{double}{b--y}
    %
    \fmfi{fermion}{l .. d .. b}
    \fmfi{fermion}{b .. a .. l}
    \fmfi{fermion}{y .. f .. t}
    \fmfi{fermion}{t .. e .. y}
    \fmfi{zigzag}{l .. x .. t}
    \fmfi{zigzag}{l .. z .. t}
    \fmfi{dashes, foreground=(1,,0.1,,0.1)}{j{right} .. n -- m .. {right}v}
    \fmfiv{d.sh=circle,d.f=empty,d.si=12}{l}
    \fmfiv{d.sh=cross, l.d=15,l.a=160,d.f=full,d.si=12}{l}
    \fmfiv{d.sh=circle,d.f=empty,d.si=12}{t}
    \fmfiv{d.sh=cross, l.d=15,l.a=20,d.f=full,d.si=12}{t}
  \end{fmfgraph*}
\end{fmffile}
 & \begin{minipage}{.75cm}\centering
     {\huge $\rightarrow$} \vspace{25mm}
 \end{minipage}
 \begin{minipage}{.2cm}
     \scalebox{4}{(} \vspace{30mm}
 \end{minipage}
 &\begin{fmffile}{1S0compVert}
 \begin{fmfgraph*}(85,90)
    \fmfipair{l',l,t,x,a,b,c,d,m,n,e,f,i,j,y,z,v}
    \fmfiequ{l'}{(0,.525h)}
    \fmfiequ{l}{(0,.5h)}
    \fmfiequ{t}{(w,.5h)}
    \fmfiequ{x}{(.5w,.8h)}
    \fmfiequ{a}{(.125w,.65h)}
    \fmfiequ{b}{(.35w,.5h)}
    \fmfiequ{c}{(.75w,.25h)}
    \fmfiequ{d}{(.125w, .35h)}
    \fmfiequ{e}{(.875w,.65h)}
    \fmfiequ{f}{(.875w,.35h)}
    \fmfiequ{y}{(.65w,.5h)}
    \fmfiequ{z}{(.75w,.25h)}
    \fmfiequ{i}{(.95w,h)}
     \fmfiequ{j}{(.3w,0)}
    \fmfiequ{v}{(.75w,h)}
    \fmfiequ{m}{(.5w,.75h)}
    \fmfiequ{n}{(.5w,.25h)}
    \fmfi{double}{l--t}
    %
    \fmfiv{d.sh=circle,d.f=empty,d.si=12}{l}
    \fmfiv{d.sh=cross, l.d=15,l.a=160,d.f=full,d.si=12}{l}
    \fmfiv{d.sh=circle,d.f=empty,d.si=12}{t}
    \fmfiv{d.sh=cross, l.d=15,l.a=20,d.f=full,d.si=12}{t}
     \fmfi{dashes, foreground=(1,,0.1,,0.1)}{j{right} .. n -- m .. {right}v}
  \end{fmfgraph*}
\end{fmffile} & \begin{minipage}{.2cm}
     \scalebox{4}{)} \vspace{30mm}
 \end{minipage} &
\begin{fmffile}{electcorr}
 \begin{fmfgraph*}(85,90)
    \fmfipair{l',l,t,x,a,b,c,d,m,n,e,f,i,j,y,z,v}
    \fmfiequ{l'}{(0,.525h)}
    \fmfiequ{l}{(0,.5h)}
    \fmfiequ{t}{(w,.5h)}
    \fmfiequ{x}{(.5w,.7h)}
    \fmfiequ{a}{(.125w,.65h)}
    \fmfiequ{b}{(.35w,.5h)}
    \fmfiequ{c}{(.75w,.25h)}
    \fmfiequ{d}{(.125w, .35h)}
    \fmfiequ{e}{(.875w,.65h)}
    \fmfiequ{f}{(.875w,.35h)}
    \fmfiequ{y}{(.65w,.5h)}
    \fmfiequ{z}{(.5w,.3h)}
    \fmfiequ{i}{(.95w,h)}
     \fmfiequ{j}{(.3w,0)}
    \fmfiequ{v}{(.75w,h)}
    \fmfiequ{m}{(.5w,.75h)}
    \fmfiequ{n}{(.5w,.25h)}
    \fmfi{zigzag}{l .. x .. t}
    \fmfi{zigzag}{l .. z .. t}
    %
    \fmfiv{d.sh=circle,d.f=empty,d.si=12}{l}
    \fmfiv{d.sh=cross, l.d=11,l.a=160,d.f=full,d.si=12}{l}
    \fmfiv{d.sh=circle,d.f=empty,d.si=12}{t}
    \fmfiv{d.sh=cross, l.d=11,l.a=20,d.f=full,d.si=12}{t}
     \fmfi{dashes, foreground=(1,,0.1,,0.1)}{j{right} .. n -- m .. {right}v}
  \end{fmfgraph*}
\end{fmffile}
\end{tabular}
\vspace{-2cm}
    \caption{Factorizing out the soft double-electric dipole contributions in the $^3S_1^{[8]}$ channel. Zigzag lines represent the chromo-electric fields. Double lines represent the color-singlet composite field.}
    \label{fig: 3S18 quark loops}
\end{figure}
Using the same techniques as before, integrating out the quark loops gives (see fig. \ref{fig: 3S18 quark loops}),
\begin{equation}
\begin{aligned}
    \Otsooct =& \frac{1}{4N_cM^2} \bra{0} \bigg[ {\cal S}_v^{\dagger,ag}\frac{1}{v\cdot\Pc}  \bigg(d^{gbc} g^2{\bfcal E}^{ b} \cdot {\bfcal E}^{c} \bigg)\bigg][{\boldsymbol{\sigma} }^j S^\dagger_0(x)]\ket{V,X_s ,X_{us}}\\
     &\times\bra{V,X_s} [ S_0(x){\boldsymbol{\sigma} }^j] \bigg[ {\cal S}_v^{ag'}\frac{1}{v\cdot\Pc}  \bigg( d^{g'b'c'} g^2{\bfcal E}^{ b'} \cdot {\bfcal E}^{c'}\bigg)\bigg]\ket{0} .
\end{aligned}
\end{equation}
After separating the Hilbert spaces, we find
\begin{equation}
\begin{aligned}
    \Otsooct =& \frac{1}{4N_cM^2}  \bra{0}[{\boldsymbol{\sigma} }^j S^\dagger_0(x)]\ket{V}\bra{V} [S_0(x) {\boldsymbol{\sigma}^j}]\ket{0}\\
    &\times \bra{0} \bigg[ {\cal S}_v^{ag}\frac{1}{v\cdot\Pc}  \bigg(d^{gbc} g^2{\bfcal E}^{ b} \cdot {\bfcal E}^{c} \bigg)\bigg]^2\ket{0}.
\end{aligned}
\end{equation}
Defining the double electric field correlator,
\begin{equation}
    \braket{{\bfcal{EE}}} =  \bra{0} \Bigg|\frac12  {\cal S}_v^{ag}\bigg[\frac{1}{v\cdot\Pc}  d^{gbc} g^2{\bfcal E}^{ b} \cdot {\bfcal E}^{c} \bigg]\Bigg|^2(x)\ket{0} , 
\end{equation}
we can write this in the compact form,
\begin{equation}
    \Otsooct = \frac{1}{N_cM^2} \braket{{\bfcal{EE}}}  \frac{3|R_V(0)|^2}{2\pi}(1+{\cal O}(\vv^2))
\end{equation}
which agrees with the pNRQCD result \cite{Brambilla:2022ayc}.
\subsection{P-wave matrix elements}
\label{sec: P-wave}
To factorize the matrix elements for quark-antiquark pairs produced in P-wave configurations we will need to be slightly more careful. In traditional NRQCD, the $\Otpz$ contributions are determined by matching onto operators of the form
\begin{equation}
    \psi^\dagger \qv^i \sigma^j T^a \chi ,
\label{eq: P-wave def}
\end{equation}
where $\qv$ is the relative three-momentum between the $Q\bar{Q}$ when they are produced at the scale $\mu \sim 2m$, which has soft scaling $\qv \sim m\vv$. Often, this operator is written with a covariant derivative instead
\begin{equation}
     \psi^\dagger \frac{i}{2}\overleftrightarrow{{\bf D}}^i \sigma^j T^a \chi.
\label{eq: P-wave def}
\end{equation}
When matching onto vNRQCD, the $Q\bar{Q}$ radiates soft gluons, which changes the relative three momentum of the $Q\bar{Q}$, so that the initial $\qv$ can be decomposed into two pieces
\begin{equation}
    \qv = \sum \pv_{s,i} + \qv' ,
\end{equation}
where $\pv_{s,i} $ is the three momentum of each soft gluon, and $\qv'$ is the relative momentum of the final $Q\bar{Q'}.$ Therefore, the NRQCD P-wave operator can be matched onto the following object in vNRQCD,
\begin{equation}
    \psi^\dagger \qv^i \sigma^j T^a \chi = \sum_{\qv'}[\psi^\dagger_{\qv'}S_v^\dagger(\frac12\overleftrightarrow{{\bfcal P}}^i -g{\bf A}_k^i) \sigma^j T^a S_v \chi_{-\qv'}] ,
\label{eq: vNRQCD P-wave def}
\end{equation}
where $\overleftrightarrow{{\bfcal P}} = {\bfcal P} - {\bfcal P}^\dagger$. Note that this operator has three contributions,
\begin{equation}
\begin{aligned}
    \sum_{\qv'}&\bigg(\psi^\dagger_{\qv'}\frac12\overleftrightarrow{{\bfcal P}}^i \sigma^j \chi_{-\qv'} [{\cal S}_v^{ab} T^b ]+ \psi^\dagger_{\qv'}  [S_v^\dagger({\bfcal P}^i -g{\bf A}_k^i)   S_v] [{\cal S}_v^{ab} T^b] \sigma^j \chi_{-\qv'}\\
    +& \psi^\dagger_{\qv'}  [ {\bfcal P}^i {\cal S}_v^{ab} T^b] \sigma^j \chi_{-\qv'}\bigg),
\end{aligned}
\label{eq: Pwave contrs}
\end{equation}
where we have used $S_v^\dagger S =1$ and the brackets indicate which fields the projector can act on. The first term is proportional to the relative momentum of the $Q\bar{Q}$ after soft radiation. This term is still in a genuine P-wave state after matching onto vNRQCD and is similar to the operators written down in ref.~\cite{Fleming:2019pzj}, which were used to study P-wave quarkonium decays in the TMD formalism. 

The last two operators project out the momentum from the soft Wilson lines. Importantly, we can identify the second term as being proportional to the chromo-electric field using eq.~(\ref{eq: gauge inv electric})
\begin{equation}
\begin{aligned}
    \sum_{\qv'}\psi^\dagger_{\qv'}  [S_v^\dagger({\bfcal P}^i -g{\bf A}_k^i)   S_v] [{\cal S}_v^{ab} T^b] \sigma^j \chi_{-\qv'}  = \sum_{\qv'}\psi^\dagger_{\qv'} (g{\bfcal E}_k^i)[{\cal S}_v^{ab} T^b] \sigma^j \chi_{-\qv'} .
\end{aligned}
\label{eq: Efield Pwave}
\end{equation}
This contribution is leading in the $\vv$ power-counting with respect to the other contributions in eq.~(\ref{eq: Pwave contrs}) because it can place the state directly in a $^3S_1^{[1]}$. Projecting out this component gives the P-wave transition operator,
\begin{equation}
    \sum_{\qv'}\psi^\dagger_{\qv'} P_{^3S_1^{[1]}}\bigg[ [g{\bfcal E}_k^i]{\cal S}_v^{ab} T^b \sigma^j \bigg]\chi_{-\qv'}  = \frac{1}{2N_c}\sum_{\qv'} \psi^\dagger_{\qv'} \big[{\cal S}_v^{ab} g{\bfcal E}_k^{i,b} \big]\sigma^j \chi_{-\qv'}  .
    \end{equation}
Hence, no further transitions are necessary and it will provide the leading, at least in theory, contribution to S-wave vector quarkonium production. The direct relation of the P-wave operator to a chromo-electric field operator was also observed by ref.~\cite{Brambilla:2022ayc} when matching onto pNRQCD. 

Note, however, that the first term in eq.~(\ref{eq: Pwave contrs})  is the only true P-wave operator at the soft scale. Therefore, there is potentially some ambiguity in the matching, and it could be the case that this contribution is the correct operator to match the $\Otpz$ onto. To contribute to S-wave vector quarkonium production, this operator will need additional chromo-electric field transitions to be placed in a $^3S_1^{[1]}$ state. We find the final contribution for the ``genuine P-wave operator" to be
\begin{equation}
\begin{aligned}
    \frac{g}{2m}\sum_{\qv'}\psi^\dagger_{\qv'}P_{^3S_1^{[1]}}\big[\{S_v^\dagger \qv^{'j} \sigma^i T^a  S_v , \frac{1}{v\cdot  \Pc}\qv' \cdot {\bfcal E}\}\big]\chi_{-\qv'} \\
    = \sum_{\qv'}\psi^\dagger_{\qv'}\frac{g}{2N_c m}\psi^\dagger_{\qv'}  \sigma^i \qv^{'j} \qv^{'k}\bigg[ {\cal S}_v^{ab} \frac{1}{v\cdot  \Pc}{\bfcal E}^{k,b}\bigg]\chi_{-\qv'}.
\end{aligned}
\label{eq: genuine Pwave transition}
\end{equation}
where the derivation is given in appendix~\ref{sec: subleading P-wave}. 

\subsubsection{Factorizing P-wave states}
We first match the $\Otpz$ matrix element onto the leading in v  vNRQCD operator derived in the previous section,
\begin{equation}
\begin{aligned}
    \Otpz =  \frac{g^2}{4N_c^2 m^2}{\cal T}^{ij,i'j'}_J\sum_X\bra{0}\chi^\dagger_{-\qv}\sigma^i \big[S_v^{\dagger,ab}{\bfcal E}^{j,b}\big] \psi_\qv\ket{V,X}\\
    \times\bra{V,X}\psi^\dagger_{\qv}  \sigma^{i'} \big[ {\cal S}_v^{ab'} {\bfcal E}^{j',b'}\big]\chi_{-\qv}\ket{0}.
\end{aligned}
 \end{equation}
where the ${\cal T}^{ij,i'j'}$ projectors are defined as \cite{Fleming:2019pzj, Brambilla:2022ayc, Echevarria:2024idp}
\begin{equation}
\begin{aligned}
    {\cal T}_0^{ij,i'j'} &= \frac13 \delta^{ij}\delta^{i'j'} , \\
    {\cal T}_1^{ij,i'j'} &= \frac12 \epsilon^{kim}\epsilon_{ki'n}\delta^{mj}\delta^{nj'} , \\
    {\cal T}_2^{ij,i'j'} & = \bigg(\frac{\delta_{im}\delta_{nj}+ \delta_{in}\delta_{jm}}{3}-\frac13 \delta_{mn}\delta_{ij}\bigg)\bigg(\frac{\delta_{i'm}\delta_{nj'}+ \delta_{i'n}\delta_{j'm}}{3}-\frac13 \delta_{mn}\delta_{i'j'}\bigg) . 
\end{aligned}
\label{eq: Pwave projectors}
\end{equation}
After inserting a color singlet composite field interaction and integrating out the quark loops (fig. {\ref{fig: 3PJ8 loops}), we find the P-wave operator reduces to the simple form
  \begin{equation}
\begin{aligned}
    \Otpz =  \frac{g^2}{4N_c m^2}{\cal T}^{ij,i'j'}_J\sum_X\bra{0}\big[{\cal S}_v^{\dagger,ab}{\bfcal E}^{j,b}\big] \big[\sigma^i S^\dagger_0(x)\big] \ket{V,X}\\
    \times\bra{V,X}\big[ S_0(x) \sigma^{i'}\big]\big[ {\cal S}_v^{ab'}{\bfcal E}^{j',b'}\big]\ket{0}.
\end{aligned}
 \end{equation}
Once again, we can factorize the Hilbert space to write this expression as
\begin{figure}
    \centering
    \begin{tabular}{ c c c c c}
    \begin{fmffile}{QL3PJ}
  \begin{fmfgraph*}(120,90)
    \fmfipair{l',l,t,x,a,b,c,d,m,n,e,f,i,j,y,z,v}
    \fmfiequ{l'}{(0,.525h)}
    \fmfiequ{l}{(0,.5h)}
    \fmfiequ{t}{(w,.5h)}
    \fmfiequ{x}{(.5w,.8h)}
    \fmfiequ{a}{(.125w,.75h)}
    \fmfiequ{b}{(.35w,.5h)}
    \fmfiequ{c}{(.75w,.25h)}
    \fmfiequ{d}{(.125w, .25h)}
    \fmfiequ{e}{(.875w,.75h)}
    \fmfiequ{f}{(.875w,.25h)}
    \fmfiequ{y}{(.65w,.5h)}
    \fmfiequ{z}{(.75w,.25h)}
    \fmfiequ{i}{(.95w,h)}
    \fmfiequ{j}{(.35w,0)}
    \fmfiequ{v}{(.7w,h)}
    \fmfiequ{m}{(.5w,.75h)}
    \fmfiequ{n}{(.5w,.25h)}
    \fmfi{double}{b--y}
    %
    \fmfi{fermion}{l .. d .. b}
    \fmfi{fermion}{b .. a .. l}
    \fmfi{fermion}{y .. f .. t}
    \fmfi{fermion}{t .. e .. y}
    \fmfi{zigzag}{l .. x .. t}
    \fmfi{dashes, foreground=(1,,0.1,,0.1)}{j{right} .. n -- m .. {right}v}
    \fmfiv{d.sh=circle,d.f=empty,d.si=12}{l}
    \fmfiv{d.sh=cross, l.d=15,l.a=160,d.f=full,d.si=12}{l}
    \fmfiv{d.sh=circle,d.f=empty,d.si=12}{t}
    \fmfiv{d.sh=cross, l.d=15,l.a=20,d.f=full,d.si=12}{t}
  \end{fmfgraph*}
\end{fmffile}
 & \begin{minipage}{.75cm}\centering
     {\huge $\rightarrow$} \vspace{25mm}
 \end{minipage}
 \begin{minipage}{.2cm}
     \scalebox{4}{(} \vspace{30mm}
 \end{minipage}
 &\begin{fmffile}{1S0compVert}
 \begin{fmfgraph*}(85,90)
    \fmfipair{l',l,t,x,a,b,c,d,m,n,e,f,i,j,y,z,v}
    \fmfiequ{l'}{(0,.525h)}
    \fmfiequ{l}{(0,.5h)}
    \fmfiequ{t}{(w,.5h)}
    \fmfiequ{x}{(.5w,.8h)}
    \fmfiequ{a}{(.125w,.65h)}
    \fmfiequ{b}{(.35w,.5h)}
    \fmfiequ{c}{(.75w,.25h)}
    \fmfiequ{d}{(.125w, .35h)}
    \fmfiequ{e}{(.875w,.65h)}
    \fmfiequ{f}{(.875w,.35h)}
    \fmfiequ{y}{(.65w,.5h)}
    \fmfiequ{z}{(.75w,.25h)}
    \fmfiequ{i}{(.95w,h)}
     \fmfiequ{j}{(.3w,0)}
    \fmfiequ{v}{(.75w,h)}
    \fmfiequ{m}{(.5w,.75h)}
    \fmfiequ{n}{(.5w,.25h)}
    \fmfi{double}{l--t}
    %
    \fmfiv{d.sh=circle,d.f=empty,d.si=12}{l}
    \fmfiv{d.sh=cross, l.d=15,l.a=160,d.f=full,d.si=12}{l}
    \fmfiv{d.sh=circle,d.f=empty,d.si=12}{t}
    \fmfiv{d.sh=cross, l.d=15,l.a=20,d.f=full,d.si=12}{t}
     \fmfi{dashes, foreground=(1,,0.1,,0.1)}{j{right} .. n -- m .. {right}v}
  \end{fmfgraph*}
\end{fmffile} & \begin{minipage}{.2cm}
     \scalebox{4}{)} \vspace{30mm}
 \end{minipage} &
\begin{fmffile}{eleccorr}
 \begin{fmfgraph*}(85,90)
    \fmfipair{l',l,t,x,a,b,c,d,m,n,e,f,i,j,y,z,v}
    \fmfiequ{l'}{(0,.525h)}
    \fmfiequ{l}{(0,.5h)}
    \fmfiequ{t}{(w,.5h)}
    \fmfiequ{x}{(.5w,.8h)}
    \fmfiequ{a}{(.125w,.65h)}
    \fmfiequ{b}{(.35w,.5h)}
    \fmfiequ{c}{(.75w,.25h)}
    \fmfiequ{d}{(.125w, .35h)}
    \fmfiequ{e}{(.875w,.65h)}
    \fmfiequ{f}{(.875w,.35h)}
    \fmfiequ{y}{(.65w,.5h)}
    \fmfiequ{z}{(.75w,.25h)}
    \fmfiequ{i}{(.95w,h)}
     \fmfiequ{j}{(.3w,0)}
    \fmfiequ{v}{(.75w,h)}
    \fmfiequ{m}{(.5w,.75h)}
    \fmfiequ{n}{(.5w,.25h)}
    \fmfi{zigzag}{l--t}
    %
    \fmfiv{d.sh=circle,d.f=empty,d.si=12}{l}
    \fmfiv{d.sh=cross, l.d=11,l.a=160,d.f=full,d.si=12}{l}
    \fmfiv{d.sh=circle,d.f=empty,d.si=12}{t}
    \fmfiv{d.sh=cross, l.d=11,l.a=20,d.f=full,d.si=12}{t}
     \fmfi{dashes, foreground=(1,,0.1,,0.1)}{j{right} .. n -- m .. {right}v}
  \end{fmfgraph*}
\end{fmffile}
\end{tabular}
\vspace{-2cm}
    \caption{Factorizing out the soft chromo-magnetic dipole contributions in the $^3P_J^{[8]}$ channel. Zigzag line represents the chromo-electric field. Double lines represent the color-singlet composite field.}
    \label{fig: 3PJ8 loops}
\end{figure}
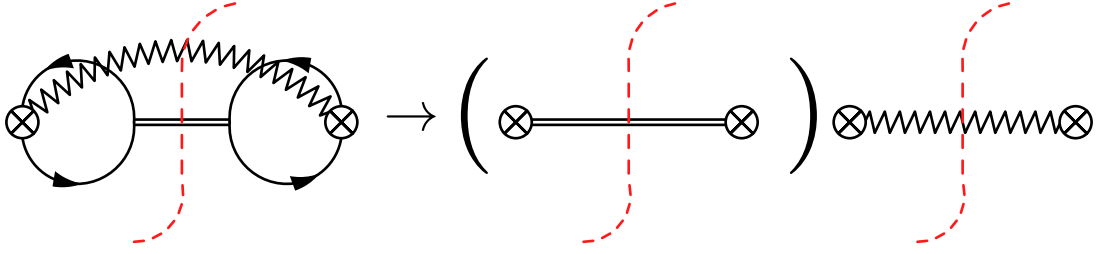
\begin{equation}
\begin{aligned}
    \Otpz = & \frac{g^2}{12N_c } {\cal T}^{ij,i'j'}_J\bra{0} \big[\sigma^i S^\dagger_0(x)\big] \ket{V}\bra{V}\big[ S_0(x) \sigma^{i'} \big]\ket{0}\\
    &\times\bra{0} \big[ {\cal S}_v^{ab} {\bfcal E}^{j,b}\big]\big[ {\cal S}_v^{ab'}{\bfcal E}^{j,b'}\big]\ket{0}, 
\end{aligned}
 \end{equation}
where we have averaged over the Cartesian indices in the second line. 
Finally, after averaging over spins and summing over polarizations, we can write the factorized P-wave matrix element as
\begin{equation}
    \Otpz = \frac{1}{36 N_c} (2J+1) \frac{3\big|R_V(0)\big|^2}{2\pi} \braket{{\bfcal E}} , 
\label{eq: leading 3PJ fact}
\end{equation}
where
\begin{equation}
    \braket{{\bfcal E}} = \bra{0} \big[ {\cal S}_v^{ab} {\bfcal E}^{j,b}\big]^2\ket{0}.
\end{equation}
%
%
Equation (\ref{eq: leading 3PJ fact}) agrees with the result determined using pNRQCD in ref.~\cite{Brambilla:2022ayc}. 

 Next, we match onto the genuine P-wave operator at the soft scale (eq.~\ref{eq: genuine Pwave transition}) and find instead 
\begin{equation}
    \Otpz^{(1)} = \frac{(m \braket{\vv^2}_V)^2}{108 N_c} (2J+1) \frac{3\big|R_V(0)\big|^2}{2\pi} \braket{{\bfcal E}}^{(1)} ,
\label{eq: gen 3PJ fact}
\end{equation}
where the result is now proportional to the binding energy of the quarkonium state ($m \braket{\vv^2}_V = M_V - 2m$) and a different electric field correlator
\begin{equation}
    \braket{{\bfcal E}}^{(1)} = \bra{0} \bigg[ {\cal S}_v^{ab} \frac{1}{v\cdot  \Pc}{\bfcal E}^{j,b}\bigg]^2\ket{0}.
\end{equation}
This derivation can be found in appendix~\ref{sec: subleading P-wave}.

\section{Discussion and applications}
\label{sec: discussion}
The purpose of this work was to show that the soft scale can be factorized from the other sectors in quarkonium production matrix elements in the vNRQCD framework which keeps soft degrees of freedom explicit Using the pNRQCD formulation, ref.~\cite{Brambilla:2022ayc} showed that important relations exist between the NRQCD production matrix elements for different S-wave quarkonium states. Here we discuss these relations in the context of the vNRQCD approach and highlight key differences in our findings.  

\subsection{Universality of the S-wave vector quarkonium LDMEs}
First, we note that the chromo-electric and chromo-magnetic field correlators in the precious section are completely universal. That is, they are independent of state and are universally defined, regardless of the S-wave vector quarkonium state being measured. Therefore, taking ratios of the LDMEs between two different vector quarkonium states, $V$ and $V'$ produces powerful constraints on the color-octet matrix elements,
\begin{equation}
\begin{aligned}
    &\braket{O^V(^3S_1^{[8]})}(\mu) = \frac{m_{Q'}^2}{m_{Q}^2} \frac{|R_V (0)|^2}{|R_{V'}(0)|^2} \braket{O^{V'}(^3S_1^{[8]})}(\mu) , \\
    &\braket{O^V(^3P_J^{[8]})}= \frac{\big|R_V(0)\big|^2}{\big|R_{V'}(0)\big|^2} \braket{O^{V'}(^3P_J^{[8]})}, \\
    &\braket{O^V(^1S_0^{[8]})}= \frac{m_{Q'}^2}{m_{Q}^2}\frac{c_F^2(m_{Q},\mu)}{c_F^2(m_{Q'},\mu)}\frac{|R_V (0)|^2}{|R_{V'}(0)|^2} \braket{O^{V'}(^1S_0^{[8]})}, \\
\end{aligned}
\label{eq: LDME relations}
\end{equation}
where the running of the matrix elements was derived in ref.~\cite{Brambilla:2022ayc}. These ratios agree exactly with the relationships first derived using pNRQCD \cite{Brambilla:2022ayc, Brambilla:2022rjd}. Importantly, if one considers instead the ``genuine P-wave contribution" from eq.~(\ref{eq: gen 3PJ fact}), then the analogous relation between the matrix elements of different states would be given by
\begin{equation}
    \braket{O^V(^3P_J^{[8]})}^{(1)}= \bigg(\frac{m_{Q} \braket{\vv^2}_V}{m_{Q'} \braket{\vv^2}_{V'}}\bigg)^2\frac{\big|R_V(0)\big|^2}{\big|R_{V'}(0)\big|^2} \braket{O^{V'}(^3P_J^{[8]})}^{(1)}.
    \label{eq: LDME Pwave Newrelation}
\end{equation}

The connections between LDMEs in eqs.~(\ref{eq: LDME relations}) and~(\ref{eq: LDME Pwave Newrelation}) are quite remarkable. In essence, they reduce the number of non-perturbative parameters for S-wave vector quarkonium production from~12 to~3. These results have already been exploited to some degree. Using matrix elements determined from a global analysis of $J/\psi$ production data in order to determine $\Upsilon(3S)$ data \cite{Brambilla:2024iqg}, where good agreement with the data is found. In future studies, these relations can be used to more precisely constrain the LDMEs of all S-wave vector quarkonium states as the number of free parameters are greatly reduced and the number of available measurements grows. Such an analysis should further significantly reduce the uncertainties on the quarkonium production matrix elements, such as those extracted via global analysis in table~\ref{tab: LDMEs}.

\subsection{Velocity scaling of the LDMEs}
Since soft gluons and soft momenta scale homogeneously as $m\vv$, we can power count the matrix elements and vacuum correlators to estimate the importance of each production mechanism. 
First, we can start with $\Otsosing$ since it is leading in the power counting and is known to scale as
\begin{equation}
    \Otsosing \sim \vv^3 .
\end{equation}
All factors of the wave function at the origin also scale as 
\begin{equation}
    |R_{V'}(0)|^2 \sim \vv^3.
\end{equation}

Now, for the color-octet production matrix elements, we need to estimate the size of the chromo-magnetic, chromo-electric, and double chromo-electric vacuum correlators. In the vNRQCD power counting,
\begin{equation}
\begin{aligned}
\braket{{\bfcal B}} &\sim \alpha_s(m\vv)\vv^2 ,\\
\braket{{\bfcal{EE}}} &\sim \alpha_s(m\vv)^2\vv^2 , \\
\braket{{{\bfcal E}}} &\sim \alpha_s(m\vv) \vv^2.\\
\end{aligned}
\end{equation}
Using the defining assumption that $\alpha_s(m\vv) \sim \vv$ in Coulombic systems, we find that the $\Oosz$ and $\Otsooct$ LDMEs scale as $\vv^6$ and $\vv^7$, respectively,
\begin{equation}
\begin{aligned}
    \Oosz \propto \braket{{\bfcal B}} |R_V(0)|^2 \sim \vv^6 , \\
    \Otsooct \propto \braket{{{\bfcal{EE}}}} |R_V(0)|^2 \sim \vv^7 \\
    \Otpz \propto \braket{{{\bfcal E}}} |R_V(0)|^2 \sim \vv^6 \\
\end{aligned}
\end{equation}

Lastly, the ``genuine P-wave contribution" that we have identified contains a correlator that scales as $\alpha_s(m\vv) \sim \vv$. However, as discussed in appendix~\ref{sec: subleading P-wave}, this matching is not proportional to the wave function at the origin, but instead,
\begin{equation}
    \big|\overline{\nabla^2R}_V^{(0)}(0)\big|^2 \sim \braket{\vv}_V^4|R_V(0)|^2 \sim \vv^7.
\end{equation}
Therefore the soft contribution to the $\Otpz^{(1)}$ matrix element must scale as $\alpha_s(m\vv)\vv^7 \sim \vv^8$ and is hence suppressed by $\vv$ with respect to the leading contributions. The scaling of each matrix element is summarized in table \ref{tab: LDME scaling}.

\begin{table}[htbp]
\centering
\begin{tabular}{c|c|c|c|c|c}
\hline
& $\begin{aligned}\Otsosing\end{aligned}$ & $\begin{aligned}\Otsooct \,\end{aligned}$ & $\begin{aligned}\Oosz\end{aligned}$ & $\begin{aligned}\Otpz/m^2\end{aligned}$ &  $\begin{aligned}\Otpz^{(1)}/m^2\end{aligned}$\\
\hline
$\vv$-scaling  & $\vv^3$ & $\vv^7$ & $\vv^6$ & $\vv^6$ & $\vv^8$\\
\hline
\end{tabular}
\caption{Scaling of LDMEs predicted from this analysis. }
\label{tab: LDME scaling}
\end{table}

 Even though the genuine P-wave contribution in eq.~(\ref{eq: gen 3PJ fact}) is suppressed in the $\vv$ power-counting with respect to eq.~(\ref{eq: leading 3PJ fact}), this does not necessarily mean that it can be neglected. A recent analysis of $J/\psi$ production data \cite{Brambilla:2024iqg} showed that the NRQCD predictions for $J/\psi$ produced at the LHCb overshoots the data for $p_T < 9$ GeV, even at moderate $6 $ GeV $\lesssim p_T $, where NRQCD factorization should still be valid. Similar results are seen in comparisons with HERA \cite{Brambilla:2024iqg}, particularly at moderate to large $z$, but also surprisingly in the $0.1 < z < 0.3$ bin. These disagreements can be entirely attributed to the P-wave color-octet mechanism, which grow rapidly as $p_T$ shrinks. If the P-wave operator were actually more suppressed than initial estimates suggest, as it would be if one were to instead match onto the genuine P-wave contribution that we have identified using eq.~(\ref{eq: gen 3PJ fact}), then this would resolve the apparent tension. Alternatively, it could be that the power counting in the theory changes as $p_T$ decreases (which is known to happen in SIDIS in the TMD formalism \cite{Echevarria:2024idp, Copeland:2025vop}), and the contribution in eq.~(\ref{eq: gen 3PJ fact}) becomes enhanced. This could cancel with the leading term in eq.~(\ref{eq: leading 3PJ fact}) to generate a smaller P-wave channel at low $p_T$. While a RG analysis (in appendix \ref{sec: subleading P-wave RG}) shows that eq.~(\ref{eq: leading 3PJ fact}) is a necessary contribution to reproduce the $^3S_1^{[8]}$ mixing with the $^3P_J^{[8]}$ under RG, a well established fact in traditional NRQCD, the clear disagreement of the P-wave mechanism with experimental measurements in ref.~\cite{Brambilla:2024iqg} shows that other contributions, such as eq.~(\ref{eq: gen 3PJ fact}) may still need to be considered. A more thorough analysis of the P-wave color-octet channel is warranted in future work. 

Lastly, the scalings derived in this section deviate from the traditional NRQCD arguments which state each color-octet LDME must scale as $\vv^7$. This is because the non-perturbative transitions considered in this paper (and the analysis of ref.~\cite{Brambilla:2022ayc}) occur via soft gluon radiation, not ultrasoft. As mentioned in sec.~\ref{sec: background}, ultrasoft transitions for the color-octet operators to $^3S_1^{[1]}$ states will give contributions that scale as $\vv^7$. This implies that the soft transitions for the $\Otsooct$ are the same size as the ultrasoft. This is something that requires further examination. In the next section, we will provide two arguments that show why the ultrasoft power-correction can be neglected so that the relations in eq.~(\ref{eq: LDME relations}) still hold. 

\subsection{Ultrasoft transitions in the matrix elements}
As we saw in the previous section, the soft transitions do not clearly lead in the $\vv$ power-counting for the $\Otsooct$ LDME with respect to possible ultrasoft transitions. Therefore, let us consider what may happen if we have ultrasoft transitions instead. 

For the $\Otsooct$ one of the ultrasoft operators that mediates the transition of a $Q\bar{Q}$ in a $^3S_1^{[8]}$ state to a $^3S_1^{[1]}$ state is the double chromo-electric field interaction, 
\begin{equation}
    \frac{1}{2m}\tilde{\psi}_\pv^{\dagger} \tilde{{\bfcal E}}_{us}^2\tilde{\psi}_\pv' .
\end{equation}
Unlike the soft emission, when the radiation is ultrasoft the $Q\bar{Q}$ pair can actually propagate before radiating off the chromo-electric fields. Therefore, to study the transition, we need to consider the diagrams where this interaction is inserted on the $Q\bar{Q}$ loop. One such insertion is shown in fig. \ref{fig: 3S18 usoft loops}.
\begin{figure}
    \centering
    \begin{tabular}{ c }
    \begin{fmffile}{QL3S18us}
  \begin{fmfgraph*}(120,90)
    \fmfipair{l',l,t,x,a,b,c,d,m,n,e,f,i,j,y,z,v,q}
    \fmfiequ{l'}{(0,.525h)}
    \fmfiequ{l}{(0,.5h)}
    \fmfiequ{t}{(w,.5h)}
    \fmfiequ{x}{(.5w,.9h)}
    \fmfiequ{a}{(.125w,.75h)}
    \fmfiequ{b}{(.35w,.5h)}
    \fmfiequ{c}{(.75w,.25h)}
    \fmfiequ{d}{(.125w, .25h)}
    \fmfiequ{e}{(.875w,.75h)}
    \fmfiequ{f}{(.875w,.25h)}
    \fmfiequ{y}{(.65w,.5h)}
    \fmfiequ{z}{(.75w,.25h)}
    \fmfiequ{i}{(.95w,h)}
    \fmfiequ{q}{(.5w,.6h)}
    \fmfiequ{j}{(.35w,0)}
    \fmfiequ{v}{(.7w,h)}
    \fmfiequ{m}{(.5w,.75h)}
    \fmfiequ{n}{(.5w,.25h)}
    \fmfi{double}{b--y}
    %
    \fmfi{fermion}{l .. d .. b}
    \fmfi{fermion}{b .. a .. l}
    \fmfi{fermion}{y .. f .. t}
    \fmfi{fermion}{t .. e .. y}
    \fmfi{gluon}{a .. x .. e}
    \fmfi{gluon}{a .. q .. e}
    \fmfi{dashes, foreground=(1,,0.1,,0.1)}{j{right} .. n -- m .. {right}v}
    \fmfiv{d.sh=circle,d.f=empty,d.si=12}{l}
    \fmfiv{d.sh=cross, l.d=15,l.a=160,d.f=full,d.si=12}{l}
    \fmfiv{d.sh=circle,d.f=empty,d.si=12}{t}
    \fmfiv{d.sh=cross, l.d=15,l.a=20,d.f=full,d.si=12}{t}
  \end{fmfgraph*}
\end{fmffile}
\end{tabular}
    \caption{Quark/antiquark loops in the $^3S_1^{[8]}$ channel with ultrasoft emissions.}
    \label{fig: 3S18 usoft loops}
\end{figure}
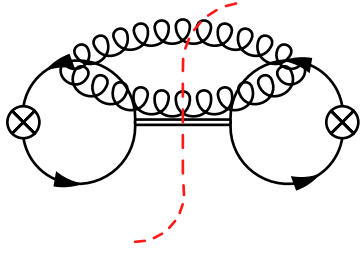

Integrating out the quark loops from the diagram yields,

\begin{equation}
\begin{aligned}
   &\frac{g_{us}^4}{N_c}  \sum_{X_s} \bra{0}\sigma^i \bigg[i\int d^3 \rv\int\frac{d^4 q}{(2\pi)^4} \frac{1}{E+\bar{p}^0+q^0 - \frac{{\bf q}^2}{2m}}\frac{1}{E+q^0 - \frac{{\bf q}^2}{2m}}\frac{1}{E -q^0 - \frac{{\bf q}^2}{2m}} \\
   & \qquad \qquad 
   \times  {\bfcal E}_{us}^2 V^{(1)}(r) e^{i\qv \cdot \rv} S_\rv^\dagger\bigg]\ket{V ,X_s}\bra{V ,X_s}  \bigg[-i\int d^3 \rv'\int\frac{d^4 q'}{(2\pi)^4} S_{\rv'}   V^{(1)}(r')  {\bfcal E}_{us}^2  e^{i\qv' \cdot \rv'} \\
   &\qquad \qquad  \times\frac{1}{E+q^{'0 }- \frac{{\bf q'}^2}{2m}}\frac{1}{E -q^{'0}  - \frac{{\bf q'}^2}{2m}} \frac{1}{E+\bar{p}^0+q'^0 - \frac{{\bf q'}^2}{2m}}\bigg]  \sigma^i  \ket{0} \,\\
   =& \frac{g_{us}^4}{N_c} \lim_{{\bf a}\to 0} \sum_{X_s} \bra{0}\sigma^i \bigg[S_\rv^\dagger \frac{1}{V^{(1)}(r) +  \bar{p}^0} {\bfcal E}_{us}^2  \bigg]_{\rv ={\bf a}}\ket{V ,X_s}\\
   &\qquad \qquad  \times\bra{V ,X_s}   \bigg[{\bfcal E}_{us}^2 \frac{1}{V^{(1)}(r') + \bar{p}^0} S_{\rv'}   \bigg]_{\rv'={\bf a}}  \sigma^i  \ket{0} , 
\end{aligned}
\label{eq: us 3S18}
\end{equation}
where we have evaluated the integral using the same techniques and employed the equation of motion for the color singlet field,
\begin{equation}
    \bigg(E_{Q\bar{Q}} - \frac{\nabla_\rv^2}{m}\bigg) S_\rv = V^{(1)}(r) S_\rv.
\end{equation}
As we saw in sec. \ref{sec: soft decouple} when considering a soft seagull vertex on the quark loop, evaluating the limit ${\bf a} \to 0$ forces eq.~(\ref{eq: us 3S18}) to 0 because the potential is singular at $\rv = 0$. This is a direct consequence of the LDMEs being defined as local operators with no radial separation, as we discussed in sec. \ref{sec: zero radial}. This same sequence occurs for all other ultrasoft transitions because the quarks and antiquark fields will always stay near their mass shell and propagate before the interaction. The zero-radial separation from the production operator will always kill these diagrams. Therefore, we argue that ultrasoft transitions can be neglected. Higher order terms in the multipole expansion of eq.~(\ref{eq: multipole expan}) will allow for operators with non-zero radial separation, but these contributions will be power-suppressed.

Formally, the matrix element in eq.~(\ref{eq: us 3S18}) is non-perturbative and its evaluation should not be taken too seriously. It is possible that non-perturbative effects will generate new scales in the matrix element and shift the radial separation to a non-zero value. Hence it may still be relevant. However, even still, ultrasoft transitions should be sub-leading because they occur at later time-scales than the soft transitions considered throughout this paper. That is, ultrasoft transitions will occur at time-scales $\tau_{us} \sim 1/m\vv^2$ while the soft transitions occur at times $\tau_s \sim 1/m\vv$. If the assumption that the soft and ultrasoft scales are widely separated is correct, $m\vv \gg m\vv^2$, then $\tau_s \ll \tau_{us}$ so the time-ordering of the matrix elements will guarantee that the soft transition will occur first. These transitions will place the state in a $^3S_1^{[1]}$ configuration. Further ultra-soft transitions will be unnecessary and power-suppressed.

\subsection{Applications to the TMD framework}
In the TMD framework, the production matrix operators have been identified to be not the LDMEs, but TMD Shape Functions (TMDShFs)~\cite{Echevarria:2024idp} and TMD Soft Transition Functions (TMDSTFs)~\cite{Copeland:2025vop}. TMDSTFs were identified to be leading in the vNRQCD power-counting. Therefore, we will apply the factorization techniques developed in this paper to the $^1S_0^{[8]}$ TMDSTF of ref.~\cite{Copeland:2025vop} to address quarkonium production in SIDIS in the TMD framework. At leading order, the gauge invariant TMDSTF is given by
\begin{equation}
\begin{aligned}
      &T^V_{^1S_0^{[8]} }(\bv_T;\mu,\eta)=\\
      &\qquad \qquad \frac{1  }{M^2N_c^2\sqrt{S(\bv_T)}}\sumint_{X_s}\bra{0}\big[{\cal S}_v^{\dagger, bc} {\cal S}_n^{\dagger ba}  \chi^\dagger_{\bf p_{\bar{Q}}}\bigg[\frac{1}{v\cdot\Pc}   gc_F{\boldsymbol{\sigma} \cdot {\bfcal B}}_s^c \bigg] \psi_{\bf p_{Q}}\big]({\bf b}_T) \ket{V,X_s}\\
      &\qquad \qquad  \times\bra{V, X_s}\big[ {\cal S}_n^{ b'a} {\cal S}_v^{b'c'}\psi^\dagger_{\bf p_{Q}}\bigg[\frac{1}{v\cdot\Pc}   gc_F{\boldsymbol{\sigma} \cdot {\bfcal B}}_s^{c'} \bigg] \psi_{\bf p_{Q}} ](0)\ket{0} .
\end{aligned}
\label{eq: TMDSTF}
\end{equation}
Note that there is a difference from ref.~\cite{Copeland:2025vop} by a factor of 4 in the overall normalization, which we have absorbed into definition of the chromo-magnetic gluon field. Importantly, the TMDSTF is process dependent because it contains soft radiation from the initial state in the form of $n$-directional Wilson lines, ${\cal S}_n$. This  is a consequence of the cross section factorization occurring for $Q\sim M_V$, since in this regime the soft-collinear effective theory \cite{Bauer:2000yr} soft scale is parametrically similar to the vNRQCD soft scale $\qv_T \sim m\vv$. It is also because the matrix element is defined in vNRQCD where the soft scale interacts with the potential sector of the quarks and antiquarks. Therefore, they cannot be disentangled from each other. 

Invoking the composite field techniques of sec. \ref{sec: soft decouple}, we can integrate out the quarks/antiquarks and re-write the matrix element as 
\begin{equation}
\begin{aligned}
    &T^V_{^1S_0^{[8]} }(\bv_T;\mu,\eta)=\\
      &\qquad \qquad \frac{1  }{M^2N_c\sqrt{S(\bv_T)}}\sumint_{X_s}\bra{0}\bigg[ {\cal S}_v^{\dagger, bc} {\cal S}_n^{\dagger ba}  \frac{1}{v\cdot\Pc}   gc_F{\boldsymbol{\sigma} \cdot {\bfcal B}}_s^c \bigg]^i ({\bf b}_T)  \big[\sigma^i S^\dagger_0(\bv_T)\big]\ket{V,X_s}\\
      &\qquad \qquad  \times\bra{V, X_s}\bigg[ {\cal S}_n^{ b'a} {\cal S}_v^{b'c'} \frac{1}{v\cdot\Pc}   gc_F{\boldsymbol{\sigma} \cdot {\bfcal B}}_s^{c'} \bigg]^j(0) \big[\sigma^j S_0(0)\big] \ket{0}.
\end{aligned}
\end{equation}
Once the quark and antiquark fields are integrated out, we know the composite field and soft Hilbert spaces factorize as in eq.~(\ref{eq: fact Hilbert}), so the TMDSTF reduces to
\begin{equation}
\begin{aligned}
    &T^V_{^1S_0^{[8]} }(\bv_T;\mu,\eta)=\\
      &\qquad \qquad  \frac{1  }{M^2N_c\sqrt{S(\bv_T)}}\bra{0} \bigg[{\cal S}_v^{\dagger, bc} {\cal S}_n^{\dagger ba}  \frac{1}{v\cdot\Pc}   gc_F{\boldsymbol{\sigma} \cdot {\bfcal B}}_s^c \bigg]^i ({\bf b}_T)  \bigg[{\cal S}_n^{ b'a} {\cal S}_v^{b'c'}\frac{1}{v\cdot\Pc}   gc_F{\boldsymbol{\sigma} \cdot {\bfcal B}}_s^{c'} \bigg]^j(0) \ket{0}\\
      &\qquad \qquad  \times\bra{0}\big[\sigma^i S^\dagger_0(\bv_T)\big]\ket{V}\bra{V}\big[\sigma^j S_0(0)\big]\ket{0}.
\end{aligned}
\end{equation}
Since the composite field only carries ultrasoft momentum we can multi-pole expand away it's dependence on $\bv_T$,
\begin{equation}
    S_0(\bv_T) \approx S_0(0) + {\cal O}(\bv_T m\vv^2).
\end{equation}
After averaging over spins and summing over polarizations, we can write the TMDSTF in a simplified form,
\begin{equation}
    T^V_{^1S_0^{[8]} }(\bv_T; \mu, \eta) = \frac{1}{3N_cM^2} \braket{{{\bfcal B}_n}(\bv_T; \mu, \eta)} \frac{3|R_V(0)|^2}{2\pi} ,
\label{eq: fact TMDSTF}
\end{equation}
where 
\begin{equation}
\begin{aligned}
    \braket{{{\bfcal B}_n}(\bv_T;\mu, \eta)} = \frac{1  }{\sqrt{S(\bv_T)}}{{\rm Tr}_c}\bra{0} \bigg[{\cal S}_v^{\dagger, bc} {\cal S}_n^{\dagger ba} g \frac{c_F}{v\cdot\Pc}   {{\bfcal B}}_s^c \bigg]^i ({\bf b}_T)  \bigg[ \bigg[{\cal S}_n^{ b'a} {\cal S}_v^{b'c'}g\frac{c_F}{v\cdot\Pc}   {{\bfcal B}}_s^a \bigg]^i(0) \ket{0}
\end{aligned}
\end{equation}
is the process-dependent chromo-magnetic vacuum correlator. 

The factorized expression in eq.~(\ref{eq: fact TMDSTF}) has two distinct advantages over the initial definition in eq.~(\ref{eq: TMDSTF}). First, all of the $\bv_T$ dependence is encoded in the vacuum matrix element $\braket{{{\bfcal B}_n}(\bv_T;\mu, \eta)}$, which, while still non-peturbative, can in principle be computed using lattice QCD. Second, the vacuum correlator $\braket{{{\bfcal B}_n}(\bv_T;\mu, \eta)}$ is {\it universal}. This implies that the following relationship for the TMDSTFs of different quarkonium states
\begin{equation}
    T^V_{^1S_0^{[8]} }(\bv_T;\mu, \eta)  = \frac{m_{Q'}^2}{m_Q^2}\frac{c_F^2(m_Q,\mu)}{c_F^2(m_{Q'},\mu)}\frac{|R_V(0)|^2}{|R_{V'}(0)|^2} T^{V'}_{^1S_0^{[8]} }(\bv_T;\mu, \eta).
\label{eq: TMDSTF relationship}
\end{equation}
where we have evolved the TMDSTFs for $V$ and $V'$ to the same scale. This is remarkably useful because, while the TMDSTFs are not process-independent, the relationship in eq.~(\ref{eq: TMDSTF relationship}) implies a {\it state-independence} for quarkonium production in SIDIS at small $\qv_T$.  Consequently, the TMDSTFs for different quarkonium states can be related to each other, dramatically improving the predictive power of the TMD framework. For example, if one measured $\Upsilon(1S)$ production at the EIC, then in principle the TMDSTFs for $J/\psi$ produced at the EIC will be constrained. Therefore the only free parameter in a TMD factorization of the $e+p\to J/\psi +X$ cross section, would be the gluon TMDs - thereby enabling a clean extraction.

\section{Conclusion}
\label{sec: conclusion}

In this paper, we investigated the similarities and differences between vNRQCD and pNRQCD with regard to the  factorization of LDMEs in quarkonium production. 
Our analysis verified \cite{Brambilla:2022ayc} and derived powerful new factorization theorems for quarkonium production matrix elements in NRQCD. To accomplish this, we used a Hubbard-Stratonovich transformation on the vNRQCD Lagrangian and argued that the soft sector decouples during production. We discussed how the color-octet LDMEs can be matched onto ``transition operators" which mediate non-perturbative the color-octet to color-singlet transition via soft gluon radiation in the vNRQCD framework. We then employed the Hubbard-Stratonovich transformation to factorize each of the color-octet LDMEs in terms of the wave function at the origin and a vacuum correlator of either chromo-electric or chromo-magnetic fields. Our analysis shows that these results permit equivalence relations between the color-octet LDMEs for different S-wave vector quarkonium LDMEs, first derived in ref.~\cite{Brambilla:2022ayc}.  Additionally, we find novel operator contributions to the color-octet P-wave mechanism and argue that these may be relevant for explaining discrepancies between data and NRQCD predictions at small $p_T$. We further power-counted the fields in the vacuum correlators and found that the LDMEs obey a slightly different scaling than what is expected in traditional NRQCD. We discussed potential ultrasoft power corrections to the results derived here and provide  arguments as to why these may be negligible. Finally, we apply these techniques to the recently derived TMDSTFs, which are the dominant operators for quarkonium production in SIDIS in the TMD framework \cite{Copeland:2025vop}. The factorized results for the TMDSTFs are perhaps even more critical than those derived for the LDMEs, because the TMDSTFs are process dependent and hence have a limited predictive ability on their own. However, using the techniques in this paper, we find a new state-independence property for the TMDSTFs in SIDIS, increasing the ability to constrain these non-perturbative production matrix elements for quarkonium production in the TMD framework. 

Moving forward these results can be used to constrain quarkonium production LDMEs in future analyses. Similar to what was done in ref.~\cite{Brambilla:2024iqg}, a global analysis of all S-wave vector quarkonium production data should be performed to constrain these universal matrix elements and the chromo-electric and chromo-magnetic field correlators. In this analysis, it will be useful to include $1/p_T^2$ power corrections for hadroproduction in the collinear factorization framework, which occur in the form of double parton fragmentation functions \cite{Fleming:2013qu, Lee:2021oqr, Kang:2014tta, Kang:2011mg}. Additionally, it may be necessary to explore additional power corrections in the $\vv$ expansion for the quarkonium production operators since it is known that these can be large for charmonium~\cite{Bodwin:1994jh, Braaten:2002fi}.

At small $p_T$, there are  many open questions about quarkonium production in the TMD framework. The $^3S_1^{[8]}$ and $^3P_J^{[8]}$ sub-leading color-octet TMDSTFs~\cite{Copeland:2025vop} still need to be derived for SIDIS and their TMD evolution kernels need to be computed. Moreover, the correct TMDSTFs and TMD factorization theorems should be derived to study quarkonium production at small $p_T$ in other processes, such as $e^+e^-$. The results from this paper will help to determine the appropriate non-perturbative production operators for the TMD formalism, which will be a critical effort if the gluon TMD PDFs in the proton are to be properly constrained.

Lastly, a number of theoretical models have been proposed to address the modification of quarkonium cross sections in hot and cold QCD matter, see~\cite{Brambilla:2010cs,Andronic:2015wma,Rothkopf:2019ipj,Boer:2024ylx,Arleo:2025oos} and references therein for discussion in the context of current and future measurements.     
At the fundamental level, interactions with a nuclear medium have been incorporated in the NRQCD Lagrangian via off-shell Glauber and Coulomb gluon exchanges~\cite{Makris:2019kap,Makris:2019ttx} and the leading and subleading corrections to vacuum dynamics were obtained for different momentum scaling of the medium quasi-particles. Phenomenological applications that take as input the dissociation of quarkonia based on broadening of the $Q\bar{Q}$ state in matter due to collisional interactions have shown promise in describing the hierarchy of nuclear attenuation of ground and excited $J/\psi$ and $\Upsilon$ states at intermediate and high transverse momenta/energies~\cite{Sharma:2012dy,Aronson:2017ymv,Boer:2024ylx}.        
This work opens new avenues to extend the first principles treatment of quarkonium production in reactions with nuclei to small transverse momenta and to address transitions between states based on microscopic dynamics.

All of these directions should be explored in future work.

\acknowledgments

The authors thank Sean Fleming, Duff Neill, Emanuele Mereghetti, and Ira Rothstein for helpful comments and discussions. M.C. and I.V. are supported by the U.S. Department of Energy through the Topical Collaboration in Nuclear Theory on Heavy-Flavor Theory (HEFTY) for QCD Matter under award no. DESC0023547. The work of M.C. and I.V. is also supported by the US Department of Energy through the Los Alamos National Laboratory. Los Alamos National Laboratory is operated by Triad National Security, LLC, for the National Nuclear Security Administration of the U.S. Department of Energy (Contract No. 89233218CNA000001). Research presented in this article was supported by the Laboratory Directed Research and Development program of Los Alamos National Laboratory under project numbers 20260377ER, 20240131ER, and 20240127ER.

\appendix

\section{Re-summing soft interactions in the vNRQCD Lagrangian }
\label{app: WAA prop}
\subsection{The $W_{AA}$ transformation}

In section \ref{sec: soft decouple}, we discuss a potential soft interaction with the composite fields, and argue that it vanishes for quarkonium production. However, this interaction could in principle contribute for processes with $r\neq 0$ and soft final states. In this appendix, we discuss this interaction further.  

At leading order, the interaction of a quark field with two soft gluons generates the amplitude in the non-relativistic limit,
\begin{equation}
\begin{aligned}
     \alpha_s(m\vv)\frac{1}{p^0 +\bar{q}_{us}-\frac{(\pv+\bar{\qv})^2}{2m}}  \frac{i}2 f^{abc} T^c U^{(0)}_{\mu\nu}(\qv, \qv') \epsilon^{\mu,a}(\qv) \epsilon^{\nu,b}(\qv') \sqrt{2m}\xi\\
     \to \alpha_s(m\vv)\sum_{\pv,\qv,\qv'}  \frac{i}2 f^{abc} T^c U^{(0)}_{\mu\nu}(\qv, \qv') A_\qv^{\mu,a} A_{\qv'}^{\nu,b} \frac{1}{i\overleftrightarrow{\partial_0} -\frac{\overleftrightarrow{{\bfcal P}}^2}{2m}} \psi_\pv , \\
\end{aligned}
\label{eq: DSG vertex}
\end{equation}
where $\bar{\qv} = (\qv - \qv')$ and the leading order coefficients, $U^{(0),ij}$, can be found in ref.~\cite{Manohar:1999xd}. Here, we define $\overleftrightarrow{\partial_0} = i\overrightarrow{\partial}_0 -i\overleftarrow{\partial_0}$ and $\overleftrightarrow{\bfcal P} = {\bfcal P} -{\bfcal P}^\dagger$. In the second line we have replaced the polarization vectors and spinors with the gluon and quark fields respectively, since this operator reproduces the interaction in the first line of eq.~(\ref{eq: DSG vertex}). Even though this contribution appears at ${\cal O}(\alpha_s)$, the contribution multiplying the quark scales as $\alpha_s(m\vv)/\vv \sim 1$ overall and hence is not suppressed. This means we can repeat the interaction an arbitrary number of times without penalty. Since gluons are bosons, we can sum over all permutations of the two gluon vertex and average over the number of insertions. This generates the all-orders structure 
\begin{equation}
    \sum_k \sum_{\rm perm.}\frac{1}{k!}\bigg(\frac{\alpha_s (m\vv)}{2}\bigg)^k \bigg( \prod_{j=1}^k\sum_{\qv_j,\qv_j'}   A_{\qv_{j}}^{\mu_j,a_j} A_{\qv'_j}^{\nu_j,b_j} \frac{if^{a_j b_j c_j}T^{c_j}  U^{(0)}_{\mu_j\nu_j}(\qv_j, \qv'_j)}{i\overleftrightarrow{\partial_0} -\frac{\overleftrightarrow{{\bfcal P}}^2}{2m}}\bigg)\psi_\pv .
\end{equation}
Define the combination that appears,
\begin{equation}
    W_{AA} = \sum_k \sum_{\rm perm.}\frac{1}{k!}\bigg(\frac{\alpha_s (m\vv)}{2}\bigg)^k \bigg( \prod_{j=1}^k \sum_{\qv_j, \qv_j'} A_{\qv_{j}}^{\mu_j,a_j} A_{\qv'_j}^{\nu_j,b_j}\frac{if^{a_j b_j c_j}T^{c_j}  U^{(0)}_{\mu_j\nu_j}(\qv_j, \qv'_j)}{i\overleftrightarrow{\partial_0} -\frac{\overleftrightarrow{{\bfcal P}}^2}{2m}}\bigg) ,
\end{equation}
as the operator encoding the all orders insertion of the double soft vertex. The $W_{AA}$ structure exponentiates 
\begin{equation}
\begin{aligned}
    W_{AA} =& \sum_{\rm perm.}\exp\Bigg[\sum_{\qv, \qv'}\frac{i\alpha_s (m\vv)}{2}    f^{a b c}T^{c}  U^{(0)}_{\mu\nu}(\qv, \qv') A_{\qv}^{\mu,a} A_{\qv'}^{\nu,b} \frac{1}{i\overleftrightarrow{\partial_0} -\frac{(\bfcal{P}-\bfcal{P}^\dagger)^2}{2m}}\Bigg]
\end{aligned}
\label{eq: WAA exp}
\end{equation}
and satisfies the properties
\begin{equation}
\begin{aligned}
    i{\partial}_\mu W_{AA} {\cal O} = \frac{\alpha_s(m\vv)}{2}\sum_{\qv,\qv'} & A_\qv^{\mu,a} A_{\qv'}^{\nu,b} \frac{i f^{abc} T^cU^{(0)}_{\mu\nu}(\qv, \qv')}{i\overleftrightarrow{\partial_0} -\frac{\overleftrightarrow{{\bfcal P}}^2}{2m}}  W_{AA}i\overleftrightarrow{\partial_\mu} {\cal O} ,\\
    f({\cal P}^\mu) W_{AA} {\cal O} =  \frac{\alpha_s(m\vv)}{2}\sum_{\qv,\qv'}  &A_\qv^{\mu,a} A_{\qv'}^{\nu,b} \frac{i f^{abc} T^cU^{(0)}_{\mu\nu}(\qv, \qv')}{i\overleftrightarrow{\partial_0} -\frac{\overleftrightarrow{{\bfcal P}}^2}{2m}} W_{AA}f(\overleftrightarrow{{\cal P}}^\mu) {\cal O} ,
\end{aligned}
\label{eq: WAA prop}
\end{equation}
where $f({\cal P}^\mu)$ is some generic function of the projector. See the following section for more details regarding these properties. Likewise, for the antiquark field we that notice that  an interaction with the double gluon vertex gives the amplitude
\begin{equation}
\begin{aligned}
     \alpha_s(m\vv)\frac{1}{-p^0  +\bar{q}^0_{us}
     -\frac{(-\pv+\bar{\qv})^2}{2m}}  \frac{i}2 f^{abc} T^c U^{(0)}_{\mu\nu}(\qv, \qv') \epsilon^{\mu,a}(\qv) \epsilon^{\nu,b}(\qv') \sqrt{2m}\eta\\
     \equiv \alpha_s(m\vv)\sum_{\pv,\qv,\qv'}  \frac{i}2 f^{abc} T^c U^{(0)}_{\mu\nu}(\qv, \qv') A_\qv^{\mu,a} A_{\qv'}^{\nu,b} \frac{1}{i\overleftrightarrow{\partial_0} -\frac{(\bfcal{P}-\bfcal{P}^\dagger)^2}{2m}} \chi_{-\pv}.\\
\end{aligned}
\label{eq: }
\end{equation}
So, after inserting an arbitrary number of these interactions and summing over all permutations, we see the antiquark also gets a factor of $W_{AA}$.

Importantly, we find that the structure $W_{AA}$ is not a unitary operator (see app. \ref{app: WAA unit}). That is, 
\begin{equation}
    W_{AA}^\dagger W_{AA} \ne 1.
\end{equation}
This is in stark contrast to Wilson lines, which are necessarily unitary operators. The non-unitarity of $W_{AA}$ occurs because soft-interactions change the label momentum of the quark fields and the non-relativistic quark propagators do not eikonalize. 
This means that we cannot field redefine this interaction away. In other words, BPS-type field redefinitions, as in sec. \ref{sec: decouple}, do not work here.

To show what we mean, consider the following field redefinition
\begin{equation}
    \psi_\pv \to W_{AA} \tilde{\psi}_\pv, ~~~~~~~~~ \chi_{-\pv} \to W_{AA} \tilde{\chi}_{-\pv} . 
\label{eq: WAA field redef}
\end{equation}
Under this transformation, the kinetic term in the vNRQCD Lagrangian becomes
\begin{equation}
\begin{aligned}
   \sum_\pv \tilde{\psi}^\dagger_\pv W_{AA}^\dagger\bigg(i\partial_0 -\frac{\bfcal{P}^2}{2m}\bigg)W_{AA}\tilde{\psi}_\pv.\\
\end{aligned}
\end{equation}
Using the properties in eq.~(\ref{eq: WAA prop}), we can combine this with the leading order double soft gluon interaction and that the combination reduces to
\begin{equation}
\begin{aligned}
  &\sum_\pv \tilde{\psi}^\dagger_\pv W_{AA}^\dagger\bigg(i\partial_0 -\frac{\bfcal{P}^2}{2m} - \alpha_s(m\vv)\sum_{\qv,\qv'}  \frac{i}2 f^{abc} T^c U^{(0)}_{\mu\nu}(\qv, \qv') A_\qv^{\mu,a} A_{\qv'}^{\nu,b}\bigg) W_{AA}\tilde{\psi}_\pv\\
  =& \sum_\pv \tilde{\psi}^\dagger_\pv W_{AA}^\dagger\bigg(\alpha_s(m\vv)\sum_{\qv,\qv'} \frac{i}2f^{abc} T^c U^{(0)}_{\mu\nu}(\qv, \qv') A_\qv^{\mu,a} A_{\qv'}^{\nu,b} \frac{1}{i\overleftrightarrow{\partial_0} -\frac{(\bfcal{P}-\bfcal{P}^\dagger)^2}{2m}} \bigg\{i\overleftrightarrow{\partial_0}-\frac{(\bfcal{P}-\bfcal{P}^\dagger)^2}{2m}\bigg\}\\
  &\times W_{AA} - \alpha_s(m\vv)\sum_{\qv,\qv'}  \frac{i}2 f^{abc} T^c U^{(0)}_{\mu\nu}(\qv, \qv') A_\qv^{\mu,a} A_{\qv'}^{\nu,b}W_{AA}+W_{AA}\bigg\{i\partial_0 -\frac{\bfcal{P}^2}{2m}\bigg\}\bigg) \tilde{\psi}_\pv\\
  &=\sum_\pv  \tilde{\psi}^\dagger_\pv \bigg[W_{AA}^\dagger W_{AA}\bigg]\bigg(i\partial_0 -\frac{\bfcal{P}^2}{2m}\bigg)\tilde{\psi}_\pv .
\end{aligned}
\end{equation}

Our calculations show that, despite simplifying to some degree, the interaction does not completely decouple in the Lagrangian.
The field definitions in eq.~(\ref{eq: }), modifies the leading terms interaction terms with the composite fields from the Hubbard-Stratonovich transformation of section \ref{sec: decouple}.
\begin{equation}
\begin{aligned}
    {\cal L}_c =&  \int d^3\rv V^{(1)}(r) \bigg(S_\rv^\dagger S_\rv - S^\dagger_\rv \sum_\pv e^{i\pv\cdot\rv} \frac{1}{\sqrt{N_c}}\big[\tilde{\chi}^\dagger_{-\pv} W_{AA}^\dagger \big] \big[W_{AA} \tilde{\psi}_\pv\big] \\
    &- S_\rv \sum_\qv e^{-i\qv \cdot \rv} \frac{1}{\sqrt{N_c}} \big[\tilde{\psi}^\dagger_\qv W_{AA}^\dagger \big]\big[W_{AA} \tilde{\chi}^\dagger_{-\qv}\big]\bigg) + {\rm ~octet ~ fields} ,
\end{aligned}
\label{eq: WAA Lagrangian}
\end{equation}
where we have omitted the $W_\rv$ Wilson lines for clarity. They can be easily restored using reparameterization invariance. Again, the $W_{AA}^\dagger W_{AA}$ factors product does not yield unity. This suggests that these factors should appear as a new interaction in the pNRQCD Lagrangian, which can be derived by integrating out the quarks and antiquarks.

\subsection{Unitarity violation for $W_{AA}$}
\label{app: WAA unit}
In this section we illustrate why the $W_{AA}$ operator is not unitary. The conjugate of $W_{AA}$ can be easily defined by taking 
\begin{equation}
\begin{aligned}
    (W_{AA})^\dagger =&\bigg( \sum_k \sum_{\rm perm.}\frac{1}{k!}\bigg(\frac{i\alpha_s (m\vv)}{2}\bigg)^k  \prod_{j=1}^k \sum_{\qv_j, \qv_j'} f^{a_j b_j c_j}T^{c_j}  U^{(0)}_{\mu_j\nu_j}(\qv_j, \qv'_j) A_{\qv_{j}}^{\mu_j,a_j} A_{\qv'_j}^{\nu_j,b_j} \\ & \qquad \qquad \qquad \qquad \qquad \qquad \qquad 
    \frac{1}{i\overleftrightarrow{\partial_0} -\frac{(\bfcal{P}-\bfcal{P}^\dagger)^2}{2m}}\bigg)^\dagger\\
    =& \sum_k \sum_{\rm perm.}\frac{1}{k!}\bigg(\frac{i\alpha_s (m\vv)}{2}\bigg)^k  \prod_{j=1}^k \sum_{\qv_j, \qv_j'} f^{a_j b_j c_j}T^{c_j}  U^{(0)}_{\mu_j\nu_j}(\qv_j, \qv'_j) \\ & 
    \qquad \qquad \qquad \qquad \qquad \qquad \qquad 
     \bigg( \frac{1}{i\overleftrightarrow{\partial_0} -\frac{(\bfcal{P}-\bfcal{P}^\dagger)^2}{2m}}\bigg)^\dagger A_{\qv_{j}}^{\mu_j,a_j} A_{\qv'_j}^{\nu_j,b_j}  .\\
\end{aligned}
\end{equation}
where we've used the fact that the seagull interaction must be Hermitian in order for the vNRQCD Lagrangian to be unitary
\begin{equation}
    \Bigg( \alpha_s(m\vv)\sum_{\qv,\qv'}  \frac{i}2 f^{abc} T^c U^{(0)}_{\mu\nu}(\qv, \qv') A_\qv^{\mu,a} A_{\qv'}^{\nu,b} \Bigg)^\dagger =  \alpha_s(m\vv)\sum_{\qv,\qv'}  \frac{i}2 f^{abc} T^c U^{(0)}_{\mu\nu}(\qv, \qv') A_\qv^{\mu,a} A_{\qv'}^{\nu,b} 
\end{equation}
From here, it is convenient to exponentiate the operator so that
\begin{equation}
\begin{aligned}
    W_{AA} =& \sum_{\rm perm.}\exp\Bigg[\sum_{\qv, \qv'}\frac{i\alpha_s (m\vv)}{2}    f^{a b c}T^{c}  U^{(0)}_{\mu\nu}(\qv, \qv') A_{\qv}^{\mu,a} A_{\qv'}^{\nu,b} \frac{1}{i\overleftrightarrow{\partial_0} -\frac{(\bfcal{P}-\bfcal{P}^\dagger)^2}{2m}}\Bigg] ,\\
    W_{AA}^\dagger =& \sum_{\rm perm.}\exp\Bigg[  \sum_{\qv, \qv'}\frac{i\alpha_s (m\vv)}{2}    f^{a b c}T^{c}  U^{(0)}_{\mu\nu}(\qv, \qv') \frac{1}{-i\overleftrightarrow{\partial_0} -\frac{(\bfcal{P}^\dagger-\bfcal{P})^2}{2m}} A_{\qv}^{\mu,a} A_{\qv'}^{\nu,b}\bigg] .\\
\end{aligned}
\label{eq: WAA exp}
\end{equation}
Pairing $W_{AA}$ with on-shell heavy quarks, this reduces to
\begin{equation}
\begin{aligned}
    &W_{AA} \tilde{\psi}_\pv \\
    &= \sum_{\rm perm.}\exp\Bigg[\sum_{\qv, \qv'}\frac{i\alpha_s (m\vv)}{2}    f^{a b c}T^{c}  U^{(0)}_{\mu\nu}(\qv, \qv') A_{\qv}^{\mu,a} A_{\qv'}^{\nu,b} \frac{1}{\bar{q}_{us}^0  -\frac{2\bar{\qv}\cdot \pv +\qv^2}{2m}}\Bigg]\tilde{\psi}_\pv , \\
    &\tilde{\psi}^\dagger_\pv W_{AA}^\dagger \\
    &= \sum_{\rm perm.}\tilde{\psi}_\pv^\dagger \exp\Bigg[ \sum_{\qv, \qv'}\frac{-i\alpha_s (m\vv)}{2}  f^{ab c}T^{c}   U^{(0)}_{\mu\nu}(\qv, \qv') A_{\qv}^{\mu,a} A_{\qv'}^{\nu,b}  \frac{1}{-\bar{q}_{us}^0  -\frac{-2\bar{\qv}\cdot \pv +\qv^2}{2m}}  \bigg] , \\
\end{aligned}
\end{equation}
where we have applied the derivatives and projectors on the fields and used $E - \pv^2/(2m) = 0$ for the on-shell non-relativistic quark fields. Clearly the factors in the arguments of $W_{AA}$ are not equivalent so we have
\begin{equation}
\begin{aligned}
    \sum_\pv \big[ \tilde{\psi}^\dagger_\pv W_{AA}^\dagger \big]\big[ W_{AA} \tilde{\psi}_\pv \big] \neq \sum_\pv \tilde{\psi}^\dagger_\pv \tilde{\psi}_\pv ,
\end{aligned}
\end{equation}
This demonstrates that $W_{AA}$ is not a unitary operator.

Interestingly, the non-unitarity can be traced to the fact that the propagator in the argument of $W_{AA}$ does not eikonalize. To see what we mean, compare with a normal time-like Wilson line,
\begin{equation}
    W_v = \sum_{\rm perm} \exp\bigg[\sum_\qv \frac{1}{v\cdot {\cal P}} A_\qv\bigg]
\end{equation}
Notice, under a dagger, the factor in the argument of the exponential becomes,
\begin{equation}
    \bigg(\sum_q \frac{1}{v\cdot {\cal P}} A^0_q\bigg)^\dagger = \sum_q\frac{1}{q^0}A_{-q}^0 = \sum_q \frac{1}{-q^0} A_q^0 = \sum_q A^0_q \frac{1}{v\cdot {\cal P}^\dagger}
\end{equation}
In this case, we say that the propagator from the off-shell quark has ``eikonalized", and under a dagger, it is allowed to pick up a minus sign. This means that the time-like Wilson line satisfies
\begin{equation}
    W_v W_v^\dagger = 1.
\end{equation}
%

%

\section{Summary of transitions}
Quark-antiquark pairs in color-octet configurations need additional operator insertions to flip their quantum numbers to match the final state quarkonium. Only at this point the heavy quarks can form a bound state via Coulomb gluon exchange. For S-wave vector quarkonium particles, we need to consider the subleading color-octet mechanisms their respective transitions. These are summarized in table~\ref{tab: trans}.

\begin{table}[htbp]
\centering
\begin{tabular}{|c|c c|}
    \hline
     LDME & Quark-antiquark initial state & Type of transition \\
     \hline
     $\Oosz$ & $\chi^\dagger_{-\qv} T^a \psi_\qv$ & Chromo-magnetic dipole\\
     \hline
     $\Otsooct$ & $\chi^\dagger_{-\qv} \sigma^i T^a \psi_\qv$ & \begin{minipage}{5cm}\vspace{.15cm}Double chromo-electric field\\ or \\Double chromo-electric dipole \end{minipage}\\
     \hline 
     $\Otpz$ & $\chi^\dagger_{-\qv} \qv^i \sigma^j T^a \psi_\qv$ & Single chromo-electric dipole \\     
     \hline
\end{tabular}
\caption{ Types of transitions necessary for to place the color-octet $Q\bar{Q}$ pairs in a $^3S_1^{[1]}$ configuration. }
\label{tab: trans}
\end{table}

The fields representing these emissions and their scalings estimates when the fields are soft and ultrasoft are defined in table \ref{tab: Dipole fields}. Note that to estimate the final scaling of the long distance matrix element one needs to not only consider the additional $\vv$ suppression from the chromo-magnetic and chromo-electric field insertions, but also consider the propagators between the vertices and the emissions. If the gluon is soft, the quark/antiquark propagator is knocked offshell and gets replaced with a factor of $1/v\cdot {\cal P}$ which scales like $1/\vv$. If the gluon is ultrasoft, then the quark/antiquark propagator scales like $1/\vv^2$.

\begin{table}[htbp]
\centering
\begin{tabular}{|c|c c c |}
    \hline
     & Operator & U-soft  & Soft \\
    \hline
     Chromo-magnetic dipole & $g \boldsymbol{\sigma} \cdot {\bfcal B}$ & $\vv^4$ & $\vv^2$\\
     \hline
     Chromo-electric dipole & $g \qv \cdot {\bfcal E}$ & $\vv^3$ & $\vv^2$ \\
     \hline
     Double chromo-electric field & $g^2 d^{abc} {\bfcal E}^a{\bfcal E}^b$ & $\vv^4$ & $\vv^2$\\
     \hline
     Quark/antiquark propagator & & $1/\vv^2$ & $1/\vv$\\
     \hline
\end{tabular}
\caption{Scaling of the transition operators.}
\label{tab: Dipole fields}
\end{table}

\section{Other soft contributions to P-waves}
\label{sec: subleading P-wave}
In sec.~\ref{sec: P-wave} we saw that the $\Otpz$ matched onto multiple operators in vNRQCD, but only one contributed to S-wave vector quarkonium production at leading order in $\vv$. The other operators were suppressed because they required further transitions to be put in a $^3S_1^{[1]}$ configuration. The operator 
\begin{equation}
    \sum_{\qv'} \psi^\dagger_{\qv'} \frac12{\bfcal P}^i \sigma^j T^b \chi_{-\qv'} S_v^{ab}
\label{eq: vNRQCD true Pwave}
\end{equation}
is particularly interesting because this is the only contribution in eq.~(\ref{eq: Pwave contrs}) that is proportional to the relative momentum of the $Q\bar{Q}$. Therefore, it can be interpreted as a genuine P-wave state at the scale $\mu \sim m\vv$ and will play a greater role in P-wave quarkonium production. As an exploratory effort, in this section we derive the transitions necessary for this operator to contribute to S-wave vector quarkonium production, which is the main topic of this paper. This effort will explicitly show that its contributions are formally subleading with respect to the dominant operator identified in sec.~\ref{sec: P-wave}, but also that there is a new relation between matrix elements.

\subsection{Tom's RPI technique for P-waves}
Since eq. (\ref{eq: vNRQCD true Pwave}) is in a true P-wave state, by conventional wisdom, it must flip to a $^3S_1^{[8]}$ via a chromo-electric dipole emission. In ref.~\cite{Fleming:2019pzj}, it was shown that gauge invariant chromo-electric dipole transitions can quickly be derived using reparameterization invariance. This is because these operators contain information about the small relative momentum of the quark/antiquark pair which can be related to the initial velocity of the heavy quarks via a reparameterization of the form,
\begin{equation}
   v^\mu \to  v_{\pm}^\mu = v^\mu \pm \frac{q^\mu}{2m} + {\cal O}\bigg(\frac{q^2}{2m}\bigg).
\end{equation}
Under this transformation, we can define new fields in vNRQCD that satisfy
\begin{equation}
\begin{aligned}
    \slashed{v}_\pm \psi_{\qv,\pm } =& \psi_{\qv, \pm} , \\
    \slashed{v}_\pm \chi_{-\qv,\pm } =& -\chi_{-\qv,\pm} , \\
    v_\pm \cdot D S_{v, \pm}  =&0.
\end{aligned}
\end{equation}
The P-wave matching to these new fields generates an operator with the same form as
\begin{equation}
    \psi^\dagger_{\qv,+}S_{v, +}^\dagger \qv \cdot \Gamma S_{v, -} \chi_{-\qv,-}.
\label{eq: RPI p-wave}
\end{equation}
By solving for the new fields in terms of the old ones~\cite{Fleming:2019pzj}
\begin{equation}
\begin{aligned}
    \psi_{\qv,\pm} = \psi_\qv \pm \frac{\slashed{q}}{4m} \psi_\qv , \\
    \chi_{-\qv, \pm} = \chi_{-\qv} \mp \frac{\slashed{q}}{4m} \chi_{-\qv} , \\
    S_{v,+} = S_v \mp\frac{g}{2m} S_v \frac{1}{v\cdot  \Pc}\qv \cdot {\bfcal E} , 
\end{aligned}
\end{equation}
we can derive the subleading operators in the original theory that are related to eq.~(\ref{eq: P-wave def}) by RPI
\begin{equation}
\begin{aligned}
    \psi^\dagger_{\qv,+}S_{v, +}^\dagger \qv \cdot \Gamma S_{v, -} \chi_{-\qv,-} = & ~\psi^\dagger_{\qv}S_v^\dagger \qv \cdot \Gamma S_v \chi_{-\qv}-\frac{\qv^i }{4m}\psi^\dagger_{\qv}S_v^\dagger \{{\boldsymbol \gamma}\cdot \qv,\Gamma \}S_v \chi_{-\qv} \\
    & + \frac{g}{2m} \psi^\dagger_{\qv}\{S_v^\dagger \qv \cdot \Gamma  S_v , \frac{1}{v\cdot  \Pc}\qv \cdot {\bfcal E}\}\chi_{-\qv}.
\end{aligned}
\label{eq: RPI operators}
\end{equation}
The first term is just the leading order operator while the last two are subleading in the power-counting. While both of the subleading operators in eq.~(\ref{eq: RPI operators}) can flip the P-wave into an S-wave, only the last operator can bring the state to a color singlet. Therefore, we identify the last term in eq.~(\ref{eq: RPI operators}) as the dominant transition operator. We can easily recognize this contribution as a gauge invariant chromo-electric dipole insertion on the leading P-wave operator, as we expected. 

For a generic $^3P_J^{[8]}$ configuration, we can replace the vertex structure with 
\begin{equation}
    \qv\cdot \Gamma \to  \sigma^i \qv^j T^a ,
\end{equation}
where contracting the cartesian indices with $\delta^{ij}$, antisymmetric, and symmetric traceless tensors will give the vertices for the $^3P_0^{[8]}$, $^3P_1^{[8]}$, and $^3P_2^{[8]}$ configurations, respectively. From here, we derive the leading transition operator in vNRQCD by projecting out the $^3S_1^{[1]}$ component in eq.~(\ref{eq: RPI operators})

\begin{equation}
    \frac{g}{2m}\psi^\dagger_{\qv}P_{^3S_1^{[1]}}\big[\{S_v^\dagger \qv^j \sigma^i T^a  S_v , \frac{1}{v\cdot  \Pc}\qv \cdot {\bfcal E}\}\big]\chi_{-\qv} = \frac{g}{2N_c m}\psi^\dagger_{\qv}  \sigma^i \qv^j \qv^k\bigg[ {\cal S}_v^{ab} \frac{1}{v\cdot  \Pc}{\bfcal E}^{k,b}\bigg]\chi_{-\qv}.
\end{equation}

\subsection{Factorizing the subleading P-wave contribution}
We begin by considering the vacuum matrix element squared of the operator derived in the previous section. This is given by
\begin{equation}
\begin{aligned}
    \Otpz^{(1)} =  \frac{g^2}{4N_c^2 m^2}{\cal T}^{ij,i'j'}_J\sum_X & \bra{0}\chi^\dagger_{-\qv}\sigma^i \qv^j \qv^k \bigg[S_v^{\dagger,ab}\frac{1}{v\cdot  \Pc}{\bfcal E}^{k,b}\bigg] \psi_\qv\ket{V,X}\\
  &  \times\bra{V,X}\psi^\dagger_{\qv}  \sigma^{i'} \qv^{j'} \qv^{k'}\bigg[ {\cal S}_v^{ab'} \frac{1}{v\cdot  \Pc}{\bfcal E}^{k',b'}\bigg]\chi_{-\qv}\ket{0}, 
\end{aligned}
 \end{equation}
and the ${\cal T}^{ij,i'j'}$ projectors are defined in eq.~(\ref{eq: Pwave projectors})
Inserting a color singlet composite field interaction and integrating out the quark loops, we find the subleading operator reduces to 
  \begin{equation}
\begin{aligned}
    \Otpz^{(1)} =  \frac{g^2}{4N_c m^2}{\cal T}^{ij,i'j'}_J\sum_X & \bra{0} \bigg[{\cal S}_v^{\dagger,ab}\frac{1}{v\cdot  \Pc}{\bfcal E}^{k,b}\bigg] \big[\sigma^i \qv^j \qv^k S^\dagger_0(x)\big] \ket{V,X}\\
   &  \times\bra{V,X}\big[ S_0(x) \sigma^{i'} \qv^{j'} \qv^{k'}\big]\bigg[ {\cal S}_v^{ab'} \frac{1}{v\cdot  \Pc}{\bfcal E}^{k',b'}\bigg]\ket{0}.
\end{aligned}
 \end{equation}
Factorizing the Hilbert space,
\begin{equation}
\begin{aligned}
    \Otpz^{(1)} =  \frac{g^2}{36N_c m^2} {\cal T}^{ij,i'j'}_J & \bra{0} \big[\sigma^i \qv^2 S^\dagger_0(x)\big] \ket{V}\bra{V}\big[ S_0(x) \sigma^{i'} \qv^2\big]\ket{0}\\
    &\times\bra{0}  \bigg[ {\cal S}_v^{ab} \frac{1}{v\cdot  \Pc}{\bfcal E}^{j,b}\bigg]\bigg[ {\cal S}_v^{ab'} \frac{1}{v\cdot  \Pc}{\bfcal E}^{j',b'}\bigg]\ket{0} ,
\end{aligned}
 \end{equation}
where we have averaged over the Cartesian indices in the first line. Notice that, unlike for the S-wave factorization, we find that the result is proportional to composite field matrix elements with two powers of the relative momentum. These matrix elements are not related to the wave function at the origin, but instead can be identified as \cite{Bodwin:1994jh}
\begin{equation}
    \bra{0} \qv^2 [\sigma^i S_0^\dagger(x)]\ket{V } = \sqrt{\frac{1}{2\pi}
    }\overline{\nabla^2R}_V^{(0)}(0) ,
\label{eq: Wave func deriv}
\end{equation}
where the overline indicates that the wave function has been renormalized to subtract off linear divergences as $r\to 0$. The exact definition in terms of the derivative on the wave function is slightly ambiguous due to the aforementioned divergences, however the point here is that the matrix element in eq.~(\ref{eq: Wave func deriv}) is power suppressed with respect to the leading order wave function at the origin. In fact, the matrix element in eq.~(\ref{eq: Wave func deriv}) can be estimated using the Gremm-Kapustin relation \cite{Gremm:1997dq},
\begin{equation}
    \overline{\nabla^2R}_V^{(0)}(0) \approx m^2\braket{\vv^2}_V R_V(0) + {\cal O}(\vv^2)., 
\end{equation}
where we defined the parameter $\braket{\vv^2}_V$ which is related to the binding energy,
\begin{equation}
    m\braket{\vv^2}_V =  (M_{V}-2m) ,
\end{equation}
to make the $\vv^2$ suppression of the operator apparent.

Therefore, after averaging over spins and summing over polarizations, we can write the factorized P-wave matrix element as,
\begin{equation}
    \Otpz^{(1)} = \frac{(m \braket{\vv^2}_V)^2}{108 N_c} (2J+1) \frac{3\big|R_V(0)\big|^2}{2\pi} \braket{{\bfcal E}}^{(1)} ,
\label{eq: 3PJ fact}
\end{equation}
with the definition
\begin{equation}
    \braket{{\bfcal E}}^{(1)} = \bra{0} \bigg[ {\cal S}_v^{ab} \frac{1}{v\cdot  \Pc}{\bfcal E}^{j,b}\bigg]^2\ket{0}.
\end{equation}
%
%
Equation (\ref{eq: 3PJ fact}) is distinct from the result in sec.~\ref{sec: P-wave} (first derived in ref.~\cite{Brambilla:2022ayc}). Notably, this result is suppressed by a factor of $\vv^2$ with respect to the dominant contribution, as expected. 

\subsection{$^3S_1^{[8]}$ and $^3P_J^{[8]}$ mixing}
\label{sec: subleading P-wave RG}
One may wonder if the leading contribution in eq.~(\ref{eq: Pwave contrs}) is the correct operator for the $\Otpz$ to be matched onto, since this contribution is no longer proportional to the relative momentum of the $Q\bar{Q}$. Depending on what one may identify as a ``true" P-wave state, there may be some ambiguity in the matching.

One argument in support of such matching is as follows: consider the evolution equations for $\Otsooct$. To derive the evolution, we extract the UV divergences from the double electric field correlator defined in sec.~\ref{sec: 3S18}. Using the same prescription as in ref.~\cite{Brambilla:2022ayc}, this is done by treating a subset of the fields as perturbative and leaving the remaining fields in a non-perturbative correlator. We find the UV divergent piece to be
\begin{equation}
\begin{aligned}
    \braket{{\bfcal{EE}}}&=g^2d^{g'bc'} d^{gbc} \int\frac{d^D\qv}{(2\pi)^D} \frac{1}{2E_q} \frac{1}{\qv^2} (\delta^{ij} -\frac{\qv^i\qv^j}{E_q^2})\bra{0} \big|  {\cal S}_v^{ag} g {\bfcal E}^{c}\big|^2(x)\ket{0}\\
    &= \frac{4\alpha_s}{3\pi}B_F \bigg(\frac{1}{\epsilon_{UV}}- \frac{1}{\epsilon_{IR}}\bigg) \bra{0} \big|  {\cal S}_v^{ac} g {\bfcal E}^{c}\big|^2(x)\ket{0}, 
\end{aligned}
\end{equation}
with $B_F = (N_c^2 -4)/(4N_c)$. Here, we have used the implicit sum over labels for the electric fields in the correlator's definition to cancel the factors of $1/2$ and evaluated the integral using dimensional regularization in $D= 3-2\epsilon$ dimensions. Therefore, the renormalization group (RG) equation for $\braket{{\bfcal{EE}}}$ is given by
\begin{equation}
    \frac{d}{d\log\mu} \braket{{\bfcal{EE}}} =  \frac{4\alpha_s}{3\pi}B_F  \bra{0} \big|  {\cal S}_v^{ac} g {\bfcal E}^{c}\big|^2(x)\ket{0}.
\end{equation}
Notice that the correlator that appears here is exactly the electric correlator from sec.~\ref{sec: P-wave}, i.e. the result from using the leading operator in eq.~(\ref{eq: Pwave contrs}), not the subleading operator proportional to the relative momentum of the $Q\bar{Q}$. Moreover, multiplying by the appropriate factors, we observe the standard RG equation for the $\Otsooct$ \cite{Echevarria:2020qjk, Fleming:2019pzj, Petrelli:1997ge, Brambilla:2022ayc}
\begin{equation}
    \frac{d}{d\log\mu} \Otsooct=  \frac{24\alpha_s}{\pi}B_F  {\braket{^3P_0^{[8]}}}.
\end{equation}
This can only be reproduced if we use the leading operator matching described in sec.~\ref{sec: P-wave}.  Again, we note that the subleading operator from app.~\ref{sec: subleading P-wave} will be more significant for the production operators for P-wave vector quarkonium states, such as $\chi_{cJ}$ and $\chi_{bJ}$ states. 

\bibliographystyle{JHEP}
\bibliography{biblio.bib}

\end{document}